\documentclass[aps,prd,twocolumn,groupedaddress,floatfix,showpacs,amsmath,amssymb]{revtex4}

\usepackage{graphicx}
\usepackage{bm}
\usepackage{dcolumn}

\begin{document}


\title{Changing universe model with applications }


\author{John C. Hodge}
\email[]{jch9496@blueridge.edu}
\affiliation{Blue Ridge Community College, 100 College Dr., Flat Rock, NC, 28731-1690}


\date{\today}

\begin{abstract}
Although the current galaxy models yield calculations consistent with much of the data, many irregularities exist, exceptions have been found to the current models, the $\Lambda$CDM model apparently fails on galaxy scales, dark matter remains elusive, the phenomena at the center of galaxies are only beginning to be addressed, and many observations and empirical relationships are unexplained.  A changing universe model (CUM) is proposed that posits the stuff of our universe is continually erupting into our universe from sources at the center of galaxies.  In this first test of the CUM, a single equation that describes the rotation velocity curve of spiral galaxies is derived.  The equation is also used to correlate the measured mass of the theorized, central supermassive black hole with other galaxy parameters.  The equation adds one parameter for each mature galaxy and one parameter for each particle (atom nucleus and smaller) species to Newtonian dynamics.  Rising, flat, and declining rotation velocity curves are explained without unknown dark matter.  Ten previously unknown parametric relationships are discovered.  Also, the CUM integrates several heretofore unrelated observations.  The CUM model is in an early development stage and has already predicted new relationships among galaxy parameters.

\end{abstract}

\pacs{98.80.Bp,98.62.Ai,98.62.Dm, 98.62.Js }

\maketitle

\section{INTRODUCTION}

The discrepancy between the Newtonian estimated mass in galaxies and galaxy clusters and the observation of star and gas kinematics is well established.  Currently, the possible explanations for this discrepancy are that a large mass $M_\mathrm{DM}$ of dark matter (DM) and dark energy exists (the $\Lambda$CDM model) or that Newtonian gravity is modified on galactic scales.  The current $\Lambda$CDM model of the structure of the universe appears successful on large scales~\cite{bahc}, but galactic scale predictions disagree with observations~\cite{bell,bosc,sell}.  The most popular modified gravity model is proposed by the Modified Newtonian Dynamics (MOND) model~\cite[and references therein]{bott}.  MOND suggests gravitational acceleration changes with galactic distance scales.  MOND appears limited to only the disk region of spiral galaxies and appears to have two falsifiers~\cite{bott}.  How MOND may be applied to cosmological scales is unclear.  However, MOND may represent an \emph{effective} force law arising from a broader force law~\cite{mcga}.  The evidence suggest the problem of a single model explaining both galactic scale and cosmological scale observations is fundamental~\cite{sell}.

An alternate explanation is the possibility there exists another force originating from the center of galaxies.  This proposal, called the changing universe model (CUM), posits the stuff of our universe is matter and a matterless scalar $\rho$ whose gradient exerts a force $\bm{F}_\mathrm{s}$ on the cross section of matter.  The $\bm F_\mathrm{s}$ is outward from galactic center.  Thus, $\bm F_\mathrm{s}$ is a repulsive force in a galaxy relative to gravity.  However, it differs from a gravity type force since it acts on the cross section of matter and it is matterless. 

The rotation velocity $v$ (km$\,$s$^{-1}$) of a particle in the plane of a spiral galaxy (hereinafter galaxy) reflects the forces acting on the particle.  An explanation of $v^2$ as a function of radius $r$ (kpc) from the center of a galaxy (RC) requires a knowledge of galaxy dynamics and an explanation of radically differing slopes.  The $v$ and, thus, the RC is measured along the major axis.  Although $v$ is non-relativistic, the result of calculations using RC models are used in cosmological calculations.  Thus, the RC is an excellent tool to evaluate galaxy models.  Further, a complete explanation of the RC must also explain black holes and observations of the center of galaxies.

\begin{figure}
\includegraphics[width=0.4\textwidth]{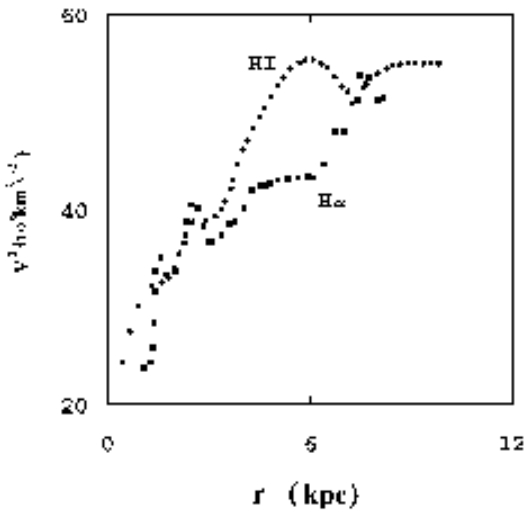}
\caption{\label{fig:0}Plot of the H{\scriptsize{I}} (filled diamonds) and H$\alpha$ (filled squares) RC for NGC4321. }
\end{figure}

\citet{batt} and \citet{sofu} provides an overview of the current state of knowledge of RC's in the outer bulge and disk regions of galaxies.  The particles most often measured in the disk region of a galaxy are hydrogen gas by H{{\scriptsize{I}}} observation and stars by observing the H$_\mathrm{\alpha}$ line.  The particles being measured in the inner bulge region are stars by observation of the of H$_\mathrm{\alpha}$, CO, and other spectral lines.  Also, the RC differs for different particles.  For example, although the H{\scriptsize{I}} and H$_\mathrm{\alpha}$ RCs for NGC4321 \cite{semp} in Fig.~\ref{fig:0} differ in the outer bulge, they approach each other in the outer disk region.  Most models of RC monitor the H{\scriptsize{I}} in the outer bulge and disk regions and ignore the differing H$_\mathrm{\alpha}$ RC.  Also, when drawing the H{\scriptsize{I}} RC, a straight line is often drawn through the center.  This practice is inconsistent with stellar observations~\cite{scho}.

The observation of rising RCs in the outer bulge and rising, flat, or declining RC's in the disk region of galaxies poses a perplexing problem.  If most of the matter of a galaxy is in the bulge region, classical Newtonian mechanics predicts a Keplerian declining RC in the disk.  

Currently, a convention on how to calculate error (e.g. random or systematic) in RCs is lacking in the literature.  One popular method is based on the assumption of symmetry.  This practice of averaging the $v$ of the approaching and receding sides of the RC is ignoring important information~\cite{conse}.  Asymmetry appears to be the norm rather than the exception~\cite{jog}.  The RC of all nearby galaxies for which the kinematics have been studied have asymmetry~\cite{jog}.  At larger distance, objects appear to become more symmetric as they become less well-resolved (H{\scriptsize{I}} resolution is typically 15 arcsec to 50 arcsec)~\cite{conse}.

\citet{hodg2} found a correlation between the size and distance of neighboring galaxies and the asymmetry in the disk of the H{\scriptsize{I}} RC.  The galaxies studied with minimal asymmetry also had rising RCs.  Therefore, rising RCs are intrinsic to galaxies.

A strong correlation between the core radius $R_c$ and the stellar exponential scale length $R_d$ was found by \citet{dona}.  The $R_d$ is derived from photometric measurements.  The $R_c$ is obtained from kinematic measurements.  If this relationship holds for both high surface brightness (HSB) galaxies, low surface brightness (LSB) galaxies, galaxies with rising RCs, and galaxies with declining RCs, modeling such a relationship is very difficult using current galaxy models.

\citet{ghez} and \citet{fer5} have observed Keplerian motion to within one part in 100 in elliptical orbits of stars that are from less than a pc to a few 1000 pc from the center of galaxies. 

The orbits of stars within nine light hours of the Galaxy center indicates the presence of a large amount of mass within the orbits~\cite{ghez, ghez3, ghez4, scho, dunn}.  To achieve the velocities of 1300 km$\,$s$^{-1}$ to 9000 km$\,$s$^{-1}$~\cite{scho} and high accelerations~\cite{ghez}, there must be a huge amount of very dense particles such as millions of black holes, dense quark stars~\cite[and references therein]{pras}, and ionized iron~\cite{wang} distributed inside the innermost orbit of luminous matter.  The mass $M_\mathrm{\bullet}$ (in M$_\odot$) within a few light hours of the center of galaxies varies from $10^6 $ M$_\mathrm{\odot}$ to $10^{10}$ M$_\mathrm{\odot}$~\cite{fer2,gebh}.  The $M_\mathrm{\bullet}$ can be distributed over the volume with a density of at least $10^{12}$ M$_\mathrm{\odot}$ pc$^{-3}$~\cite{ghez,ghez2,ghez4,dunn}.  The orbits of stars closest to the center are approximately 1,169 times~\cite{scho} the Schwartschild radius of the theorized supermassive black hole (SBH)~\cite{ghez}.  However, such large mass crowded into a ball with a radius of less than 60 AU must either quickly dissipate, must quickly collapse into a SBH~\cite{korm, mago}, or there must exist a force in addition to the centrifugal force holding the mass from collapse.  \citet{moua} suggested there is some extended mass around Sgr A.  Also, models of supermassive fermion balls wherein the gravitational pressure is balanced by the degeneracy pressure of the fermions due to the Pauli exclusion principle are not excluded~\cite{bili}.  A strong repulsive force at the center of galaxies would explain the shells of shocked gas around galactic nuclei~\cite[page 595]{binn}\cite{koni}, the apparent inactivity of the central object~\cite{baga,baga2,naya,zhao}, and the surprising accuracy of reverberation mapping as a technique to determine $M_\mathrm{\bullet}$~\cite{merr2}.  However, discrepancies have been noted among the methods used to measure $M_\bullet$~\cite[and references therein]{gebh2,merr2}.

The first published calculation of the slope of the $M_\bullet$ to velocity dispersion $\sigma_\mathrm{v}$ (in km$\,$s$^{-1}$) curve ($M_\bullet - \sigma_\mathrm{v}$ relation) varied between 5.27$\pm$0.40~\cite{fer6} and 3.75$\pm$0.3~\cite{gebh}.  Reference~\cite[and references therein]{trem} suggested the range of slopes are caused by systematic differences in the velocity dispersions used by different groups.  However, the origin of these differences remains unclear.

\citet{fer4} found the ratio of the $M_\mathrm{\bullet}$ to $M_\mathrm{DM}$ of the theorized dark matter halo around galaxies was a positive value that decreased with halo mass.  However, if the intrinsic rotation curve is rising~\cite{hodg2}, the effect of the force of $ M_\mathrm{DM}$ in the equations implies the effect of the center object must be repulsive.  Such a repulsive force was called a ``wind'' by \citet{shu, silk}.  The ``wind'' (a gas) exerted a repulsive force acting on the cross sectional area of particles.  Therefore, denser particles such as black holes move inward relative to less dense particles. 

A multitude of X-ray point sources, highly ionized iron, and radio flares without accompanying large variation at longer wavelengths have been reported near the center of the Milky Way \cite{baga, baga2, baga3, binn, genz, zhao, wang}.

\citet{mclu} found that the mass in the central region is 0.0012 of the mass of the bulge.  \citet{fer5} reported that approximately 0.1\% of a galaxy's luminous mass is at the center of galaxies and that the density of SBH's in the universe agrees with the density inferred from observation of quasars.  \citet{merr} found similar results in their study of the $M_\mathrm{\bullet} - \sigma_\mathrm{v}$ relation.  \citet{fer4} found a tight relation between rotation velocity $v_\mathrm{c}$ in the outer disk region and bulge velocity dispersion $\sigma_\mathrm{c}$ ($v_\mathrm{c} - \sigma_\mathrm{c}$) which is strongly supporting a relationship of a center force with total gravitational mass of a galaxy.  \citet{wand} showed the $M_\mathrm{\bullet}$ of AGN galaxies and their bulge luminosity follow the same relationships as their ordinary (inactive) galaxies, with the exception of narrow line AGN.  \citet{grah,grah2,grah3} found correlations between $M_\mathrm{\bullet}$ and structural parameters of elliptical galaxies and bulges.  Either the dynamics of many galaxies are producing the same increase of mass in the center at the same rate or a feedback controlled mechanism exists to evaporate the mass increase that changes as the rate of inflow changes as suggested by \citet{merr2}.  The RCs imply the dynamics of galaxies differ so the former explanation is unlikely.

In accordance with the Principle of Fundamental Principles (see Appendix~\ref{sec:princ}), a search was made for a physical process in the Newtonian physical domain that may model and explain the observed data of spiral galaxies.  Such a system was found in the fractional distillation process (FDP).  A FDP has a Source of heat (Space energy) and material input (matter) at the bottom (center of galaxies) of a container.  The heat of the FDP is identified with a constituent of our universe that is not matter and is identified as Space (in units of ``que'', abbreviated as ``q'', - a new unit of measure) whose density $\rho$ is a potential energy.  The term ``Space'' will mean the amount of Space in ques.  In the FDP, temperature decreases with height from the Source.  The opposing forces of gravity and the upward flow of gas acting on the cross section of molecules in an FDP cause the compounds to be lighter with height and lower temperature.  Heavy molecules that coalesce and fall to the bottom are re-evaporated and cracked.  The heat equation with slight modification rather than gas dynamics modeled the flow of heat (Space energy).  Temperature decline of the FDP was modeled as dissipation of matter and Space energy into three dimensions.  Thus, the Changing Universe Model (CUM) was born.

This paper posits that there is a Source of strength $\epsilon$ at the center of each galaxy.  The Source is a source of all matter and of Space.  The $\rho$ forms a scalar potential field.  Energy is conserved in the emission of matter/energy and Space from the Source.  The unit of measure of $\epsilon$ is in the amount of que released per second which is proportional to the amount of mass released per second, $\epsilon_\mathrm{m}$ in units of M$_\mathrm{\odot}$ s$^{-1}$.  The application of the CUM to the RC of galaxies explains observations and empirical relations.  Also, ten new relationships among galaxy parameters were predicted and confirmed.  This, the first test of the CUM, successfully explains RC observations and suggests an alternative to DM.  The CUM, at this early stage of model development, is bringing together heretofore apparently unrelated observations.  

The object of this article is to develop and test the CUM by suggesting a galaxy model consistent with the CUM, by explaining observations of the RC and the centers of galaxies, and by deriving relations among galaxy parameters.  In section~\ref{sec:model} the CUM model is developed from the fundamental principles, which are stated in Appendix~\ref{sec:princ}.  An equation for the RC is developed from the energy equations of the CUM, applied to galaxies, and compared with galaxy observations in Section~\ref{sec:gala}.  The posited description of particles is used to explain several phenomena involving observations of the center of galaxies.  Other parameter relationships are presented in Section~\ref{sec:other}.  Section~\ref{sec:lines} shows the interger relatinships developed in the previous sections are not random.  The results are discussed in Section~\ref{sec:disc}.  Section~\ref{sec:conc} lists the conclusions.

\section{\label{sec:model}CUM MODEL}

The assumed Fundamental Principles are stated in Appendix~\ref{sec:princ}.  Throughout the text, capitalizing the first letters of the name of the Fundamental Principle denotes these principles.

The Principles of Change and Repetition requires that the three dimensions (3D) be created from two dimensions (2D).  The creation of 3D from zero dimensions (0D) is not only unlikely but forbidden.  The universe begins with a Change that may be a rupture of 0D.  The Change itself is a beginning of time.  The 0D ruptures to create one dimensional (1D) points and Space between the points on a line.  The 0D is a Source of 1D and continually ejects Space into one end of the 1D line.  The other end must expand into the void.  Thus, the 1D line is continually growing.  The 1D line has only one direction.  Therefore, the 1D line is straight.  The 1D energy is $\tau_\mathrm{1} t=\rho_\mathrm{1} l$ where $\tau_\mathrm{1}$ is the point tension, $t$ is time elapsed between point eruption, $\rho_\mathrm{1}$ is the line Space density, and $l$ is the length between points.  The first transition creates length and time 
\begin{equation}
l=ct
\label{eq:2},
\end{equation}
where $c=\tau_\mathrm{1} / \rho_\mathrm{1}$ is the constant of proportionality.  

When points in 1D rupture, the 2D Space is inserted into 2D at many points along the 1D line.  The rupture can have no component remaining in 1D.  Thus, a right angle is defined and the metric coefficients equal one (a Cartesian plane).  Accordingly, 2D is flat.  The 2D energy for each transition is $\tau_\mathrm{2} l=\rho_\mathrm{2} S$ where $\tau_\mathrm{2}$ is the line tension, $l$ is the line length, $\rho_\mathrm{2}$ is the area Space density, and $S$ is the area.

A 2D Sink is at a place in 2D that has Space tightly contained in a line.  The line is a one-dimensional border in 2D.  If a line with a ``line tension'' compressing Space is deformed perpendicular to the line, an arc of height $h$ is formed.  The geometric constant of the line is its length $l$.  If a condition of $l=\pi h$ is reached, a bubble in 2D is formed.  The end points of the line are coincident.  This rupture causes Space to be removed from the 2D plane.  This 2D Sink becomes a Source in 3D Space.  The line becomes an object in 3D called a hod.  The hod is the smallest piece of matter from which all other matter is constructed.  The hod is supple so that its shape may change in 3D.    The edge of the 2D plane expands into the void.  Hence, the distance between the points that rupture into 3D is increasing.  The expansion is recurrent and is in 2D, not 3D.  Because the rupture into 3D is orthogonal to 2D, 3D is flat (Cartesian three space dimensions), also.  The $l$ becomes area $\alpha$.  The Space area in 2D becomes the volume $V$ of Space density $\rho$ in 3D.  One que is the amount of Space released by one hod in the transition.  

Since nothing exists in 3D except because of the transitions, the energy for a transition is 
\begin{equation}
\tau\alpha = \rho_\mathrm{I} \, V_\mathrm{I}
\label{eq:1},
\end{equation}
where $\rho_\mathrm{I}$ and $V_\mathrm{I}$ are the initial $\rho$ and $V$ at the  transition, respectively, and $\tau$ is the proportionality constant called surface tension.

Similarly, corollary II of the Fundamental Principle implies a 3D Sink may be formed.

Succeeding ruptures at the same point adds Space and hods to the universe.  Since we observe there is Space and $\rho$ in our neighborhood, the Space and hods then flow from the Source and creates the galaxy around the Source.

The total energy transferred into 3D from Eq.~(\ref{eq:1}) is 
\begin{equation}
N_\mathrm{t} \, \tau \alpha = N_\mathrm{t} \, \rho_\mathrm{I} \, V_\mathrm{I}
\label{eq:3},
\end{equation}
where $N_\mathrm{t}$ is the total number of hods in the universe.

In the observed universe, the 2D plane is common for all Sources.  As the Space from a Source expands into the void, it encounters other Space from other Sources.  Since all Space originates from one 0D Source, the merged Space is causally correlated and coherent.  However, 2D Space is distinct from 3D Space.

In 2D, the surface was a tightly bound line on the outside of Space.  In 3D, Space is on the outside of a surface.  The $\bm{\tau}$ on the 2D Space was high enough to contain the Space.  Now $\bm{\tau}$ acts to attract 3D Space.  Since the hod formed in the extreme pressure of the Source, the $\bm{\tau}$ maximum ability to act on Space exceeds the pressure of Space on the hod away from the Source.  Since $\tau$ and $\alpha$ are constant, as the extent of the volume $V$ increases so $\rho$ must decrease.  Therefore, the hod applies its energy, $\tau\alpha$, on the Space immediately surrounding the hod.  Since the hod contained the Space before release and Space was at a much higher $\rho$ than is surrounding the hod in 3D, all the hod's tensional energy is exerted on Space.  A hod is a 2D discontinuity in the continuity of 3D Space.  These discontinuities are surfaces in Space with an edge not containing loops. 

Hods are discrete.  The mathematics of integer numbers can then be used to count hods.  In addition, hods can combine, one with another.  The number of hods and the structure of the combinations of hods create the different particles and forces, except gravity, of the universe.  Therefore, the language of hods is integer numbers, structures, counting, subtraction, and summation.

Space is continuous, perfectly elastic, and compressible.  The Space continuously fills the region it occupies except for the hods.  Transformation of Space over time is continuous.  Hods and the outer edge of Space as it expands into the void are discontinuities in Space.  Space really can meet the mathematical criteria of considering an arbitrarily small volume.  Therefore, the language of Space is real numbers, tensors, field theories, differentiation, and integrals.

The three interactions of concern are that each hod exerts a force on Space, Space exerts a force on the hod, and Space exerts a force on adjacent Space.  A hod is not attracted or repelled by other hods through a field or other action at a distance device.  Since $\tau$ can act only perpendicular to the hod's surface, the force Space exerts on the hod is perpendicular to the surface of the hod.

There are three dimensions in our universe.  Counting is a characteristic of hods.  Therefore, that our universe is three-dimensional is a characteristic of the hods and not of Space.

There are two concepts of distance.  Distance can be by reference to a fixed grid.  Traditional methods have used Space as the fixed grid.  In this model, distance is a multiple of the average diameter a hod.  Since $\alpha$ is a constant and the hod may be deformed, the hod average diameter will be considered a constant throughout the universe beyond the Source and in moving and accelerating reference frames.  The measure of distance will be a ``ho'', which is defined as the average diameter of the hod.  The CUM uses ho distance to form a grid in a flat universe with
\begin{equation}
1 \, \mathrm{ho} = 2 \,( \frac{\alpha}{\pi})^{1/2}
\label{eq:4}.
\end{equation}

Another concept of distance is the amount of Space between two surfaces.  By analogy, this type of distance in gas is the amount of gas between two surfaces.  Increasing the amount of gas between the two surfaces increases the distance.  One way to do this is by movement of the surfaces.  Another way is to add more gas from a source.  If the medium is compressible, the distance measured by the amount of gas varies with the change in density of the medium.  The CUM model allows Space to be compressible.  Therefore, the amount of Space between two surfaces can change if the $\rho$ (que ho$^{-3}$) of the Space between two surfaces changes.

The terms distance, length, area, and volume will refer to ho measurements.  The unit of measure of distance is usually expressed in km, pc, kpc, or Mpc.  If the measurement is taken by means of angular observation or time-between-events observations such as using Cepheid variables that do not depend on the amount of Space, the units of pc are proportional to the ho unit.  Since the proportionality constant is unknown, the units of pc will be used.  However, if the measurement such as redshift depends on the amount of Space, the units of pc should be used with caution. 

Since the hods are discrete, $\epsilon$($t$) is a digital function in time.  The hod exerts a tremendous $\tau$ to hold the Space in its 2D surroundings.  Upon transition, the Space immediately around the hod experiences an instantaneous addition to the amount of Space in the 3D universe.  This step function creates a shockwave in the surrounding Space.  Space then flows out into the void causing the $V$ of Space to increase and a $\rho$ variation in the $V$.

If Space has no relaxation time and no maximum gradient, the eruption of Space from a Source would spread out instantly so that $\rho$ would be virtually zero everywhere.  Since we observe Space in our region of the universe, the Anthropic Principle implies Space flow must have a relaxation time and a maximum $\rho$ gradient (${\bm\nabla}_\mathrm{m}\rho$) that is a constant of the universe.  Bold face type denotes a vector.

Since Space has a relaxation time, the mirror property in hods is inertia $\iota$ per $\alpha$.  A hod is an object with an area $\alpha$, a surface tension per unit area $\bm{\tau}$, and inertia per unit area $\iota$.  The $\alpha$, the magnitude of $\bm{\tau}$, and the $\iota$ are the same for all hods and are universal constants.

The $\bm{\tau}$ and $\iota$ furnish a resistance to deformation on the hod.  Since $\alpha$ is a constant, a deformation of a hod causes the edge to change shape.

In the region around an active Source, ${\bm\nabla}\rho = {\bm\nabla}_\mathrm{m}\rho$.  Space expands into the void where $\rho=0$ and ${\bm\nabla}\rho=0$.  Therefore, at some radius $r_\mathrm{sc}$ from the Source, $\rho$ becomes another function of time and distance from the Source.  At $r_\mathrm{sc}$, ${\bm\nabla}\rho$ must be continuous, ${\bm\nabla}\rho = {\bm\nabla}_\mathrm{m}\rho$ and ${\bm\nabla}^2\rho > 0$.  If $\epsilon$($t$) changes, then $r_\mathrm{sc}$ changes.

As the Space increases $V$, Space increases between the hods.  The hods move and attain a velocity d$\bm d_{ij}$/d$t$ where $\bm d_{ij}$ is the vector distance from the $i^\mathrm{th}$ hod to the  $j^\mathrm{th}$ hod.  The hod's $\iota$ inhibits the movement.  Thus, a feedback condition is established wherein the d$\bm d_{ij}$/d$t$ is balanced by $\iota$.  Therefore, the hods movement is proportional to the total inertia $\iota \alpha$ of the hod and to the velocity of the hod.  The momentum $\bm p_{ij}$ between the of the $i^\mathrm{th}$ and $j^\mathrm{th}$ hods is defined as
\begin{equation}
\bm p_{ij} \equiv K_\mathrm{p} \, \iota \alpha \, \frac{\mathrm{d} \bm d_{ij}}{\mathrm{d} t}
\label{eq:5},
\end{equation}
where $K_\mathrm{p}$ is the proportionality constant and $\equiv$ means ``is defined as''.

The energy $T_{ij}$ expended by the Space expansion of $V$ upon the hod is
\begin{equation}
T_{ij} = \int \bm p_{ij} \bullet \mathrm{d} \bm d = K_\mathrm{p} \iota \alpha \int \frac{\mathrm{d}}{\mathrm{d} t} \left| \frac{\mathrm{d} d_{ij}}{\mathrm{d} t} \right | ^2 \, {\mathrm{d} t}
\label{eq:5a}.
\end{equation}

Thus, 
\begin{equation}
T_{ij} = K_\mathrm{t} \,K_\mathrm{p} \,\iota \alpha \,\left| \frac{\mathrm{d} d_{ij}}{\mathrm{d} t} \right| ^2
\label{eq:6},
\end{equation}
where $K_\mathrm{t}$ is the proportionality constant.

The $T_{ij}$ is a scalar quantity and is the kinetic effect of the $i^\mathrm{th}$ hod on the  $j^\mathrm{th}$ hod.  The total $T_{ij}$ energy $T_j$ exerted on the $j^{\mathrm{th}}$ hod by all other hods is 
\begin{equation}
T_{j} = \sum^{N_\mathrm(t)}_{ i \neq j \,i=1} T_{ij} 
\label{eq:6a}.
\end{equation}

Because the amount of energy added by the movement of $\iota \alpha$ must be the negative of energy due to position and by the inertia which inhibits hod movement, the Principle of Negative Feedback applies.  As Space is added between hods, the energy the hods exert on each other must increase.  Since the $\rho$ term of Eq.~(\ref{eq:1}) does not apply to the hod, the loss of energy must be due to $\tau \alpha$.  Call this energy between the of the $i^\mathrm{th}$ and $j^\mathrm{th}$ hods the potential energy $U_{ij}$.  By the Principle of Negative Feedback, $U_{ij}$ is proportional to $\tau \alpha$ and is a function $f(d_{ij})$ of $d_{ij}$ between hods 
\begin{equation}
U_{ij} = - K_\mathrm{u} \,\tau \alpha \, f(d_{ij}) , \quad i \neq j
\label{eq:7}
\end{equation}
where $K_\mathrm{u}$ is the proportionality constant and the negative sign is because the hods exert an attraction to each other.  

The total $U_{ij}$ energy $U_j$ exerted on the $j^{\mathrm{th}}$ hod by all other hods is
\begin{equation}
U_{j} = \sum^{N_\mathrm{t}}_{ i \neq j \,i=1} U_{ij}
\label{eq:6b}.
\end{equation}

The $j^\mathrm{th}$ hod's $\iota \alpha$ inhibits the changing $\bm p_{ij}$.  The inertia of the hods results in work $W$ done on the hod in time d$t$ 
\begin{equation}
W_{j} = - \int_\mathrm{b}^\mathrm{e} \frac{\mathrm{d} }{\mathrm{d} t} \, \sum^{N_\mathrm{t}}_{ i \neq j \,i=1}\bm p_{ij} \bullet \mathrm{d} \bm d
\label{eq:6c},
\end{equation}
\begin{equation}
W_{j} = - K_\mathrm{p} \, \iota \alpha \,  \left( d_\mathrm{e} \frac{\mathrm{d} ^2 d_{\mathrm{e}ij}}{\mathrm{d} t ^2}-d_\mathrm{b} \frac{\mathrm{d} ^2 d_{\mathrm{b} ij}}{\mathrm{d} t ^2} \right)
\label{eq:6d},
\end{equation}
where subscript ``b'' and subscript ``e'' indicate the beginning and ending positions in the time period, respectively.

Therefore, the total energy equation in 3D is Eq.~(\ref{eq:3}) with $T_{j}$ and $U_{j}$ is
\begin{equation}
N_\mathrm{t} \, \tau \alpha \, + \,  \sum^{N_\mathrm{t}}_{i \neq j \,j=1}   (T_{j}\, + \, U_{j} ) \, - \, \int \! \! \! \int \! \! \! \int_\mathrm{Space}\, \rho \, \mathrm{d} V =0
\label{eq:8}.
\end{equation}

Implied in Eqs.~(\ref{eq:3}) and (\ref{eq:8}) is 
\begin{equation}
\sum^{N_\mathrm{t}}_{i \neq j \,j=1} \, \sum^{N_\mathrm{t}}_{i=1} K_\mathrm{t} \,K_\mathrm{p} \,\iota \alpha \,\left( \frac{\mathrm{d} d_{ij}}{\mathrm{d} t} \right) ^2 = \sum^{N_\mathrm{t}}_{i \neq j \,j=1} \sum^{N_\mathrm{t}}_{i=1} K_\mathrm{u} \,\tau \alpha \, f(d_{ij})
\label{eq:8a}.
\end{equation}

``Flatness'' is a fundamental condition of the creation of matter and Space.  Therefore, the Cartesian coordinate system with four equal angles to define a complete circle is the form of the ``flat'' measuring system.

Note, $\bm d_{ij}$ is between hods.  The process of summing over all hods provides a reference frame similar to Mach's principle.  Also, since there is distance between galaxies, the $U_{ij}$ and $T_{ij}$ of hods in different galaxies includes a heritage from 1D and 2D.  

\subsection{\label{sec:ener}Energy continuity equation}

Consider an arbitrary volume $\bigtriangleup V$ bounded by a surface $\bigtriangleup s$.  Space will be in $\bigtriangleup V$.  Hods, Sources, and Sinks may be in $\bigtriangleup V$.  Equation~(\ref{eq:6}) implies $T_{ij} = T_{ji}$.  Equation~(\ref{eq:7}) implies $U_{ij} = U_{ji}$.  Therefore, the total energy $E_\mathrm{v}$ in $\bigtriangleup V$ is 
\begin{eqnarray}
E_\mathrm{v} &=& N \tau \alpha + \int \! \! \! \int \! \! \! \int_{\bigtriangleup \mathrm{V}} \, \rho \mathrm{d} V \nonumber\\*
&+& \sum_{j=1}^N ( T_{j} + U_{j} + W_{j}) \nonumber\\*
&+& \sum_{k=1}^{N_\epsilon} \epsilon_k + \sum_{l=1}^{N_\eta} \eta_l 
\label{eq:11},
\end{eqnarray}
where $N$ is the number of hods in $\bigtriangleup V$, the integration is over $\bigtriangleup V$, $N_\epsilon$ is the number of Sources in $\bigtriangleup V$, $N_\eta$ is the number of Sinks in $\bigtriangleup V$, $\epsilon_k$ is the strength of the $k^\mathrm{th}$ Source, and $\eta_l$ is a negative number which is the strength of the $l^\mathrm{th}$ Sink.  

The classical development of an energy continuity equation includes the assumption that energy is neither created nor destroyed.  In the CUM, an arbitrary volume may include creation into our universe of energy (Sources) or the destruction from our universe of energy (Sinks).

Following the usual procedure for the development of an energy continuity equation such as the heat equation for the temporal evolution of temperature; remembering hods have no volume and, therefore, the region can be made regular and simply connected by mathematically drawing cuts around the hod discontinuities; and using the Principle of Minimum Potential Energy and the Principle of Feedback and Minimum Action gives the integral form of the energy continuity equation 
\begin{eqnarray}
\int  \! \! \! \int \! \! \! \int C \frac{\mathrm{d}}{\mathrm{d} t} E_\mathrm{v} \mathrm{d} V  &=& S \int \! \! \! \int \bm{\nabla}\rho \bullet \bm{n} \mathrm{d} s \nonumber \\*
&+& S_\mathrm{u} \int \! \! \! \int \, \bm {\nabla} \left| \sum^{N}_{i \neq j \,j=1} U_j \right| \bullet \bm{n} \, \mathrm{d} s \nonumber\\*
&+& S_\mathrm{t} \int \! \! \! \int \, \bm{\nabla} \left| \sum^{N}_{i \neq j \,j=1} T_j \right| \bullet \bm{n} \, \mathrm{d} s \nonumber\\*
&+& S_\mathrm{w} \int \! \! \! \int \, \bm{\nabla} \left| \sum^{N}_{i \neq j \,j=1} W_j \right| \bullet \bm{n} \, \mathrm{d} s \nonumber\\*
&+&  \int \! \! \! \int  \left[ \tau \alpha N_\mathrm{tuw}(t) \bm v_\mathrm{n}\right] \bullet \bm n \, \textrm{d} s \nonumber\\*
&+&  \int \! \! \! \int \! \! \! \int (\sum_{k=1}^{N_\epsilon} \epsilon_k + \sum_{l=1}^{N_\eta} \eta_l  ) \, \mathrm{d} V
\label{eq:20},
\end{eqnarray}
where: (1) The $S$ is a constant of proportionality.  Call $S$ the conductivity.  $S$ is the ability of energy to flow and is assumed to be independent of position or direction.  $S$ is a factor that slows the flow of Space.  The idea of their being some kind of viscosity or friction associated with Space would require the expenditure of energy.  Since Equation (\ref{eq:8}) lacks such a term, Space energy dissipates without loss of total energy.  Therefore, there is a relaxation time associated with Space.  This means that the stress created by the sudden entry of Space upon a transition must relax asymptotically with time.  The compression of the transition is dissipated by the flow of Space into the void.
(2) The $S_\mathrm{u}$ is a constant of proportionality.  Call $S_\mathrm{u}$ the potential conductivity.  $S_\mathrm{u}$ is the ability of energy to flow and is assumed to be independent of position or direction.
(3) The $S_\mathrm{t}$ is a constant of proportionality.  Call $S_\mathrm{t}$ the kinetic conductivity.  $S_\mathrm{t}$ is the ability of kinetic energy to flow and is assumed to be independent of position or direction.
(4) The $S_\mathrm{w}$ is a constant of proportionality.  Call $S_\mathrm{w}$ the work conductivity.  $S_\mathrm{w}$ is the ability of work energy to flow and is assumed to be independent of position or direction.
(5) The $N_\mathrm{tuw} =  N + \sum_{j=1}^N ( T_{j} + U_{j} + W_{j}) \vert _{T_{j} , U_{j},  W_{j}}$ is the number of hods which leave the $\bigtriangleup V$ with the energy associated with each hod and $\vert _{T_{j} , U_{j}, W_{j}}$ means that $ T_{j}$ , $ U_{j}$, and $ W_{j}$ are constant with respect to the differentiation.
(6) The $\bm v_\mathrm{n}$ is the velocity vector of the $N_\mathrm{tuw}$.

Applying Gauss's divergence theorem to the surface integrals of Equation~(\ref{eq:20}), and noting that the volume could be chosen to be a region throughout which the integrand has a constant sign yields the differential form of the energy continuity equation 
\begin{eqnarray}
\frac{\textrm{d} E_\mathrm{v}}{\textrm{d} t} & = & D^2 \, \bm{\nabla}^2 \rho  + \frac{S_\mathrm{u}}{C} \bm{\nabla}^2 \left| \sum^{N}_{i \neq j \,j=1} U_{j} \right| \nonumber\\*
&+& \frac{S_\mathrm{t}}{C} \bm{\nabla}^2 \left| \sum^{N}_{i \neq j \,j=1} T_{j} \right| + \frac{S_\mathrm{w}}{C} \bm{\nabla}^2 \left| \sum^{N}_{i \neq j \,j=1} W_{j} \right| \nonumber\\*
&+& \bm{\nabla} \bullet \left[ \tau \alpha N_\mathrm{tuw}(t) \bm v_\mathrm{n}\right] \nonumber\\*
&+& C^{-1} \, (\sum_{k=1}^{N_\epsilon} \epsilon_k + \sum_{l=1}^{N_\eta} \eta_l  ) 
\label{eq:21},
\end{eqnarray}
where $D^2=S/C$ and is called the diffusivity constant of Space.

The presence of Sources and Sinks leads to a non-homogeneous equation.  Note the similarity of this equation to the diffusion equation for the temporal evolution of the concentration of chemical species in a fluid and the heat equation for the temporal evolution of distributions of temperature.  Like the molecular diffusivity and thermal diffusivity, $D^2$ has its origins in the dynamics and interactions that transfer energy.

\subsection{\label{sec:forc}Forces}

The $\rho$ terms and $U_{j}$ term in Eq.~(\ref{eq:11}) change by spatial movement and are called impelling forces since they cause an energy change.  The $N \tau \alpha$ term, the $W_j$ term, and $T_{j}$ term in Eq.~(\ref{eq:11}) change by temporal differences and are called nurturing forces since their movement carries energy. 

\subsubsection{\label{sec:spac}Space}

Define a volume $V$ containing Space for the distances being considered and throughout which the terms of Eq.~(\ref{eq:21}) except $\rho$ terms are constant.  The force of Space at a point $\bm{F_\mathrm{sp}}$ is defined as
\begin{equation}
\bm{F_\mathrm{sp}} \, \equiv \, - \, D^2 \, \bm{\nabla}\rho 
\label{eq:31},
\end{equation}
where the negative sign means the force is directed opposite to increasing $\rho$.

Because there is no shear stress in Space, the force exerted by Space is exerted only normal to surfaces.  Consider a cylinder of cross section $\mathrm{d} s$ around a regular, simply connected volume with the ends of the cylinder normal to $\bm{F_\mathrm{sp}}$.  The $\mathrm{d} s$ has a difference of force $\bigtriangleup \bm{F_\mathrm{s}}$ on each end of the cylinder.  Allow the height of the cylinder to become arbitrarily small.  The $\bigtriangleup \bm{F_\mathrm{s}} \rightarrow \mathrm{d} \bm{F_\mathrm{s}}$ and 
\begin{equation}
\mathrm{d} \bm{F_\mathrm{s}} \, = \, - \, D^2 \, (\bm{n} \, \bullet \, \bm{\nabla} \rho ) \, \mathrm{d} \bm{s}
\label{eq:32},
\end{equation}
where $\bm{n}$ is the outward unit normal of $\mathrm{d} \bm{s}$.

Integrating Eq.~(\ref{eq:32}) over a surface, the Space force on the surface $\bm{F_\mathrm{s}}$ becomes
\begin{equation}
\bm{F_\mathrm{s}} \, = \, - \, D^2 \, \int \! \! \! \int \, \bm{\nabla}\rho \, \bullet \, \bm{n} \, \mathrm{d} \bm{s}
\label{eq:33}.
\end{equation}

The force of Space on a surface is proportional to the cross section of the surface perpendicular to $\bm{\nabla} \rho$.

The $\rho$ due to Sources and Sinks can be calculated from Eq.~(\ref{eq:21}).  If all terms of Eq.~(\ref{eq:21}) are constant except the $\rho$, $\epsilon$, and $\eta$ terms, the average $\eta$ of each Sink remains constant over a long period of time, and the average $\epsilon$ of each Source, each galaxy, remains constant over a long period of time, then Eq.~(\ref{eq:21}) can be solved for a steady-state condition.  The range of distance from Sources and Sinks will be assumed to be sufficient such that the transition shock is dissipated and $r > r_\mathrm{sc}$.  We can consider $\rho$ at a point to be the superposition of the Space effects of each Source and Sink.  Therefore, for $\rho$ from the $j^\mathrm{th}$ galaxy and $k^\mathrm{th}$ Sink at the $i^\mathrm{th}$ point ($\rho_{i}$), Eq.~(\ref{eq:21}) becomes 
\begin{subequations}
\label{eq:34}
\begin{eqnarray}
\frac{\mathrm{d}}{\mathrm{d} t} \, (r_{ij} \rho_{i}) \, = \, D^2 \, \nabla^2 \, (r_{ij} \rho_{i} \, ), \label{eq:34:a}\\*
\frac{\mathrm{d}}{\mathrm{d} t} \, (r_{ik} \rho_{i}) \, = \, D^2 \, \nabla^2 \, (r_{ik} \rho_{i} \, ) \label{eq:34:b}
\end{eqnarray}
\end{subequations}
in spherical coordinates, where $r_{ij}$ is the distance from the $j^\mathrm{th}$ galaxy Source and $r_{ik}$ is the distance from the $k^\mathrm{th}$ Sink to the $i^\mathrm{th}$ point, and $\rho_{i}$ depends entirely on $r_{ij}$ and $r_{ik}$.

The boundary conditions are:
\begin{subequations}
\begin{eqnarray}
\rho_{i} (r_{ij},0) \,&=&\,  \mathrm{a \, function \, of \, distance, \, only}, \label{eq:a} \\*
\rho_{i} (r_{ij},t) \, &\rightarrow & \, 0, \, \mathrm{as} \, r \rightarrow \, \infty , \label{eq:b}\\*
\rho_{i} (r_{ik},0) \,&=&\,  \mathrm{a \, function \, of \, distance, \, only}, \label{eq:c} \\*
\rho_{i} (r_{ik},t) \, &\rightarrow & \, 0, \, \mathrm{as} \, r \rightarrow \, \infty. \label{eq:d} 
\end{eqnarray}
\end{subequations}

This is analogous to the problem of heat flow in a spherical solid with a continuous Source of strength $\epsilon_j$ at $r_{ij}=0$ or with a continuous Sink of strength $\eta_k$ at $r_{kk}=0$.  By superposition the total $\rho_i  (x,y,z)$ may be calculated~\cite{cars261} for $r_{ij}$ and $r_{ik}$ outside the $ \bm{\nabla_\mathrm{m}} \rho$ volume of the Sources and Sinks as
\begin{equation}
\rho_{i} (r) \, = \, \frac{1}{4 \pi D^2} \, \left( \sum_{j=1}^\mathrm{all \, galaxies} \, \frac{\epsilon_{j}}{r_{ij}} - \sum_{k=1}^\mathrm{all \, Sinks} \, \frac{ | \eta_{k} |}{r_{ik}} \right) 
\label{eq:35}.
\end{equation}

At $r_{ij}=r_\mathrm{sc}$ of the $j^\mathrm{th}$ galaxy , the effect of the other Sources and Sinks is very small.  Therefore, 
\begin{equation}
\bm{\nabla} \rho_{i} = \frac{\epsilon_{j}}{ 4 \pi D^2} \bm{\nabla} r_{\mathrm{sc}j}^{-1} = \bm{\nabla_\mathrm{m}} \rho
\label{eq:36},
\end{equation}
where $r_{\mathrm{sc}j}$ is the $r_{\mathrm{sc}}$ for the $j^\mathrm{th}$ galaxy.

At the $j^\mathrm{th}$ Source, the nearly instantaneous transition of the hods and Space into 3D creates a shock wave zone of radius $r_{\mathrm{sw}j}$, where the transition forced $\bm{\nabla} \rho$ is greater than $ \bm{\nabla_\mathrm{m}} \rho$.  The assumption of constant D is no longer valid.  In the region $r_{\mathrm{sw} j} < r_{ij} < r_{\mathrm{sc} j}$, $\epsilon_j $ can be considered to be a time average of the nearly digital transitions and $\bm{\nabla} \rho_i = \bm{\nabla_\mathrm{m}} \rho$.

At the Source of the $j^\mathrm{th}$ galaxy $r_{ij}=0$, $\rho_{i} (r_{ij}=0)$ is a maximum value $\rho_{\mathrm{max} j}$ and $\bm{\nabla} \rho_{i}$ and $\nabla^2 \rho_{i}$ are discontinuous.  For periods of the time required for Space to flow to $r_{\mathrm{sc} j}$, $\epsilon_j$ can be considered a constant.  All the Space from the Source is flowing through the surface of the sphere of radius $r_{\mathrm{sc} j}$.  Since the volume of a sphere $\propto r^3$ and $\epsilon$ is a volume per unit time, $\epsilon \propto r$.  Further, this proportionality holds for any radius determined by $\epsilon$.  If $\epsilon_j $ increases, $r_{\mathrm{sc} j}$ increases proportionally as suggested by Eq.~(\ref{eq:36}).  Therefore, 
\begin{equation}
r_{\mathrm{sc}j} \, = \, \frac{K_\mathrm{sc}}{\left | \bm{\nabla_\mathrm{m}} \rho \right | } \, \epsilon_j 
\label{eq:37},
\end{equation}
where $K_\mathrm{sc}$ is the proportionality constant which depends on how $D$ varies in the $\bm{\nabla_\mathrm{m}} \rho$ zone around the Source and $\rho_{\mathrm{max} j} = K_\mathrm{sc} \epsilon_j$.

The force $\bm {F_\mathrm{ts}}$ exerted on an assembly of hods by the Space force due to Sources and Sinks is calculated by combining Eq.~(\ref{eq:33}) and (\ref{eq:35}) and integrating over the cross section of the assembly yields 
\begin{equation}
\bm F_{\mathrm{ts} i} \, = \, G_\mathrm{s} \,m_\mathrm{s} \, \, \left( \sum_{j=1}^\mathrm{all \, galaxies} \, \frac{\epsilon_{j}}{r_{ij}^3} \bm{r}_\mathrm{ij}- \sum_{k=1}^\mathrm{all \, Sinks} \, \frac{ | \eta_{k} | }{r_{ik}^3} \bm{r}_\mathrm{ik} \right) 
\label{eq:48},
\end{equation}
where $m_\mathrm{s}$ is the Space effective cross section of the assembly of hods and $G_\mathrm{s} = ( 4 \pi )^{-1}$.

\subsubsection{\label{sec:inte}Interaction of Space and hods}

The hod in 2D is transformed into 3D because of the energy pressure of surrounding hods.  The transition releases this energy into 3D.  The $\tau$ acts on Space uniformly across the surface of the hod and exerts an attractive pressure on Space normal to the surface of the hod.  Since the hods in 2D are closed loops in a plane, the hods arrive in 3D as surfaces oriented flat-to-flat.  At the transition, $V_\mathrm{I} = 2 d_\mathrm{I} \alpha$ where $d_\mathrm{I}$ is the initial distance that Space extends from a hod's surface.  At the radius $r_\mathrm{h}$ of the hod from the Source, Space has expanded and the $\rho_i (r_\mathrm{h}) < \rho_\mathrm{I}$.  However, within $0 < r_\mathrm{h} < r_\mathrm{sw}$, $\bm{\nabla} \rho = \bm{\nabla_\mathrm{m}} \rho$ and $\rho_\mathrm{h} < \rho_i$ where $\rho_\mathrm{h}$ is the maximum Space density the hod can attract to its surface.  All the Space energy in the shock wave zone is within a zone with $\bm{\nabla} \rho = \bm{\nabla_\mathrm{m}} \rho$.  Therefore, all the $\rho_\mathrm{I} V_\mathrm{I}$ is ``bound'' to the hods $\rho_\mathrm{I} V_\mathrm{I} = \tau \alpha = \tau_\mathrm{b} \alpha$, where $\tau_\mathrm{b}$ is the amount of surface tension used to bind Space to the hod.

Define a Cartesian coordinate system in a volume $V$ with the origin at the center of a circular hod with diameter $a$, with the $z$-axis perpendicular to the plane of the hod, and with the $x$-axis and $y$-axis in the plane of the hod.

At a radius $r_\mathrm{h}$ from the Source outside the shock wave zone, the maximum distance $\bm d_\mathrm{h}$ (ho) from the hod surface along the z axis within which $\bm{\nabla} \rho = \bm{\nabla_\mathrm{m}} \rho$ is directed away from the hod surface has a value 
\begin{equation}
d_\mathrm{h} = \frac{\rho_\mathrm{h} - \rho_\mathrm{d} }{ \left | \bm{\nabla_\mathrm{m}} \rho \right |}, \, \rho_\mathrm{d} =\rho_i (d_\mathrm{h}) < \rho_\mathrm{h}, \, r_\mathrm{sw} < r_\mathrm{h}
\label{eq:41c},
\end{equation}
where $\rho_\mathrm{d} $ is the Space density at $d_\mathrm{h}$.

Hods that are within $2 d_\mathrm{h}$ of each other form an assembly of hods within a volume with a $\rho =\rho_\mathrm{d}$ equipotential surface.  For $r_\mathrm{sw} < r_\mathrm{h} \leq r_\mathrm{sc}$, $\bm{\nabla} \rho = \bm{\nabla_\mathrm{m}} \rho$ everywhere.

For $ r_\mathrm{h} > r_\mathrm{sc} $, volumes more than $d_\mathrm{h}$ from a hod surface have $\bm{\nabla} \rho < \bm{\nabla_\mathrm{m}} \rho$, $\rho = \rho_i$, Eq.~(\ref{eq:41c}) applies, and $\tau_\mathrm{b} \alpha < \tau \alpha$.  Therefore, there is an amount of ``free'' surface tension energy $\tau_\mathrm{f}$ such that
\begin{subequations}
\label{eq:40a}
\begin{eqnarray}
\tau \alpha = & \tau_\mathrm{b} \alpha + \tau_\mathrm{f} \alpha &\label{eq:40aa}\\*
\tau_\mathrm{f} = & d_\mathrm{h} (\rho_\mathrm{h} - \rho_\mathrm{d}), & \, r_\mathrm{h}> r_\mathrm{sc}, \label{eq:40ab}\\*
\tau_\mathrm{b} = & 2 d_\mathrm{I} - \tau_\mathrm{f}, & \label{eq:40ac}\\*
\tau_\mathrm{f} = & 0, &r_\mathrm{h} \leq r_\mathrm{sc}. \label{eq40ad} 
\end{eqnarray}
\end{subequations}

The $\tau_\mathrm{f}$ is exerted on $\rho_i$.  For the remainder of this Paper, unless otherwise stated, the region under consideration is with $\rho_i < \rho_\mathrm{h}$.  Those hods that are oriented flat-to-flat emerge from the $r_\mathrm{sc}$ zone as assemblies of hods.  Other hods become independent.

Since the energy at the Source upon transition is high and since we observe Space to be more extensive than the hods, $0 < d_\mathrm{h} \ll 1$ ho.  

Consider the method of conjugate functions of a complex variable with a transformation
\begin{equation}
t \, = \,\frac{a}{2} \, \cosh ( w )
\label{eq:23},
\end{equation}
where $t=x+iz$ and $w=u+iv$.  Therefore, 
\begin{eqnarray}
\frac{x^2}{\cosh^2 u} \,+\, \frac{z^2}{\sinh^2 u} \, = \, \frac{a^2}{4}
\label{eq:24},\\*
\frac{x^2}{\cos^2 v} \,-\, \frac{z^2}{\sin^2 v} \, = \, \frac{a^2}{4}
\label{eq:25}.
\end{eqnarray}
Geometrically, the strip in the $w$-plane between the lines $v=0$ and $v=\pi$ transforms into the entire $t$ plane.  The line at $u=0$ between $v=0$ and $v= \pi$ transforms to the surface of the hod.  The lines $v=$constant are transformed into confocal hyperbolae lines which are streamlines of the $\rho_{ij}$ field where $\rho_{ij}$ is the $\rho$ at the $j^\mathrm{th}$ point due to the $i^\mathrm{th}$ hod.  The lines $u$=constant are transformed into confocal ellipses with the foci at the ends of the hod.  Since $\tau$ acts normally to the surface of the hod, $u$=constant are equipotential lines of $\rho_{ij}$ caused by $\tau$ [see Eq.~(\ref{eq:7})].

If the hod is circular around the $z$-axis, the $\rho_{ij}$ streamlines and equipotential lines are also circular around the $z$-axis.  The $\rho_{ij}$ equipotential lines form oblate spheroids with the minor axis along the $z$-axis.

Consider a point at a large distance from the hod, $r \gg a$, the $u$ of Eq.~(\ref{eq:24}) is very large.  Equation~(\ref{eq:24}) becomes $x^2+z^2=d^2$ where $d$ is the distance from the center of the hod.  At large $d$ the $\rho_{ij}$ equipotential lines are concentric spheres.  The $\rho_{ij}$ streamlines are the radii of the spheres.

Since the equipotential volume surrounded by $\rho_\mathrm{d}$ is an oblate spheroid, the distance $d_\mathrm{x}$ (in ho) in the plane of the hod to $\rho =\rho_\mathrm{d}$ is 
\begin{equation}
d_\mathrm{x} \, = \sqrt{d_\mathrm{h}^2 + 0.25} - 0.5
\label{eq:41a},
\end{equation}
which is much smaller than $a$.

Therefore, the $\rho =\rho_\mathrm{d}$ equipotential may be approximated as a cylinder with height $2 d_\mathrm{h}$ and end area $= \alpha$.

If $\bm{\nabla} \rho_\mathrm{s} > 0$ across a hod, the $\bm{\nabla}\rho$ on one side of the hod at an infinitesimal distance d$d$ from the $\rho=\rho_\mathrm{d}$ surface $\bm{\nabla}\rho_\mathrm{s1}(d + \mathrm{d} r)$ differs from the other side $\bm{\nabla}\rho_\mathrm{s2}(d - \mathrm{d} r)$.  Since the $\bm{\nabla} \rho$ is limited to $\bm{\nabla}_\mathrm{m} \rho$, the force of Space from Eq.~(\ref{eq:32}) is transferred to the hod surface.  Therefore, the net force d$\bm{F_\mathrm{s}}$ of Space on an element of surface d$s$ of the hod is
\begin{subequations}
\label{eq:43}
\begin{eqnarray}
\mathrm{d} \bm{F_\mathrm{s}} \, &=& \, D^2 (\bm{n} \bullet \bm{\nabla} \rho_\mathrm{s2} ) \mathrm{d} \bm{s} \, - \, D^2 (\bm{n} \bullet \bm{\nabla} \rho_\mathrm{s1} ) \mathrm{d} \bm{s}, \label{eq:43a}\\*
\mathrm{d} \bm{F_\mathrm{s}} \, & \approx & \, - D^2 (\bm{n} \bullet \bm{\nabla} \rho_\mathrm{{s}} ) \mathrm{d} \bm{s}, \label{eq:43b} 
\end{eqnarray}
\end{subequations}
where $\bm{\nabla} \rho_\mathrm{{s}}$ is considered approximately constant over the small $2 d$ distance.  This approximation fails in volumes close to $r_\mathrm{sc}$.

Unless otherwise stated, when distances greater than one ho are implied, $d \ll 1$ ho and $\rho = \rho_\mathrm{s}$ will be considered to be at the hod surface. 

Since the hod is supple, the application of d$\bm {F_\mathrm{s}}$ on d$\bm s$ of the hod in the x-y plane will cause d$\bm s$ to move normal to the hod surface.  If d$\bm {F_\mathrm{s}} = 0$ everywhere else, a deformation, d$u$, of the hod surface normal to hod surface will occur.  Since $\alpha$ and $\tau$ are constant, the edge of the hod must change shape.  A small deformation allows movement in only the z direction.  The edge stays in the x-y plane.  However, since edges of a free, individual hod are not constrained and $\tau$ is very strong, the hod membrane does not oscillate normal to the hod surface as it would for a stretched membrane.  Therefore, d$\bm {F_\mathrm{s}}$ will cause the hod to tend to conform to the $\bm{\nabla} \rho_\mathrm{s}$ streamlines.  Since Space has a relaxation time, the movement of hods will cause the streamlines to be continually shifting.  A slight delay, caused by $\iota$ and by $S$, in adjusting to a change in the streamline will allow interesting effects.  If $\bm{\nabla} \rho_\mathrm{s} \, \bullet \, \bm n \, \neq \, 0$ for a brief instant on only one edge of the hod and if the other edge experiences a force parallel to the surface $\bm{\nabla} \rho_\mathrm{s} \, \bullet \, \bm n \, = \, 0$.  The hod will travel along the streamline at the maximum allowed velocity.  If the streamline the hod is following has a very sharp bend, such as at the edge of another hod where the hod is much stronger than anywhere except near the Source, the inertia energy of the hod will let it continue.  For an instant, $\bm{\nabla} \rho_\mathrm{h} \, \bullet \, \bm n \, \neq \, 0$.  The hod will slow or even reverse direction.  This allows the transfer of energy between hods.

If the $\bm{\nabla} \rho$ field of Space varies little over the surface of the hod, the deformations become a general shaping of the hod causing the hod to tend to conform to the force lines, or streamlines, of $\rho$.  The hod will move to make $\bm{\nabla} \rho \bullet \bm n \, = \, 0$ on the surface of the hod.

If $\bm{\nabla} \rho \bullet \bm n \, = \, 0$ from some direction, the hod will experience zero Space impelling force from that direction.  If the $\bm{\nabla} \rho$ field changes little over a range of several hod diameters, small deformations with $\bm{\nabla} \rho \bullet \bm n \, \neq \, 0$ will cause the hod to have a forward inertia.

\subsubsection{\label{sec:grav}Gravity}

Call $\bm{F_{\mathrm{g} k}}$ the force of gravity on the $k^\mathrm{th}$ hod, 
\begin{equation}
\bm{F_{\mathrm{g} k}} \, \equiv \, - \, \frac{S_\mathrm{u}}{C} \,\sum_{k \neq j \, j=1}^\mathrm{N_t} \, \bm{\nabla} U_{kj}
\label{eq:22},
\end{equation}
where the negative sign means the force is directed opposite to increasing $U$ and the sum is over the other ($j^\mathrm{th}$) hods.

Define a volume $V$ containing Space for the distances being considered and throughout which the terms of Eq.~(\ref{eq:21}) except the $ U$ terms are constant.  By the Principle of Superposition, this allows the examination of just one force.  Equation~(\ref{eq:21}) can be solved for a steady-state condition.  We can consider $\rho$ at a point to be the superposition of the effects of each hod.  Therefore, Eq.~(\ref{eq:21}) at the $i^\mathrm{th}$ point due to the $j^\mathrm{th}$ hod becomes
\begin{equation}
\frac{\mathrm{d}}{\mathrm{d} t} \, (d_{ij} U_{j}) \, = \, \frac{S_\mathrm{u}}{C} \, \nabla^2 \, (d_{ij} U_{j} \, ) 
\label{eq:34b}
\end{equation}
in spherical coordinates.

The boundary conditions are:
\begin{subequations}
\label{eq:33as}
\begin{eqnarray}
U_{i} (d_{ij},0) \,&=&\,  \mathrm{a \, function \, of \, distance, \, only}, 
\label{eq:33as:a}\\*
U_{i} (d_{ij},t) \, &\rightarrow & \, 0 \, \mathrm{as} \, r \rightarrow \, \infty. \label{eq:33as:b}
\end{eqnarray}
\end{subequations}

Analogus to Section~\ref{sec:spac}, 
\begin{equation}
U_j = - \frac{K_\mathrm{u} C }{4 \pi S_\mathrm{u} } \frac{\tau_\mathrm{f} \alpha}{d_{ij}} 
\label{eq:34a}.
\end{equation}

The $| U_{ij} |$ is highest in a volume with a given $\tau_\mathrm{f} > 0$ value when the $k^\mathrm{th}$ and $j^\mathrm{th}$ hods have a shared $\rho=\rho_\mathrm{d}$ oblate spheroid surface and are orientated flat-to-flat.  Since $ d_{ij}$ is the distance to the center of the hod, $ d_{ij} \neq 0 $ and a singularity problem is non-existent.  

Chose two assemblies of hods where $N_\mathrm{t}$ is the number of hods in one assembly and $N_\mathrm{s}$ is the number of hods in the other assembly.  Let each assembly have a very small diameter compared to the distance $\bm r_\mathrm{ts}= <\sum_{j=1}^{N_\mathrm{t}} \sum_{k=1}^{N_\mathrm{s}} \bm d_{kj}>$ where $ < \, >$ means ``average of''.  The force of gravity of the hods in an assembly is greater on each other than the force of gravity of hods in the other assembly.  Therefore, the hods of an assembly will act as a solid body relative to the other assembly of hods.  Because hods act on Space, Space acts on Space and Space acts on hods, only the local $\tau_\mathrm{f}$ for each hod determines the $N \, \tau_\mathrm{f} \alpha$ on the hod.  Therefore, the force of gravity $\bm{F_\mathrm{g}}$ on each assembly due to the other assembly is 
\begin{equation}
\bm{F_\mathrm{g}} \,= - \frac{ (K_\mathrm{u} \alpha)^2}{ 4 \pi r_\mathrm{ts}^3} \, ( N_\mathrm{s} \tau_\mathrm{fs} ) \, ( N_\mathrm{t} \tau_\mathrm{ft} ) \, {\bm{r}}_\mathrm{ts} 
\label{eq:29},
\end{equation}
where  $\tau_\mathrm{fs}$ is the $\tau_\mathrm{f}$ for the assembly with $N_\mathrm{s}$ hods and $\tau_\mathrm{ft}$ is the $\tau_\mathrm{f}$ for the assembly with $N_\mathrm{t}$ hods.

For the simplified case of two symmetric distribution of hods with a distance $\bm{r}_\mathrm{12}$ between them and with $\tau_\mathrm{fs} = \tau_\mathrm{ft}$ 
\begin{equation}
\bm{F_\mathrm{g}} \,= - G \,\frac{m_\mathrm{g1} \, m_\mathrm{g2}}{ r_\mathrm{12}^3} \, {\bm{r}}_\mathrm{12}
\label{eq:30},
\end{equation}
where $G \, = \, ( 4 \pi )^{-1} \, (K_\mathrm{u} \alpha)^2$, $m_\mathrm{g1} = N_\mathrm{t} \tau_\mathrm{ft} $, and $m_\mathrm{g2} = N_\mathrm{s} \tau_\mathrm{fs}  $.  The experimental determination of $m_\mathrm{g1}$ and $m_\mathrm{g2}$ must use a method depending on $U_{ij}$ and no other factor such as motion which will involve $\iota \alpha$ forces.

Note, in volumes within a sphere with a radius of $r_\mathrm{sc}$ from a galaxy's Source, $\tau_\mathrm{f} = 0$ and, therefore, $\bm{F_\mathrm{g}} =0$ and $m_\mathrm{g}=0$.  In volumes with $ r_\mathrm{sc} < r_\mathrm{h}$, $\tau_\mathrm{f}$ increases as $r_\mathrm{h}$ increases.

\subsection{\label{sec:part}Particles}

Define a particle to be an assembly of hods surrounded by a $\rho=\rho_\mathrm{d}$ equipotential surface (ES) that forms a connected volume around the hods in the particle within which Space is totally bound to the hods by $\tau_\mathrm{b}$.  Therefore, $\rho_\mathrm{s}$ outside the ES equals $\rho_\mathrm{d}$ within the ES of the particle.  If the ES is simply connected, the particle is called a simple particle and the distance for gravity potential calculations is to the center of the particle.  The potential on a surface element of the ES is determined by the distribution of hods inside the surface.  However, the volume occupied by the bound Space is the same for each hod.  Therefore, the hods are uniformly distributed over the volume inside a simply connected ES.  Since the hods of a simple particle have a minimum distance between them, the hods are tightly bound and respond as a single entity.  By the Principle of Minimum Potential Energy, the hods of a simple particle have a more stable structure with a maximum $N \tau_\mathrm{b} \alpha$ per unit surface area ratio (P/S).

Hods are unable to join outside the Source region to form a particle but may be able to join an existing particle.  The existing particle must have sufficient $N (\tau_\mathrm{f} \alpha )$ to overcome the $\iota \alpha$ of the joining hod.  Also, this implies there is a minimum number $N_\mathrm{min}$ of hods which must be combined before other hods may join it outside the Source region. 

When the distance between the flat of the hods $\leq 2 d_\mathrm{h}$, the $\bm{\nabla} \rho$ between them will change sign and become ${\bm\nabla}_\mathrm{m}\rho$ which will cause a net repulsive force between hods.  Therefore, the distance between the flat of the hods will stabilize at a distance $2 d_\mathrm{h}$ when the repulsion due to ${\bm\nabla}_\mathrm{m}\rho$ will equal $F_\mathrm{{g}}$.  If the distance between the flat of the hods $ > 2 d_\mathrm{h}$, the hods will be independent.

Also, since $d_\mathrm{h} \ll 1$ ho, hods may form in a column arrangement.  Call this type of structure Family 1 (F1) Type 1 (T1) structures with the length $L = 2 N_\mathrm{c} d_\mathrm{h} $ where $N_\mathrm{c}$ is the number of hods in the F1T1 structure.  The potential can become large enough that hods can be added without being in a Source $r_\mathrm{sc}$ zone.  The P/S of the F1T1 structure will be least when 2$N_\mathrm{min}d_\mathrm{h} = 1$ ho The addition of hods to the column also produces stable, cylinder structures since each additional hod is oriented flat-to-flat.  Therefore, the addition of hods to column length decreases the P/S.  However, for a large number of hods, the outer hods are easier to remove.  The direction parallel to the surface has $\bm{\nabla} \rho \bullet \bm n \, = \, 0$ so $F_\mathrm{s} = 0$ and maximal speeds can be achieved in a direction parallel to the surface.  This structure is a photon.  The direction along the axis has a one-hod surface.  Hence, $F_\mathrm{s} $ along the cylinder axis is as if only one hod were present.

If the inertia and external forces cause the hods in the F1T1 to start to loose overlap, the sharp $\bm{\nabla} \rho$ field at the edges of hods will push them back.  The total overlap condition is a minimum potential energy position.

Any deformation in the surface from a ripple in Space will tend to bend the hod surfaces to lose total overlap.  Since this causes a force to reestablish overlap, this structure is much more rigid than a single hod.  Because this structure is rigid, a $\bm{\nabla} \rho$  across the structure can impart a torque on the particle.

As more hods are added, the $N_\mathrm{c} ( \tau \alpha )$ increases.  The energy of the photon h$\nu$ is 
\begin{equation}
\mathrm h \nu \, = \, N_\mathrm{c} ( \tau \alpha )
\label{eq:46}.
\end{equation}

Ripples in Space will cause ripples in the surface of the outer hod.  This will slow the photon structure and cause a shearing $F_\mathrm{s}$ on the outer hod.  Thus, if $N_\mathrm{c} \, > \,N_\mathrm{min}$ in a photon, the photon can loose hods or gain hods.

Two cylindrical photons may join as side-by-side cylinders caused by $F_\mathrm{g}$.  Parallel to the surface, $\bm{\nabla} \rho_\mathrm{s} \bullet \bm n \, = \, 0$ for both. 

Also, two cylindrical photons may join in a T formation if $L > 1$ ho as the Principle of Change requires.  By extension, if $L \approx 2$ ho, there may be other cylinders added to form II, III, etc. structures.  This requires the photons in the vertical columns to have approximately, if not exactly, the same number of hods.  Otherwise, one or more of the photons will experience a $\bm{\nabla} \rho \bullet \bm n \, \neq \, 0$.  There will still be a direction where $\bm{\nabla} \rho \bullet \bm n \, = \, 0$.  Call this structure a F1, Type 2 (T2) structure.  Perhaps this is the structure of a lepton. 

Consider a F1T2 structure made of several columns and a cross member at each end.  If an end column does not overlap the cross member column, an extremely unstable condition exists.  The end column would be freed.  If a cross member is longer than the width of the row of columns, the particle could acquire another column.  The most stable structure would be when $L$ is an integer multiplier $I$ of the average diameter of a ho 
\begin{equation}
L= 2 I ( \frac{\alpha}{\pi} )^{1/2}
\label{eq:46a}.
\end{equation}

Similarly, a third photon can join the T structure so that the three are mutually orthogonal, as the Principle of Repetition allows.  However, $\bm{\nabla} \rho \bullet \bm n \, \neq \, 0$  in all directions.  Call this structure a F1, Type 3 (T3) structure.  A cross cylinder on a cross cylinder of a F1T2 structure makes the F1T3 box structure very dense and very stable since there is little ``free'' surface (unattached hod faces) compared to the $\tau_\mathrm{f} \alpha$ energy contained and P/S is minimal.  Perhaps this is the structure of quarks, quark stars, and black holes.

In small structures, the F1T3 cube has the least P/S.  However, this allows ``corners''.  In larger structures, a rounding effect such as found in crystal structures produces a more spherical ES which will have a lower P/S.  This could be a mechanism to fracture some combinations (particles) and cause faceting in all combinations like in crystals. 

\begin{figure}
\includegraphics[width=0.4\textwidth]{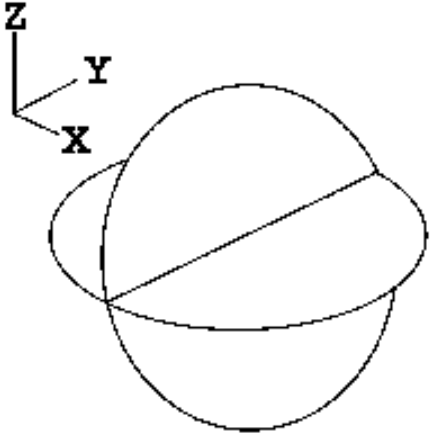}
\caption{\label{fig:1} Diagram showing the structure of the Family 2 (Muon family) fundamental particles.}
\end{figure}

Another possible structure of hods with a position of least P/S and stable ${\bm\nabla} \rho$ field is with each inserted through and normal to each other (see Fig.~\ref{fig:1}).  Call the particles formed with this structure Family 2 (F2) particles.  The particle so formed has a direction along the line where they join where $\bm{\nabla} \rho \bullet \bm n \, = \, 0$  can occur.  These may be the Muon family of particles.

\begin{figure}
\includegraphics[width=0.4\textwidth]{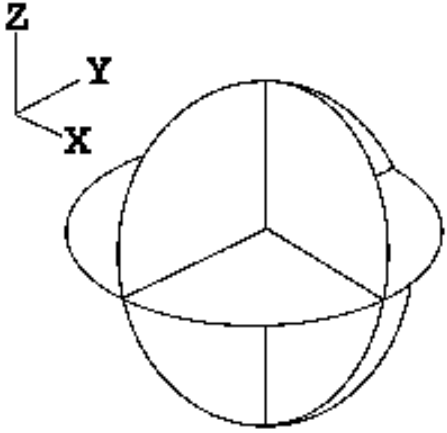}
\caption{\label{fig:2} Diagram showing the structure of the Family 3 (Tau family) fundamental particles.}
\end{figure}

By a similar argument, with less probability, three hods may come to form a particle with three mutually orthogonal hods (see Fig.~\ref{fig:2}).  Call the particles formed with this structure Family 3 (F3) particles.  These may be the Tau family of particles.

However, a $4^\mathrm{th}$ mutually orthogonal hod construction in 3D outside the immediate area of a Source is not feasible since one of the hod's surfaces will face another hod surface.

Therefore, because the hod defines 3D, there are three, and only three, families of particles.

In our volume of the universe, particles of the same type have consistent mass.  Particles become larger from gravitational attraction of additional hods as the Principle of Repetition allows.  The observed consistent mass of like particles implies the existence of the Principle of Negative Feedback to limit the size.  For T2 and T3 particles, if $L > I (\alpha / \pi)^{1/2}$ from Eq.~(\ref{eq:46a}), attracted hods may be added to the particle.  As noted previously, the attracted hod will be traveling at the maximum speed which implies a maximum $T$.  Thus, the attracted hod has energy great enough to break the gravitational bond and facet the outlying hods from the particle.  If $L < I (\alpha / \pi)^{1/2}$, the energy of the attracted hod may cause the column to facet. 

In a volume with slow changing $\rho_i$, $\rho_\mathrm{d} = \rho_i $.  Therefore, $L$ is a function of $\rho_i$.  In a volume with rapidly changing $\rho_i$, $\rho_\mathrm{d}$ may lag the change in $\rho_i$ because $\tau_\mathrm{b}$ may be absorbing or releasing Space.  Such may be the situation for neutron stars, possible quark stars, and black holes, which exist near the Source of the Galaxy. 

If the particles are to have more hods (energy/mass) as the Anthropic Principle requires, there must be another binding mechanism less sensitive to faceting as the Principle of Change requires.  The flat side of the end hods on a photon would be attracted to and could be bound to other particles.  If quarks are the largest Family ``type'' particles, baryons must be quarks and leptons joined by another means such as directly to each other or by photons.  The particle will be in a connected volume, but not a simply connected volume.  By the Principle of Repetition, the atom is electrons, neutrons, and protons joined by photons.  The photon is similar to a rod in these structures.

\subsection{\label{sec:Sink}Sink characteristics}

If a Source ceases to erupt or if some matter is repelled to the low $\bm{\nabla} \rho$ distances between Sources, gravity will cause the hods to coalesce.  In a volume where $\bm{\nabla} \rho$ becomes near zero, the attraction of hods will cause a high hod density.  If the number of hods is greater than in a black hole in this intergalactic volume, the intense force on the supple hods may cause the hods to form spheres like bubbles.  With increased pressure from a huge number of hods, the hod can go only out of our universe and turn into a four dimensional 4D object like 2D turned to 3D.

Since the Sink requires mass to form, the Sink's age is considerably less than galaxies.  Therefore, the delay before matter and Space is sent into the 4D allows 3D to accumulate mass.

The Sink requires a minimum mass to send hod bubbles into 4D.  Therefore, the mass $M_\mathrm{sink}$ around the Sink has an upper limit.  If the Sink is sending hods to 4D, then a negative feedback condition is formed and the rate $\eta$ of transition of the hods is dependant on the rate of accretion of hods.  Therefore, $\eta \propto M_\mathrm{sink}$.

The $\bm \nabla \rho$ acts in the same direction as the gravity force.  The Sink can send hods to 4D if the hods are close.  Therefore, the $\bm \nabla \rho = \bm \nabla_\mathrm{m} \rho $ around the Sink like around the Source.

\section{\label{sec:gala}GALAXY MODEL}

\subsection{General}

\begin{figure}
\includegraphics[width=0.4\textwidth]{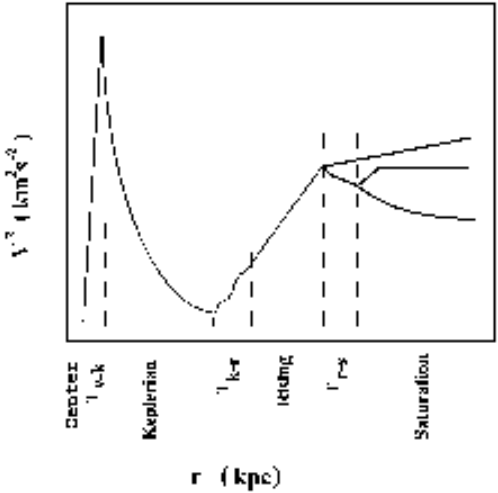}
\caption{\label{fig:3} Diagram depicting the regions of a galaxy in relation to a hypothetical rotation curve.  The lines in the Saturation Region depict the different shapes possible.}
\end{figure}

For the purpose herein, a galaxy's RC is divided into five regions~\cite{binn} as the $r$ from the Source increases (diagramed in Fig.~\ref{fig:3}), a center region (CR)~\cite{silk,baga,baga2}, a Keplerian region (KR)~\cite{ghez,fer5}, a rising region (RR) which is in the outer bulge region, a saturation region (SR) which is in the disk region, and a detached region (DR) which is outside the galaxy (herein, collectively referred to as the ``lettered regions'').  The distinct nature of each region of the RC implies different physics exists among the regions.  Therefore, some of the parameters of the equation of motion have suddenly become more or less significant.  Each of the lettered regions is separated by a transition region in which the physics is changing from one region to the next.  The CR and KR are separated by a transition region T$_\mathrm{c-k}$~\cite{scho, ghez,ghez2,wang}.  The KR and RR are separated by a transition region T$_\mathrm{k-r}$ in which the RC has a slight decline followed by a rapid rise.  A similar transition region T$_\mathrm{r-s}$ appears between the RR and SR.  Particles in DR are not in orbit around the galaxy center.  

Dividing into regions with different, predominant physics has been done for models of other phenomena. For instance, the physics of semiconductors uses such a model to describe differing current-voltage characteristics of the differing regions of transistor operation~\cite{grov}.

The $\tau_\mathrm{f} = 0$ out to $r_\mathrm{sw}$ and only unbound hods and photons exist near the Source.  By Eq.~(\ref{eq:37}) the size of $\epsilon $ is proportional to $r_\mathrm{sc}$ and, therefore, to the diameter of the Source region.  As the hods and photons travel outward beyond $r_\mathrm{sc}$, $\tau_\mathrm{f} > 0$ and particles form.

As the particles' radius $r$ from the Source increases, larger particles can form.  Outward from the Source, photons coalesce to form larger particles.  In the Big Bang model, the physics of the formation of particles is outward in time.  In the CUM, the physics of formation of subatomic particles is outward in distance from the Source.  Eventually, hydrogen forms and is pushed outward by the Space force.  As hydrogen coalesces, larger atoms are formed which fall back to the center as the gravitational force becomes larger than the Space force.  Gravity attracts the densest particles into the center.  As the particle approaches $r_\mathrm{sc}$, the declining $\tau_\mathrm{f}$ and $ L$ causes the T2 and T3 particles to disintegrate to photons and re-radiate outward.  

Consider a $\bigtriangleup V$ containing a test particle with $N$ hods such as a star or hydrogen gas in a galaxy.  The inertial mass $m_\iota$ is $N\iota \alpha$, the gravitational mass $m_\mathrm{g}$ is $N\tau_\mathrm{f} \alpha$, and the cross section mass $m_\mathrm{s}$ is the effective cross section subject to the Space force of the hods in $\bigtriangleup V$.  For simplicity, the number of hods in the particle is assumed to be constant over time.  Since stars are changing the number of hods by radiation emission and are changing the $ m_\mathrm{s}/m_\mathrm{\iota}$ ratio through changing elemental composition, this assumption is most nearly true when the $\bigtriangleup V$ contains hydrogen gas.  Therefore, the H{\scriptsize{I}} RC is preferred, where possible, to reflect the forces influencing $v$.  

Assume the particle in $\bigtriangleup V$ originated in the galaxy ($d_\mathrm{b} =0$) and the particle in $\bigtriangleup V$ is in equilibrium ($E_\mathrm{v}=N \tau_\mathrm{f} \alpha$).  The coordinate system center is placed at the Source and is aligned in a constant orientation relative to our view.  The velocity and rotation of the galaxy relative to the universe is not part of $v$.  Combining Eqs.~(\ref{eq:35}), (\ref{eq:34a}), (\ref{eq:6a}), and (\ref{eq:6d}) into Eq.~(\ref{eq:11}), rearranging terms, and considering only radial terms yields 
\begin{equation}
v^2 = r \ddot{r} + \frac{m_\mathrm{g}}{m_\mathrm{\iota}} \frac{G M}{r} -  \frac{m_\mathrm{s}}{m_\mathrm{\iota}} \frac{G_\mathrm{s} \epsilon}{r} -  \bm r  \bullet \bm a_\mathrm{o} 
\label{eq:49},
\end{equation}
where the double dots above the $r$ mean the second derivative with respect to time, $\ddot{r}$ is the radial acceleration of the test particle, $G$ is the gravitational constant, $G_\mathrm{s}$ is the cross section constant, $M$ is the mass of the galaxy inside a sphere with a center at the Source and a radius of $r$, $\epsilon$ is the Source strength of the galaxy, 
\begin{subequations}
\label{eq:50}
\begin{eqnarray}
\bm a_{\mathrm{o}i} &=& \sum_{j=1}^\mathrm{ other \, galaxies} \frac{ m_\mathrm{g} G M_{j} - m_\mathrm{s} G_\mathrm{s} \epsilon_{j}}{m_\mathrm{\iota} r_{ij}^3} \bm r_{ij} \nonumber \\*
& &  + \sum_{k=1}^\mathrm{ Sinks} \frac{ m_\mathrm{g} G M_{k} + m_\mathrm{s} G_\mathrm{s} | \eta_{k} | }{m_\mathrm{\iota} r_{ik}^3} \bm r_{ik}, \label{eq:50:a} \\* 
\bm a_{\mathrm{o}} &=& \sum_{i=1}^\mathrm{N} \bm a_{\mathrm{o}i}, \label{eq:50:b} 
\end{eqnarray}
\end{subequations}
$M_{j}$ is the total mass of the $j^\mathrm{th}$ galaxy, $\epsilon_{j}$ is the Source strength of the $j^\mathrm{th}$ galaxy, $\bm r_{ij}$ is the distance from the $i^\mathrm{th}$ test particle to the $j^\mathrm{th}$ galaxy's Source, $M_{k}$ is the total mass of the $k^\mathrm{th}$ Sink, $\eta_{k}$ is the Sink strength of the $k^\mathrm{th}$ Sink, $ m_\mathrm{\iota}$ is the inertial mass of the test particle, and $\bm r_{ik}$ is the distance from the $i^\mathrm{th}$ test particle to the $k^\mathrm{th}$ Sink. 

If $\bm a_\mathrm{o} \neq 0$ in a galaxy, then the center of mass may be slightly displaced from the Source.  The $v$ is measured along the major axis from our view.  Therefore, the $\bm r$ is directed along the major axis.  The form of Eq.~(\ref{eq:50}) for baryons is of a Newtonian force with an effective mass less than the total mass in a galaxy.  Also, the force on F1T1 (photons) particles passing by the galaxy differs because of the orientation of the particle relative to the galaxy.

In our part of the Galaxy, $m_\mathrm{\iota} = m_\mathrm{g}$ to within one part in 10$^{11}$~\cite{will}.  However, the measurement of $ m_\mathrm{g}$ depends on $\bm F_\mathrm{g}$ and, therefore, $\tau_\mathrm{f}$ that depends on $\rho$.  The ratio of the diameter of the earth divided by the distance from the center of the Galaxy suggests much greater precision is required for earth bound measurements.  Therefore, the $ m_\mathrm{g} / m_\mathrm{\iota}$ ratio will be explicitly stated in this Paper where applicable.

The notation convention used herein is as follows.  The italic, normal size letter will denote the parameter, as usual.  The upper case $M$ means the total mass inside the radius denoted by the sub letters.  The sub letters denote the radius range by region identification.  The use of ``max'' or ``min'' in the sub letters will denote the maximum value or minimum value the parameter has in the region, respectively.  For example, the radius $r_\mathrm{k}$ (kpc) in the KR which is between the radius $r_\mathrm{tckmax}$ at the maximum extent of the T$_\mathrm{c-k}$ region and the radius $r_\mathrm{kmax}$ at the maximum extent of the KR region.  Hence, $r_\mathrm{tckmax} < r_\mathrm{k} \leq r_\mathrm{kmax}$.  The notation $M_\mathrm{kmax}$ means the total mass inside the sphere with radius $r_\mathrm{kmax}$.  This is the total mass of the CR, T$_\mathrm{c-k}$, and KR regions.

The CUM suggests the change in physics among the lettered regions derive from a change in the particle $m_\mathrm{s} / m_\mathrm{\iota}$, in the forces ($\bm {F_\mathrm{ts}}$ and $\bm {F_\mathrm{tg}}$), and in $ \bm r  \bullet \bm a_\mathrm{o}$.  The lettered regions change is shared by all galaxies and, hence, may be used to derive relations among galaxies.

The following sections discuss each region.  Some sections have data that can be compared to the equations developed.

\subsection{\label{crkr}CR to KR}

The CUM posits that the radius of the CR is less than a few light hours around a Source at the center of a galaxy and is less than $r_\mathrm{sc}$.  At a radius of less than $r_\mathrm{sw}$, there is no $v$ and the problem of a singularity in Eq.~(\ref{eq:49}) at $r = 0$ is nonexistent.  Since $\tau_\mathrm{f}$ decreases as the $r$ decreases below $r_\mathrm{sc}$, the gravitational term is decreasing and the $\epsilon$ term dominates to repel particles. 

The unbound hods and photons in the CR are repelled away from the Source.  The ${\epsilon} /r $ close to the Source is large.  Therefore, only exceptionally dense particles with very small $m_\mathrm{s} / m_\mathrm{\iota}$ are closest to the Source.  Black holes which may be F1T3 particles and other very dense matter fall into the T$_\mathrm{c-k}$ of galaxies.  All other matter is repelled outward.  Black holes form a shell around the CR~\cite{binn, silk}.

Black holes may increase $m_\mathrm{s} / m_\mathrm{\iota}$ by collision and merger or may loose angular momentum such that they fall into the $r < r_\mathrm{sc}$ zone.  The result of the $L \rightarrow 0$ and $\tau_\mathrm{f} \rightarrow 0$ is a multitude of X-ray point sources without accompanying large variation at longer wavelengths near the center of the Milky Way.  Note the force of gravity must be reduced to allow the photons to separate in a burst.  The diffuse component of the X-ray sources may be other low $m_\mathrm{s} / m_\mathrm{\iota}$ particles such as iron nuclei or neutron stars with F1T3 centers~\cite[and references therein]{pras}.  At and near $r_\mathrm{sc}$ the brightness of the radiation will increase relative to star luminosity due to the reclaiming of the dense particles.

If mass does fall into the T$_\mathrm{c-k}$, the higher mass $M_\mathrm{tckmax}$ in the  T$_\mathrm{c-k}$, causes more infall into the CR.  The resulting increase in re-radiated photons causes a decrease in $M_\mathrm{tckmax}$.  From Eq.~(\ref{eq:49}), the $\epsilon$ determines the stable value of $M_\mathrm{tckmax}$.  This is the observed feedback controlled mechanism required by \citet{merr2} as proposed by the CUM. 

\subsection{\label{kr}KR}

The start of the KR is caused by a sudden and rapid change in particle type.  Almost all the very dense particles ( low $m_\mathrm{s}/m_\mathrm{\iota}$) are in the T$_\mathrm{c-k}$ shell.  As $r_\mathrm{k}$ increases, the number of dense particles per unit radius decreases, the resident particles's $m_\mathrm{s}/m_\mathrm{\iota}$ term is greater, and the $M_\mathrm{tckmax}$ is much larger than the gravitational effects of particles in the KR upon each other.  The observed Keplerian motion of particles in the KR~\cite{ghez,fer5} implies the mass $M_\mathrm{k}$ within a sphere of radius $r_\mathrm{k}$ is approximately $M_\mathrm{tckmax}$.  Also, since $F_\mathrm{g} \propto \tau_\mathrm{f}$ of the test particle, the observed Keplerian motion of particles in the KR implies the $\tau_\mathrm{f}$ of the test particle is nearly constant in the orbit of each test particle.  

The $m_\mathrm{s}/m_\mathrm{\iota}$ term is causing lighter and denser particles to move out of the KR.  The $m_\mathrm{s} / m_\mathrm{\iota}$ for hydrogen and the lighter elements is much higher than the massive particles and the $m_\mathrm{g}$ is lower.  Thus, for hydrogen in the KR the predominant term is the $\epsilon$ term which pushes hydrogen and lighter particles out of the KR.  The only lighter elements in the KR are gravitationally bound to the denser particles.  Black holes are either in or on their way to T$_\mathrm{c-k}$.  Therefore, Eq.~(\ref{eq:49}) for particles in the KR becomes 
\begin{equation}
v_\mathrm{k}^2 = r_\mathrm{k} \ddot{r}_\mathrm{k} + \frac{m_\mathrm{g}}{m_\mathrm{\iota}} \frac{G M_\mathrm{eff} }{r_\mathrm{k}} - K_\mathrm{k} 
\label{eq:51},
\end{equation}
where 
\begin{equation}
M_\mathrm{eff} \approx M_\mathrm{tckmax} - \frac{G_\mathrm{s}}{G} \frac{m_\mathrm{s}}{m_\mathrm{g}} \epsilon 
\label{eq:52},
\end{equation}
$v_\mathrm{k}$ is the $v$ for particles in the KR, and $K_\mathrm{k}$ is a relatively small constant accounting for the $ \bm a_\mathrm{o} $ term.  An upper case K followed by a sub number or letter denotes a constant.

The $\ddot{r}_\mathrm{k}$ in Eq.~(\ref{eq:51}) has a value to cause elliptical orbits for the particles which stay in the KR. 

As $r_\mathrm{k}$ increases, $v_\mathrm{k}\rightarrow 0$ in a Keplerian decline.  Since differing particles have different $m_\mathrm{s} / m_\mathrm{\iota}$ ratios such as stellar material and hydrogen, the minimum $v_\mathrm{k}$ for different particle species is reached at different $r_\mathrm{k}$.

The sign of $\bm r_\mathrm{k} \bullet \bm a_\mathrm{o}$ is opposite on opposite sides of the galaxy.  Also, the distance between $r_\mathrm{k}$ on one side to $r_\mathrm{k}$ on the other side is of the order of a few kpc.  Compared to the other terms, $\bm r_\mathrm{k} \bullet \bm a_\mathrm{o}$ is small.  Therefore, the residual of the $\bm r_\mathrm{k} \bullet \bm a_\mathrm{o}$ from each side of the galaxy is very small.  Therefore, the luminosity curve as a function of $r_\mathrm{k}$ will be only slightly different on each side of the center.  Therefore, averaging the rotation velocity from side to side reduces the effect of relatively small $\bm r_\mathrm{k} \bullet \bm a_\mathrm{o}$ term.

Since the $M_\mathrm{tckmax}$ is concentrated in a small volume, the KR is spherical.  Galaxies are usually inclined relative to our view.  Because KR is spherical, the diameter of the KR is measured as the distance along the major axis of the ellipse (from our view).

Since the $m_\mathrm{s} / m_\mathrm{\iota}$ particle type differs with radius, this Paper posits that in the high force volume of the KR the $m_\mathrm{s} / m_\mathrm{\iota}$ particle types are in strata and the strata's radius depends on $\epsilon$.  If each $m_\mathrm{s} / m_\mathrm{\iota}$ particle type has a different mass to luminosity ratio, the surface brightness $I$ versus $r$ curve ($I-r$ curve) along the major axis has discontinuities or inflection points.  Define $r_\mathrm{\epsilon d}$ (in pc) as the average radius along the $I-r$ curve in each direction from the Source to the first measured discontinuity or inflection point.  Therefore, Eq.~(\ref{eq:37}) suggests $r_\mathrm{sc} \propto r_\mathrm{\epsilon d}$ and 
\begin{equation}
\epsilon = K_\mathrm{\epsilon d} r_\mathrm{\epsilon d}
\label{eq:53},
\end{equation}
where 
$K_\mathrm{\epsilon d}$ is the proportionality constant. 

Further, the abrupt change in $m_\mathrm{s} / m_\mathrm{\iota}$ particle type occurs in all galaxies.  Therefore, $ r_\mathrm{\epsilon d}$ may be used to compare parameters among galaxies.

Evaluating Eqs.~(\ref{eq:49}) and (\ref{eq:53}) at $r_\mathrm{\epsilon d}$, solving for $r_\mathrm{\epsilon d}$ with the net effect $\bm r_\mathrm{k} \bullet \bm a_\mathrm{o} \approx 0$ when values from each side of the galaxy are averaged yields 
\begin{equation}
K_\mathrm{\epsilon d} = \frac{m_\mathrm{\iota}}{m_\mathrm{s} G_\mathrm{s}} \left[ v^2 (r_\mathrm{\epsilon d}) - \frac{m_\mathrm{g} G M (r_\mathrm{\epsilon d})}{m_\mathrm{\iota} r_\mathrm{\epsilon d} - r_\mathrm{\epsilon} \ddot r_\mathrm{\epsilon}} \right] 
\label{eq:54},
\end{equation}
where $ M (r_\mathrm{\epsilon d})$ is the mass within a sphere of radius $r_\mathrm{\epsilon d}$ and $v(r_\mathrm{\epsilon d})$ is the rotation velocity of particles at $ r_\mathrm{\epsilon d}$.

The term in the braces is the difference between the measured $ v^2 (r_\mathrm{\epsilon d})$ and the Newtonian expectation.    

Substituting Eq.~(\ref{eq:53}) into Eq.~(\ref{eq:52}) yields
\begin{equation}
M_\mathrm{eff} = M_\mathrm{tckmax} - \frac{G_\mathrm{s}}{G} K_\mathrm{\epsilon d} \frac{m_\mathrm{s}}{m_\mathrm{g}} r_\mathrm{\epsilon d} + K_\mathrm{ieff}
\label{eq:55},
\end{equation}
where $K_\mathrm{ieff}$ is the intercept of the $M_\mathrm{eff} - r_\mathrm{\epsilon d}$ plot.  Theoretically, $K_\mathrm{ieff} =0$ and is a result of measurement error.

\subsubsection{\label{MBH}$M_\bullet\,-\,r_\mathrm{\epsilon d}$ data}

The kinematic measure of $M_\bullet$ uses the relation $r \sigma_\mathrm{v}/G$.  Therefore, from Eq.~(\ref{eq:51}) $M_\bullet \approx M_\mathrm{eff}$, where the mass in the KR is relatively much smaller than the mass in the T$_\mathrm{c-k}$. 

To test Eq.~(\ref{eq:55}), $M_\mathrm{\bullet}$ data using kinematic methods and distance data for 20 sample galaxies were taken from \citet{merr2}.  The data for the 20 sample galaxies are listed in Table~\ref{tab:1}.  The data points are the measured values for the galaxies for which the SBH sphere of influence ($r_\mathrm{h}$ in \citet{merr2}) has been resolved.  The morphology data were obtained from the NED database~\footnote{The NED database is available at: http://nedwww.ipac.caltech.edu.  The data were obtained from NED on 5 May 2004.}.  The sample includes E, SO, and S type galaxies.
\begin{table*}
\caption{\label{tab:1} Data of the galaxies in Fig.~\ref{fig:4a}. }
\begin{ruledtabular}
\begin{tabular}{llllrllc}
{Galaxy\footnotemark[1]}
&{Morphology}
&{\phantom{000000}HST}&
&{\phantom{0}Distance\footnotemark[2]} 
&{\phantom{0}$r_\mathrm{\epsilon d}$\footnotemark[3]}
&{\phantom{000}$M_\bullet$\footnotemark[4]}
&{line\footnotemark[5]} \\ 
& &image&filter\phantom{0}&Mpc\phantom{0}&\phantom{0}pc& \phantom{000}$10^{8} M_\odot$&$n$ \\
\hline
I1459&E3; LINER&U2BM0102T& F555W&30.3&\phantom{0}99.5$\pm$14.7&\phantom{0}4.6\phantom{00}$\pm$ 2.8&2\\
N0221&cE2&U2E20309T& F555W&\phantom{0}0.8&\phantom{00}3.7$\pm$  0.4&\phantom{0}0.039$\pm$ 0.009&1\\
N2787&SB( r)0+  LINER&U39S0304R& F555W&\phantom{0}7.5&\phantom{0}14.6$\pm$ 3.6&\phantom{0}0.41\phantom{0}$^{+0.04}_{-0.05}$&2\\
N3031&SA(s)ab: LINER Sy1&U32L0105T& F547M&\phantom{0}3.9&\phantom{00}5.5$\pm$  1.9&\phantom{0}0.68\phantom{0}$^{+0.07}_{-0.13}$&3\\
N3115&S0-&U2J20B04T& F555W&\phantom{0}9.8&\phantom{0}49.1$\pm$\phantom{0}4.8&\phantom{0}9.2\phantom{00}$\pm$ 3.0&4\\
N3245&SA(r)\^{ }0\^{ }&U3MJ1103R& F547M&20.9&\phantom{0}35.5$\pm$10.1&\phantom{0}2.1\phantom{00}$\pm$ 0.5&3\\
N3379&E1: LINER&U2J20F03T& F555W&10.8&\phantom{0}51.7$\pm$\phantom{0}5.2&\phantom{0}1.35\phantom{0}$\pm$ 0.73&2\\
N3608&E2  LINER&U2BM0502T& F555W&23.6&\phantom{0}56.5$\pm$11.4&\phantom{0}1.1\phantom{00}$^{+1.4}_{-0.3}$&2\\
N4258&SAB(s)bc;LINER Sy1.9&U6712101R& F547M&7.2&\phantom{0}12.1$\pm$\phantom{0}3.5&\phantom{0}0.39\phantom{0}$\pm$ 0.034&2\\
N4261&E2-3: LINER&U2I50204T& F547M&33.0&189.4$\pm$16.0&\phantom{0}5.4\phantom{00}$\pm$ 1.2&2\\
N4342&S0-&U32Q0206T& F555W&16.7&\phantom{0}36.9$\pm$ 8.1&\phantom{0}3.3\phantom{00}$^{+1.9}_{-1.1}$&3\\
N4374&E1: LERG LINER&U34K0102T& F547M&18.7&240.2$\pm$ 9.1&17\phantom{.000}$^{+12}_{-6.7}$&3\\
N4473&E5&U3071503T& F555W&16.1&\phantom{0}49.1$\pm$ 7.8&\phantom{0}0.8\phantom{00}$^{+1.0}_{-0.4}$&1\\
N4486&cE0&U2900104T& F547M&16.7&\phantom{0}32.1$\pm$ 8.1&35.7\phantom{00}$\pm$10.2&6\\
N4564&E6&U3CM7202R& F702W&14.9&\phantom{0}26.0$\pm$ 7.2&\phantom{0}0.57\phantom{0}$^{+0.13}_{-0.17}$&2\\
N4649&E2&U2QO0301T& F555W&17.3&\phantom{0}52.7$\pm$ 8.4&20.6\phantom{00}$^{+5.2}_{-10.2}$&5\\
N5128&S0pec&U4100106M& F555W&\phantom{0}4.2&\phantom{0}33.2$\pm$ 2.0&\phantom{0}2.4\phantom{00}$^{+3.6}_{-1.7}$&3\\
N5845&E&U3070903T& F555W&28.5&167.3$\pm$13.8&\phantom{0}2.9\phantom{00}$^{+1.7}_{-2.7}$&2\\
N6251&E:LERG Sy2&U2PQ0701T& F555W&104\phantom{.0}&206.9$\pm$50.4&\phantom{0}5.9\phantom{00}$\pm$ 2.0&2\\
N7052&E&U2P90108T& F547M&66.1&400.3$\pm$32.1&\phantom{0}3.7\phantom{00}$^{+2.6}_{-1.5}$&1\\
\end{tabular}
\end{ruledtabular}

\footnotetext[1]{``I'' means IC and ``N'' means NGC..}
\footnotetext[2]{The distance is as stated in the reference for $M_\bullet$. }
\footnotetext[3]{The $r_\mathrm{\epsilon d} $ (kpc).  The error is $\pm$ 0.1 arcsec.}
\footnotetext[4]{The $M_\bullet$ was obtained from \citet{merr2}.}
\footnotetext[5]{The line is the ``n'' listed in Table~\ref{tab:2}.}

\end{table*}

\begin{figure*}
\includegraphics[width=7in, height=8.5in]{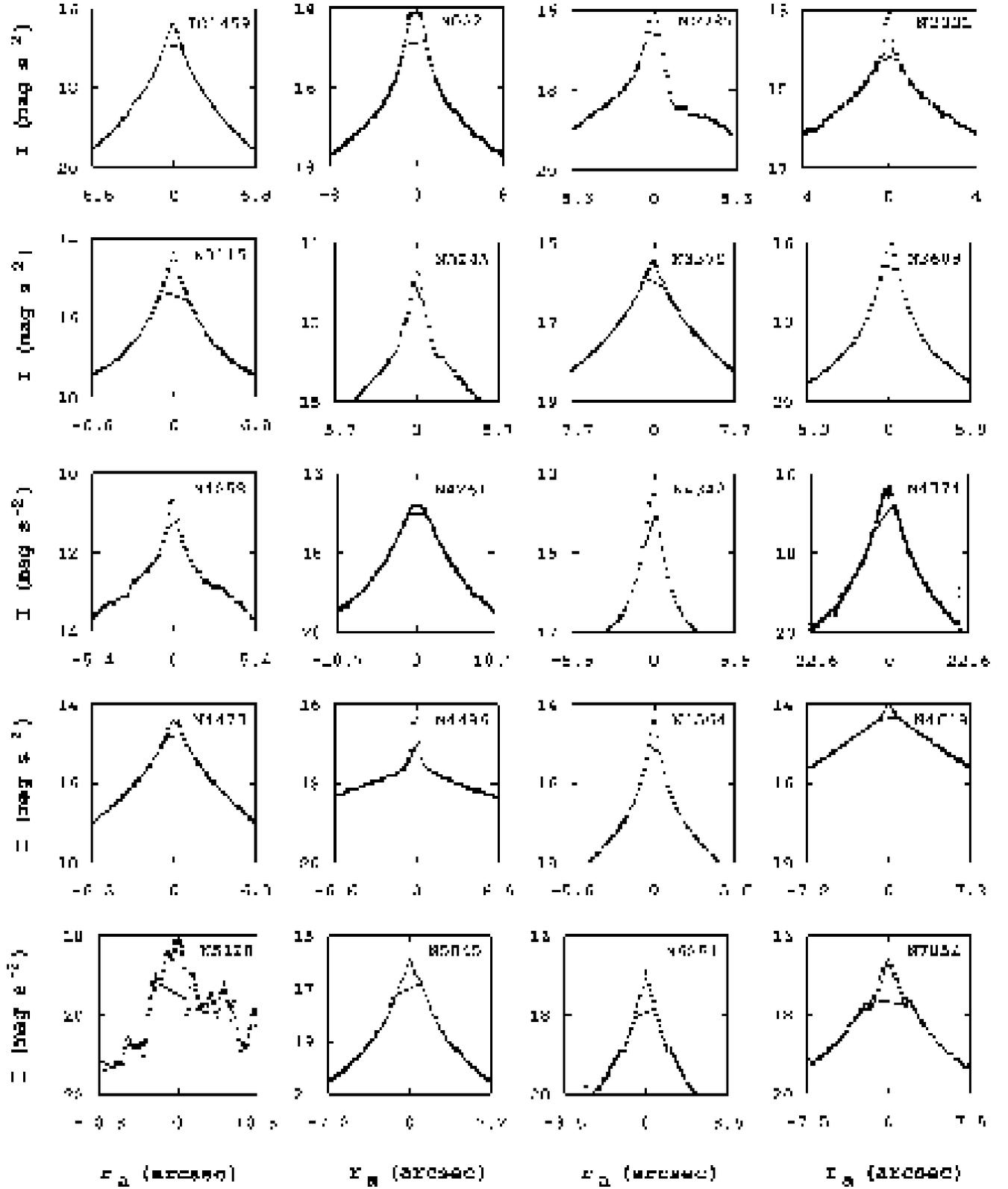}
\caption{\label{fig:4a} Plots of the $I-r$ profile for the sample galaxies.  The straight line is drawn between the chosen pixels.  The slope on the line indicates the asymmetry. }
\end{figure*}

The $r_\mathrm{\epsilon d}$ data were calculated from the distance in Refs.~\cite{merr2} and from the average angular (observed) radius $r_\mathrm{a}$ (arcsec).  The $r_\mathrm{a}$ for each sample galaxy was measured from Hubble Space Telescope (HST) images~\footnote{Based on photographic data obtained from MAST at http://archive.stsci.edu/hst/search.php.} listed in Table~\ref{tab:1}.  A Visual Basic program (see Appendix~\ref{sec:program}) was written to extract the $I-r$ curve and calculate $r_\mathrm{a}$.  The selection of each image was based upon being in the V band, upon having a sufficient area of the center of the galaxy in the image so $r_\mathrm{a}$ on both sides of the galaxy could be measured, and with sufficient exposure time to have the $I$ value~\footnote{The surface brightness was calculated using the VEGAMAG system as described in http:// www.stsci.edu/instruments/wfpc2\_dhb/wfpc2\_ch52.html.  The ZEROPOINT for the F450W, F547M, F555W, F606W, F658N, and F702W filters are 22.007, 21.689, 22.571, 22.919, 18.154, and 22.466, respectively.} at $r_\mathrm{a}$ sufficiently large.  The $I-r$ curve obtained from the HST images using the WFPC2 instrument are shown in Fig.~\ref{fig:4a}.  NGC 4607 which is listed in \citet{merr2} lacked a satisfactory image.  Since the goal of the program is to select a pixel at a discontinuity, image analysis techniques such as neighbor pixel averaging, azimuthal averaging, and combining more than one image into a single data set were avoided.  The Visual Basic program has alternate means to deal with peak (center) saturation, bad high pixel values, and bad low pixel values. 

The straight line in each $I-r$ curve in Fig.~\ref{fig:4a} is drawn between the pixels the Visual Basic program selected to calculate $r_\mathrm{a}$.  Note, the line is at differing $I$ levels among the galaxies and the slope of the line indicates asymmetry in $I$ value for the calculation of $r_\mathrm{a}$ in the same image.  Therefore, the selection of $r_\mathrm{a}$ is not based on selecting an isophotal value.  The $r_\mathrm{a}$ is not an isophotal based effective radius.  

\begin{figure}
\includegraphics[width=0.4\textwidth]{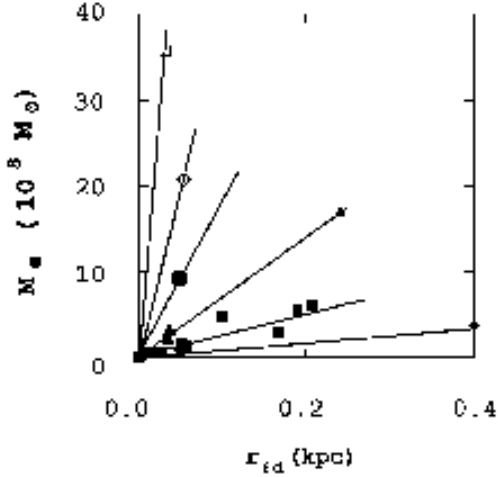}
\caption{\label{fig:4} Plot of the measured mass of the theorized supermassive black hole $M_\mathrm{\bullet}$ ($10^8 \, M_\mathrm{\odot}$) at the center of galaxies versus the distance along the major axis of a HST image to the first discontinuity or inflection point $r_\mathrm{\epsilon d}$ (pc) using the distance stated in \citet{merr2}.  The filled diamonds, filled squares, filled triangle, filled circles, open diamond, and open square are data points along a linear $M_\mathrm{\bullet}$ - $r_\mathrm{\epsilon d}$ line using data from Table~\ref{tab:1} and correspond to $n=1$, $n=2$, $n=3$, $n=4$, $n=5$, and $n=6$ as listed in Table~\ref{tab:2}, respectively.  The error bars (see Table~\ref{tab:1}) were omitted for clarity.}
\end{figure}

Figures~\ref{fig:4} shows plots of $M_\mathrm{\bullet}$ versus $r_\mathrm{\epsilon d}$ of the selected galaxies.  The straight lines were drawn by (1) identifying the half of the galaxies furthest from the origin, (2) drawing lines from the origin to the identified galaxy points, (3) noting six lines could be matched to the identified galaxies, (4) the galaxies were organized into classifications according to their proximity on the graph to a line, (5) finding the slope and intercept using the least squares method for each classification group of galaxies.  The differing slopes of the lines marked by filled diamonds, filled squares, filled triangle,``X'', star, and ``+'' suggest either differing $m_\mathrm{s} / m_\mathrm{g}$ values, differing $K_\mathrm{\epsilon d}$ values, or differing $M_\mathrm{tckmax}$ in Eq.~(\ref{eq:55}) among the galaxy classifications.  That the data points are on a straight line implies the $m_\mathrm{s} / m_\mathrm{g}$ ratio (particle type), $K_\mathrm{\epsilon d}$, or $M_\mathrm{tckmax}$ is discrete for an $\epsilon$ value.  Since the lines of n=1 to n=3 (the lines with more than one point) nearly intersect at ($r_\mathrm{\epsilon d}$,$M_\mathrm{\bullet}$)=(0,0) and $M_\mathrm{tckmax} > 0$, Eq.~(\ref{eq:55}) implies 
\begin{equation}
M_\mathrm{tckmax}  = K_\mathrm{\bullet \epsilon} \epsilon = K_\mathrm{\bullet \epsilon} K_\mathrm{\epsilon d} r_\mathrm{\epsilon d}
\label{eq:56},
\end{equation}
where $K_\mathrm{\bullet \epsilon}$ is a proportionality constant.  The $K_\mathrm{k}$ of Eq.~(\ref{eq:51}) produces a small offset in the intercept of the lines of Eq.~(\ref{eq:55}).

Therefore, 
\begin{equation}
M_\bullet \approx M_\mathrm{eff} = K_\mathrm{seff} r_\mathrm{\epsilon d} + K_\mathrm{ieff}
\label{eq:55ab},
\end{equation}
where 
\begin{eqnarray}
K_\mathrm{seff} = K_\mathrm{\bullet \epsilon} K_\mathrm{\epsilon d} - \frac{G_\mathrm{s}}{G} K_\mathrm{\epsilon d} \frac{m_\mathrm{s}}{m_\mathrm{g}} 
\label{eq:55ac}
\end{eqnarray}
and $ K_\mathrm{seff}$ is the slope of the $ M_\bullet - r_\mathrm{\epsilon d}$ plot.

\begin{table}
\caption{\label{tab:2} Data of the lines in Figures~\ref{fig:4}. }
\begin{ruledtabular}
\begin{tabular}{lllll}
{$n$\footnotemark[1]}
&{line\footnotemark[2]}
&{Corr.\footnotemark[3]} 
&{$K_\mathrm{seff}$\footnotemark[4] \footnotemark[6]}
&{$K_\mathrm{ieff}$\footnotemark[5]} \footnotemark[6]\\
& & & 10$^8$ M$_\odot$ kpc$^{-1}$&10$^8$ M$_\odot$\\
\hline
1&filled diamonds&0.99+&\phantom{000}9.0$\pm$\phantom{0}0.3&0.11$\pm$0.07\\
2&filled squares&0.92&\phantom{00}27\phantom{.0}$\pm$\phantom{0}6&0.07$\pm$0.8\\
3&filled triangles&0.99+&\phantom{00}70\phantom{.4}$\pm$\phantom{0}2&0.1\phantom{0}$\pm$0.3\\
4&filled circle& -&\phantom{0}187& 0\\
5&open diamond& -&\phantom{0}391& 0\\
6&open square& -&1113& 0
\end{tabular}
\end{ruledtabular}
\footnotetext[1]{The integer denoting the place of the line in the order of increasing $K_\mathrm{seff}$.}
\footnotetext[2]{The identification of the line symbol in Figures~\ref{fig:4}.}
\footnotetext[3]{The correlation coefficient.}
\footnotetext[4]{The least squares fit slope of the $M_\mathrm{\bullet}- r_\mathrm{\epsilon d}$ lines.}
\footnotetext[5]{The least squares fit intercept of the $M_\mathrm{\bullet}- r_\mathrm{\epsilon d}$ lines.}
\footnotetext[6]{All lines are calculated as the least squares best fit for points and ($r_\mathrm{\epsilon d}$,$M_\mathrm{\bullet}$) = (0,0).}
\end{table}

For a spherically symmetric particle, $m_\mathrm{s}$ increases with the square of the radius and $m_\mathrm{\iota}$ increases with the cube of the radius.  For a given density, larger particles have lower $m_\mathrm{s}/m_\mathrm{\iota}$.  Hydrogen is found in the outer regions of a galaxy and heavier elements are found in the inner regions.  Stars larger than our sun are found both inward and outward of us.  Therefore, hydrogen has a higher $m_\mathrm{s}/m_\mathrm{\iota}$ than iron.  Black holes have a much lower $m_\mathrm{s}/m_\mathrm{\iota}$ than other stellar particles.  Therefore, the largest cross-section area on which $F_\mathrm{s}$ acts is on the area of nuclei rather than on atoms, compounds, or larger assemblies of matter.  Gravitationally bound assemblies of matter of differing sizes and the same relative elemental composition will experience the same $M_\mathrm{eff}$.  Also, assemblies of matter with differing relative elemental composition will have differing RCs.  Differing $m_\mathrm{s}/m_\mathrm{\iota}$ cause differing $v^2$ at a given $r$.  For instance, the H$_\mathrm{\alpha}$ (found near hot stars) and H{\scriptsize{I}} RCs are different at small $r$ because the particles are very different.  However, in the disk as $r$ increases, the stellar elemental composition is of successively lighter elements.  Therefore, the H$_\mathrm{\alpha}$ and H{\scriptsize{I}} RCs become similar at larger $r$.

The slopes among the lines in Table~\ref{tab:2} obey the relation 
\begin{equation}
\log_{10} \left ( \frac{ S_\mathrm{lope}}{ U_\mathrm{nit}} \right ) = K_\mathrm{sn} n + K_\mathrm{in} 
\label{eq:57},
\end{equation}
where $S_\mathrm{lope}$ is the slope of the lines from the parameter relationship; $U_\mathrm{nit}$ is the units of measure of $S_\mathrm{lope}$; and $K_\mathrm{sn}$ and $K_\mathrm{in}$ are the slope and intercept of the linear relation of Eq.~(\ref{eq:57}), respectively.  For the $M_\bullet - r_\mathrm{\epsilon d}$ relationship, $S_\mathrm{lope} = K_\mathrm{seff}$, $U_\mathrm{nit}$ is 10$^8$M$_\odot$ kpc$^{-1}$, $K_\mathrm{sn} = 0.41 \pm 0.01$, and $K_\mathrm{in} = 0.58 \pm 0.04$  at one standard deviation (1 $\sigma$).  The $n$ is an integer of one to six depending on the line from Table~\ref{tab:2}.  The correlation coefficient and F test \footnote{The one-tailed probability that the variances of two arrays are not significantly different as calculated by the Microsoft Excel program.} for Eq.~(\ref{eq:57}) are 0.999 and 0.992, respectively.  Note, the lowest slope corresponds to n=1 of Eq.~(\ref{eq:57}) and the lines of Figures~\ref{fig:4} intersect at a point.  Note, $S_\mathrm{lope}$ is nearly independent of the distance used to calculate the $M_\bullet$ and $r_\mathrm{\epsilon d}$ if the same distance is used for both.  The values of log$_{10}$(e)$\approx 0.434$ is within the 2.3 $\sigma$ of $\log_{10} (K_\mathrm{sn} )$.  Therefore, restating Eq.~(\ref{eq:57}) as a strong Principle of Repetition yields
\begin{equation}
K_\mathrm{seff} = K_\mathrm{se} \, e^n 
\label{eq:57a},
\end{equation}
where $K_\mathrm{se} = (3.7 ^{+0.0} _{- 0.9}) \times 10^8 \, M _\odot \,$kpc$^{-1}$ at 1 $\sigma$.

Combining Eqs.~(\ref{eq:55ab}) and (\ref{eq:57a}) and setting $K_\mathrm{iff}=0$ yields 
\begin{equation}
M_\bullet = K_\mathrm{se} \, e^n \, r_\mathrm{\epsilon d}
\label{eq:57ac}.
\end{equation}

The standard deviation of the $\chi^2$ function are the error bars of $M_\bullet$ listed in Table~\ref{tab:1}.  The $\chi ^2 = 17$ and the probability value of the $\chi ^2$ $P_\chi \approx 60\%$.  The data does not invalidate Eq.~(\ref{eq:57ac}) at the significant 5\% level.  The $\chi ^2 $ and $P_\chi $ were calculated using only the galaxy data (not the ($r_\mathrm{\epsilon d}$,$M_\mathrm{\bullet}$) = (0,0) point).  The $ M_\bullet$ is a function of two parameters, $ r_\mathrm{\epsilon d}$ and $n$.  The $ r_\mathrm{\epsilon d}$ is a parameter of each galaxy, the $n$ is a galaxy classification parameter, and the $e$ implies the Principle of Repetition is applicable to the $M_\bullet$ parameter.  Given a set of parameters (n, $ r_\mathrm{\epsilon d}$), a unique value of $M_\bullet$ can be calculated.

A feature of Fig.~\ref{fig:4} is the lack of data points in the upper right of the plot (galaxies with relatively high $r_\mathrm{\epsilon d}$ and high $M_\bullet$ values).

\subsection{RR}

In the KR as $r_\mathrm{k}$ increases to its maximum value ($r_\mathrm{kmax}$), the rotation velocity $v_\mathrm{kmin}$ decreases to its minimum value and the Space and gravitational forces balance.  This implies 
\begin{subequations}
\label{eq:57aa}
\begin{eqnarray}
M_\mathrm{eff} & \rightarrow & 0, \label{eq:57aa:a} \\
v^2_\mathrm{kmin} & \rightarrow & r_\mathrm{kmax} ( {\ddot{\bm r}} _\mathrm{kmax} - \bm{r}_\mathrm{kmax} \bullet \bm{a}_\mathrm{o}). \label{eq:57aa:b}
\end{eqnarray}
\end{subequations}

Therefore, in $T_\mathrm{k-r}$ the motion of the test particle is primarily governed by the force from external galaxies.  Particles such as hydrogen are being ejected outward from the KR with a large radial velocity can cross the $T_\mathrm{k-r}$.  Particles such as dense stars in nearly circular orbits on a slow inward journey may lack the necessary radial velocity to cross the $T_\mathrm{k-r}$.  The $T_\mathrm{k-r}$ is a centrifugal barrier to inward movement of particles.  Therefore, external galaxies can have a large influence on the mass distribution and total mass in a galaxy. 

If $v_\mathrm{kmin} = 0$ in a galaxy, then the CR and KR are decoupled from the RR and SR.  The $v$ at $r > r_\mathrm{kmax}$ is as if there were no mass inside $r_\mathrm{kmax}$ and, for non-relativistic calculations, the particles in the RR and SR of an intrinsic galaxy ($\bm a_\mathrm{o} =0$) are subject to only Newtonian forces.

Two changes in the physics occur to cause the RR.  The first change in the first RR (FRR) is to change the particle elemental species which changes $m_\mathrm{s}/m_\iota$ while maintaining a spherical region shape.  The second change in the second RR (SRR) is to change the d$r$ increment shape from a spherical shell to a cylindrical shell while continuing the changing $m_\mathrm{s}/m_\iota$.  The resulting equations depend on the $v$ reaching a minimum value ( d$v/$d$r =0$) of the RC in the KR for the particle species examined.  

That shapes of astronomical particle assemblies are either spheroidal or disk is well established.  This Paper takes the disk shape as having cylindrical symmetry.  The well-known peanut shaped profile of a galaxy bulge is a varing height, cylindrical symmetry with a spheroid core.

\subsubsection{First RR}

The rotation velocity $v_\mathrm{r}$ of a particle in the RR is greater than $v_\mathrm{kmin}$.  If the radius $r_\mathrm{tkr}$ of a particle's orbit in the T$_\mathrm{k-r}$ decreases to less than $r_\mathrm{kmax}$, the attractive force decreases because $M$ is decreasing.  Thus, the effect of $\epsilon$ which is spherically symmetric about the center of the galaxy will cause the particle to accelerate to a greater orbit for a given rotation velocity.  Therefore, elliptical orbits of particles in the T$_\mathrm{k-r}$ are inhibited and $\ddot r_\mathrm{tkr} \approx 0$.  Remember one of the simplifying assumptions was the stars are stable in material composition rather than changing $m_\mathrm{s} /m_\mathrm{\iota}$.  Changing $m_\mathrm{s} /m_\mathrm{\iota}$ would result in a slow inward spiral.  Therefore, the radius $r_\mathrm{frr}$ of the orbits of the particles in the FRR become nearly circular and $\ddot r_\mathrm{frr} \approx 0$.  Thus, the mass in an elemental volume is $D_\mathrm{frr} 4 \pi r^2_\mathrm{frr} \mathrm{d} r$ where $D_\mathrm{frr}$ is the density of matter at $r_\mathrm{frr}$.  The $D_\mathrm{frr}$ is approximately constant because the particles are of the same species.  Therefore, the $M$ in Eq.~(\ref{eq:49}) is 
\begin{equation}
M_\mathrm{frr} (r_\mathrm{frr}) = \frac{4}{3} D_\mathrm{frr} \pi r_\mathrm{frr}^3 + M_\mathrm{\Delta frr}
\label{eq:58},
\end{equation}
where
\begin{equation}
M_\mathrm{\Delta frr} \equiv M_\mathrm{tkrmax} - \frac{4}{3} D_\mathrm{frr} \pi r_\mathrm{tkrmax}^3 
\label{eq:59};
\end{equation}
$M_\mathrm{frr} (r_\mathrm{frr})$ is the mass within $r_\mathrm{frr}$ as a function of $ r_\mathrm{frr}$; $ M_\mathrm{tkrmax}$ is the total mass in the CR, T$_\mathrm{c-k}$, KR, and T$_\mathrm{k-r}$; and $ r_\mathrm{tkrmax}$ is the radius of the T$_\mathrm{k-r}$.

Substituting Eq.~(\ref{eq:58}) into Eq.~(\ref{eq:49}) and taking $\ddot r \approx 0$ yields 
\begin{eqnarray}
v_\mathrm{frr}^2 &=& \frac{m_\mathrm{g}}{m_\mathrm{\iota}} \frac{4}{3} G D_\mathrm{frr} \pi r_\mathrm{frr}^2 \nonumber \\*
& & + \frac{m_\mathrm{g} G M_\mathrm{\Delta frr} - m_\mathrm{s} G_\mathrm{s} \epsilon}{m_\mathrm{\iota} r_\mathrm{frr}} \nonumber \\*
& & -  \bm r_\mathrm{frr} \bullet \bm a_\mathrm{o}(r_\mathrm{frr}) 
\label{eq:60},
\end{eqnarray}
where $v_\mathrm{frr}$ is the rotation velocity of a particle at $r_\mathrm{frr}$.

The $r_\mathrm{frr}^{-1}$ term was small at $r=r_\mathrm{kmax}$ and is decreasing as $r_\mathrm{frr}$ increases.  Therefore, the $r_\mathrm{frr}^{-1}$ term is relatively much smaller than the $r_\mathrm{frr}^2$ term.  Since the $m_\mathrm{s}$ term for hydrogen differs from the $m_\mathrm{s}$ term for stellar material, the RCs will differ.  By averaging the opposite sides to reduce the $\bm a_\mathrm{o}$ term as before yields
\begin{subequations}
\label{eq:61}
\begin{eqnarray}
v_\mathrm{frr}^2 &\approx& \frac{m_\mathrm{g}}{m_\mathrm{\iota}} \frac{4}{3} G D_\mathrm{frr} \pi r_\mathrm{frr}^2 + K_\mathrm{2}, \, \, \, \, \label{eq:61;a} \\*
& \equiv & K_\mathrm{frrslope} r_\mathrm{frrmax}^2 + K_\mathrm{frrintercept}, \label{eq:61:b}
\end{eqnarray}
\end{subequations}
where $K_\mathrm{frrslope}$ and $K_\mathrm{frrintercept}$ are the slope and intercept of the $v_\mathrm{frr}^2$-$r_\mathrm{frr}^2$ curve.

\begin{figure}
\includegraphics[width=0.4\textwidth]{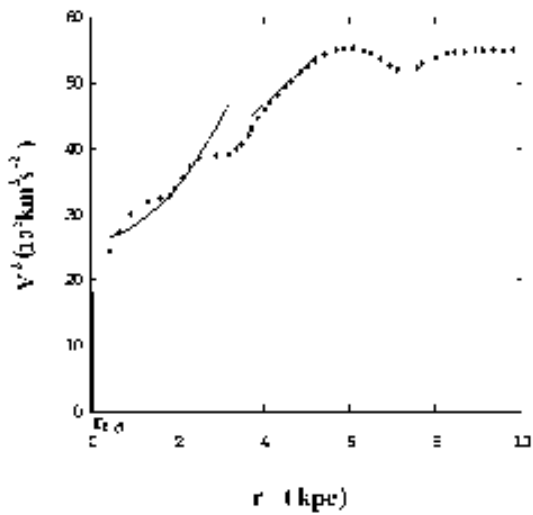}
\caption{\label{fig:6} Plot of rotation velocity $v^2 (10^3$ km$ ^2 \,$s$^{-2} )$ versus $r$ (kpc) for NGC 4321~\cite{semp} using the distance to the galaxy calculated using Cepheid variables to calculate $r$.  A linear curve has been fitted to the SRR. A $v^2$ -$r^2$ curve has been fitted to the FRR.}
\end{figure}

Figure~\ref{fig:6} shows the H{\scriptsize{I}} rotation curve of NGC 4321~\cite{semp}.  It has the rotation velocity plotted well into the bulge.  The curved line is in the FRR and is a plot of $v^2_\mathrm{frr}$ versus $r_\mathrm{frr}$ according to Eq.~(\ref{eq:61}).

\subsubsection{Second RR}

The end of the FRR and the beginning of the SRR is caused by a change in the $m_\mathrm{s} /m_\mathrm{\iota}$ factor.  The Space force starts to become significant because of $m_\mathrm{s} /m_\mathrm{\iota}$ relative to the gravitational force.  As the $r$ increases, the particle orbits become rotationally flattened and remain circular~\cite[pages 723-4]{binn}.  The spherical symmetry is ended.  If either the rotation velocity $v_\mathrm{srr}$ in the SRR, the Space force, or the gravitational force changes, the acceleration $\ddot r_\mathrm{srr}$ in the SRR will change to restore balance and the circular orbits.  Thus, the Principle of Negative Feedback applies and $\ddot r_\mathrm{srr} \approx 0$.  The orbits of the particles are nearly circular except for changes over a long period in the $\epsilon$; for changes in the elemental composition and mass of the particles such as by photon emission; and for changes in the $\bm r \bullet \bm a_\mathrm{o} $ term.  Some of the galaxy's matter, especially the lighter matter with high $m_\mathrm{s}/m_\mathrm{\iota}$, is moving radially.  Most of a matter in the SRR is in a stable, radial position.  Light material such as hydrogen and heavier elements differ in the mechanics defining their orbits due to the Space force.  

The mass in an elemental volume d$r$ of the SRR can be modeled as a cylinder shell of height $H_\mathrm{srr}$~\cite[page 724]{binn} and with density $D_\mathrm{srr}$.  The mass in an elemental volume of the SRR is $D_\mathrm{srr} H_\mathrm{srr} 2 \pi r_\mathrm{srr} \rm d r_\mathrm{srr}$ where $r_\mathrm{srr}$ is the radius of a particle's orbit in the SRR.  Unlike the inner regions, the $m_\mathrm{s}/ m_\mathrm{\iota}$ of particles vary more slowly with increasing $r_\mathrm{srr}$.  The height profile of the bulge is a measure of the amount of mass versus $m_\mathrm{s}/ m_\mathrm{\iota}$ particle type.  The interplay of the terms of Eq.~(\ref{eq:49}) determine the orbital radius of a particle.  Thus, the mass $M_\mathrm{srr}$ within the $r_\mathrm{srr}$ is 
\begin{equation}
M_\mathrm{srr}= \left( D_\mathrm{srr} H_\mathrm{srr} 2 \pi \right)r_\mathrm{srr}^2 + M_\mathrm{\Delta r}
\label{eq:62},
\end{equation}
where 
\begin{equation}
M_\mathrm{\Delta r} \equiv M_\mathrm{frrmax} - \left( D_\mathrm{srr} H_\mathrm{srr} 2 \pi \right)r_\mathrm{frrmax}^2
\label{eq:63};
\end{equation}
$ r_\mathrm{frrmax}$ is the $r$ to the outer edge of the FRR; and $ M_\mathrm{frrmax}$ is the mass in the CR, T$_\mathrm{c-k}$, KR, T$_\mathrm{k-r}$, and FRR.

Inserting Eq.~(\ref{eq:62}) into Eq.~(\ref{eq:49}) and with circular orbits yields 
\begin{eqnarray}
v_\mathrm{srr}^2 &=& \left(\frac{m_\mathrm{g}}{m_\mathrm{\iota}} G D_\mathrm{srr} H_\mathrm{srr} 2 \pi \right)r_\mathrm{srr} \nonumber \\*
& & + \frac{m_\mathrm{g} G M_\mathrm{\Delta r} - m_\mathrm{s} G_\mathrm{s} \epsilon}{m_\mathrm{\iota} r_\mathrm{srr}} \nonumber \\*
& & - \bm r_\mathrm{srr} \bullet \bm a_\mathrm{o}  
\label{eq:64}.
\end{eqnarray}

Since the particles in the SRR have larger $m_\mathrm{s} / m_\mathrm{\iota}$ and since the $r_\mathrm{srr}^{-1}$ term is small at $r=r_\mathrm{kmax}$, the $r_\mathrm{srr}^{-1}$ term is small compared to the $r_\mathrm{srr}$ term.  Equation~(\ref{eq:64}) suggest $ v_\mathrm{srr}^2 $ is nearly independent of $\epsilon$ and depends on galaxy parameters $ M_\mathrm{\Delta r}$, $D_\mathrm{srr}$ and $H_\mathrm{srr}$.  By averaging from side-to-side of the galaxy as before, the $\bm r_\mathrm{srr} \bullet \bm a_\mathrm{o} $ term is reduced.  Galaxies with other close galaxies in the $v$ plane of rotation, close galaxies with high $\epsilon$ values, or several close galaxies asymmetrically arranged, the residual of the side-to-side variation of the $\bm r_\mathrm{srr} \bullet \bm a_\mathrm{o} $ term may be relatively significant.  Asymmetry in the RC reflects this condition~\cite{hodg2}.  Therefore, 
\begin{equation}
v_\mathrm{srr}^2= K_\mathrm{sr} r_\mathrm{srr} + K_\mathrm{ir} 
\label{eq:65},
\end{equation}
where $K_\mathrm{sr}$ and $K_\mathrm{ir}$ are the slope and intercept of the linear $v_\mathrm{srr}^2 - r_\mathrm{srr}$ relationship, respectively.  The linear relationship of Eq.~(\ref{eq:65}) for NGC4321 is the straight line plotted in Fig.~\ref{fig:6}.

\begin{figure*}
\includegraphics[width=\textwidth]{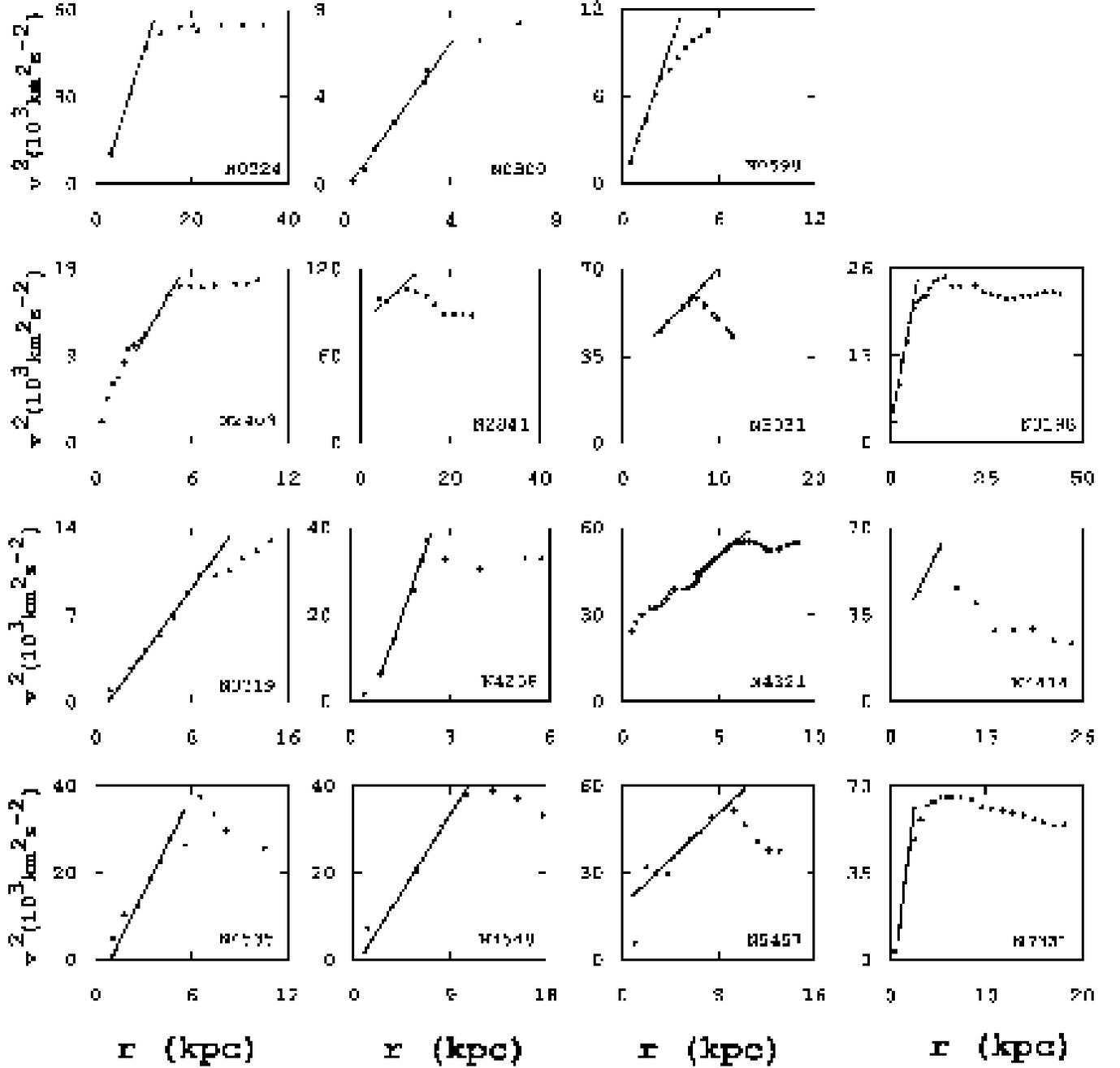}
\caption{\label{fig:8}Plots of the rotation velocity $v^2$ (10$^3$ km$^2 \,$s$^{-2}$) versus radius $r$.  The straight lines mark the SSR.}
\end{figure*}

To test Eq.~(\ref{eq:65}), galaxies with a distance $D_\mathrm{c}$ calculated using Cepheid variables (the ``$D_\mathrm{z}$'' of \citet{free}) and with a published H{\scriptsize{I}} RC which included data points in the SRR were chosen.  Figure~\ref{fig:8} shows the data points for each RC of the 15 selected galaxies.  The straight lines in Fig.~\ref{fig:8} are found by fitting a 0.99 or higher correlation coefficient, straight line to the first three data points immediately before the knee of the RC which indicates the transition region T$_\mathrm{r-s}$ between the RR and SR.  NGC 4414 had only two data points before the knee.  NGC 3198 and NGC 0224 have several data points between the knee and the straight line of which no three are on a 0.99 correlation straight line.

The data for the 15 selected galaxies are shown in Table \ref{tab:3}.  The morphology type data was obtained from the NED database.  The morphology type code $t$ and luminosity class code $lc$ data was obtained from the LEDA database~\footnote{The LEDA database is available at: http://leda.univ-lyon1.fr.  The data were obtained from LEDA on 5 May 2004.}.  The number ``No.'' of data points establishing the SRR, the values of the maximum radius $r_\mathrm{rmax}$ (kpc) of the last data point on the SRR straight line, the maximum rotation velocity $v_\mathrm{rmax}$ (km$\,$s$^{-1}$) of the last data point on the SRR straight line, $K_\mathrm {sr}$, and $K_\mathrm {ir}$ were obtained from the plots of Fig.~\ref{fig:8}.  The ``Ref.'' lists the references for each RC.

\begingroup
\squeezetable
\begin{table*}
\caption{\label{tab:3}Data for the chosen galaxies.}
\begin{ruledtabular}
\begin{tabular}{llllcllrrrrc}
Galaxy &morphology\footnotemark[1] &t\footnotemark[2]&lc\footnotemark[3]&RC\footnotemark[4]&$D_\mathrm{c}$\footnotemark[5] &No.&$r_\mathrm{rmax}$&$v_\mathrm{rmax}$\footnotemark[6]&$K_{\mathrm{sr}}$\footnotemark[7]&$K_{\mathrm{ir}}$\footnotemark[8]&Ref.\\
\hline
NGC 0224 &SA(s)b  &3 &2 &F &\phantom{0}0.79&3&10.3$\pm$3.7&217$\pm$22&5 100$\pm$\phantom{ 10}200&-5 900$\pm$11 000&\cite{gott}\\
NGC 0300 &SA(s)d &6.9 &6 &R &\phantom{0}2.00&6&3.1$\pm$2.2&72$\pm$13&1 700$\pm$\phantom{ 10}100&-400$\pm$\phantom{ 00}400&\cite{car3}\\
NGC 0598 &SA(s)cd &6 &4 &R &\phantom{0}0.84&5&2.4$\pm$1.1&85$\pm$12&1 400$\pm$\phantom{ 10}300&4 000$\pm$\phantom{0}4 000&\cite{corb}\\
NGC 2403 &SAB(s)cd &6 &5 &F &\phantom{0}3.22&3&4.5$\pm$0.9&124$\pm$10&2 860$\pm$\phantom{ 100}80&2 000$\pm$\phantom{0}1 000&\cite{bege}\\
NGC 2841 &SA( r)b;LINER Sy &3 &1 &F &14.07\footnotemark[9]&3&8.3$\pm$2.4&324$\pm$18&26 400$\pm$\phantom{ 10}700&-8 000$\pm$\phantom{0}4 000&\cite{bege, fill}\\
NGC 3031 &SA(s)ab;LINER Sy1.8 &2.4 &2 &D &\phantom{0}3.63&7&7.4$\pm$0.8&243$\pm$14&3 900$\pm$\phantom{ 10}200&30 000$\pm$\phantom{0}5 000&\cite{rots}\\
NGC 3198 &SB(rs)c &5.2 &3 &F &13.80&4&5.0$\pm$1.2&134$\pm$15&3 100$\pm$\phantom{ 10}300&2 000$\pm$\phantom{0}3 000&\cite{vana}\\
NGC 3319 &SB(rs)cd;HII &5.9 &3.3 &R &13.30&6&8.7$\pm$1.7&101$\pm$ 5&1 300$\pm$\phantom{ 10}100&-1 000$\pm$\phantom{0}4 000&\cite{moor}\\
NGC 4258&SAB(s)bc;LINER Sy1.9&4&3.9&D&\phantom{0}7.98&3&2.3$\pm$0.7&197$\pm$26&36 000$\pm$10 000&-43 000$\pm$30 000&\cite{krui2}\\
NGC 4321 &SAB(s)bc;LINER HII&4 &1 &D &15.21&6&5.8$\pm$0.8&233$\pm$14&5 100$\pm$\phantom{ 10}500&20 000$\pm$10 000&\cite{semp}\\
NGC 4414&SA(rs)c? LINER&5.1&3.6&D&17.70&2&6.4$\pm$2.9&247$\pm$45&6 000\phantom{$\pm$00 300}&20 000\phantom{$\pm$09 000}&\cite{brai}\\
NGC 4535 &SAB(s)c &5 &1.9 &D &15.78&3&4.6$\pm$1.2&166$\pm$12&7 400$\pm$\phantom{ 10}300&-7 000$\pm$17 000&\cite{chin}\\
NGC 4548&SBb(rs);LINER  Sy&3.1&2&D&16.22&3&10.6$\pm$2.9&195$\pm$12&3 700$\pm$\phantom{ 10}300&-1 000$\pm$17 000&\cite{guha}\\
NGC 5457&SAB(rs)cd&5.9&1&D&\phantom{0}6.70&4&8.5$\pm$1.3&228$\pm$12&4 000$\pm$\phantom{ 10}400&20 000$\pm$20 000&\cite{bosm2}\\
NGC 7331 &SA(s)b;LINER &3.9 &2 &F &14.72&3&2.0$\pm$0.6&211$\pm$20&11 000$\pm$ 2 000&23 000$\pm$14 000&\cite{mann}\\
\end{tabular}
\end{ruledtabular}

\footnotetext[1] {Galaxy morphological from the NED database.}
\footnotetext[2] {Galaxy morphological type code from the LEDA database.}
\footnotetext[3] {Galaxy luminosity class code from the LEDA database.}
\footnotetext[4] {Galaxy's HI rotation curve type according to slope in the outer SR region.  R is rising, F is flat, and D is declining.}
\footnotetext[5] {The distance $D_\mathrm{c}$ (Mpc) to the galaxy from~\cite{free} unless otherwise noted.}
\footnotetext[6] {The maximum rotation velocity $v_\mathrm{rmax}$ (km$\,$s${-1}$), the radius $r_\mathrm{rmax}$ (kpc) of $v_\mathrm{rmax}$, and the No. of points on the straight line from the curves in Fig.~\ref{fig:8}.  The error of $r_\mathrm{rmax}$ and $v_\mathrm{srr}$ is $\pm 5 \%$ of the value plus the difference of the chosen data values and the next data point value. }
\footnotetext[7] {Least squares slope of $v^2_{\mathrm{r}}$ - $r_{\mathrm{r}}$ for each galaxy in the SRR of HI rotation curve (km$^2 \,$s$^{-2} \,$kpc$^{-1}$).}
\footnotetext[8] {Least squares intercept of $v^2_{\mathrm{r}}$ - $r_{\mathrm{r}}$ for each galaxy in the SRR of HI rotation curve (km$^2 \,$s$^{-2}$).}
\footnotetext[9] {The $D_\mathrm{c}$ is from~\cite{macr}.}
\end{table*}
\endgroup

This sample has low surface brightness (LSB), medium surface brightness (MSB), and high surface brightness (HSB) galaxies; galaxies with a range of the 21-cm line width $W_{20}$ at 20 percent of peak of from 120 km s$^{-1}$ to 607 km$\,$s$^{-1}$; includes LINER, Sy, HII and less active galaxies; galaxies which have excellent and poor agreement between distances $D_{tf}$ calculated using the Tully-Fisher relation (TF) and $D_c$; a $D_c$ range of from 0.75 Mpc to 17.39 Mpc; field and cluster galaxies; and galaxies with rising, flat, and declining RCs in the SR.

The CUM predicts a $v_\mathrm{frr}^2 - r_\mathrm{frr}^2$ relationship if a galaxy has spherical symmetry in a FRR.  At least five of the galaxies in Fig.~\ref{fig:8} have data points sufficiently far inward to show the $v_\mathrm{frr}^2 - r_\mathrm{frr}^2$ relation and don't.  The CUM suggests these galaxies lack a FRR (spherical symmetry).  

As $r_\mathrm{srr}$ increases, $m_\mathrm{s} /m_\mathrm{\iota}$ increases and the predominant particles have less mass and decreasing density.  Decreasing amount of mass in a given cylindrical shell and the rotation flattening effect causes $H_\mathrm{srr}$ to decrease to a disk.  This is end of the RR and the beginning of the transition region T$_\mathrm{r-s}$ between the RR and SR.

\subsubsection{RR data}

Because this change is caused by the change in elemental types which depends on the $\epsilon$ of a galaxy, the end of the RR is occurring under similar conditions in all galaxies.  Thus, the mass $M_\mathrm{rmax}$ of the galaxy inside a sphere with a radius of $r_\mathrm{rmax}$ can be used to compare parameters among galaxies. 

Evaluating Eq.~(\ref{eq:49}) at $r_\mathrm{rmax}$ and rearranging terms yields 
\begin{equation}
r_\mathrm{rmax} v^2_\mathrm{rmax} = \frac{m_\mathrm{g}}{m_\mathrm{\iota}} G M_\mathrm{rmax} -  \frac{m_\mathrm{s}}{m_\mathrm{\iota}} G_\mathrm{s} K_\mathrm{\epsilon d} r_\mathrm{\epsilon d} + K_\mathrm{irr}
\label{eq:74},
\end{equation}
where
\begin{equation}
K_\mathrm{irr} = r_\mathrm{rmax}^2 \ddot r_\mathrm{rmax}- r_\mathrm{rmax} \bm r_\mathrm{rmax} \bullet \bm a_\mathrm{o}
\label{eq:75}.
\end{equation}

The $ r_\mathrm{rmax} v^2_\mathrm{rmax}$ term is proportional to the effective mass $M_\mathrm{effr}$ inside a sphere with a radius of $r_\mathrm{rmax}$.  Since $\ddot r_\mathrm{rmax} \approx 0$, $ K_\mathrm{irr}$ is a relatively small value.  However, in addition to the close-galaxy caveat, a very young galaxy or a galaxy with significant infall of matter such as colliding galaxies can cause a deviation in $K_\mathrm{irr}$.

Since $r_\mathrm{rmax}$ is in the RR and not in T$_\mathrm{r-s}$ or SR, $v_\mathrm{rmax}$ can be found in rising, flat and falling RCs.  The gas spectral lines such as H$_\mathrm{\alpha}$ reflects the motion of a hot star and not hydrogen gas.  The elemental content of hot stars changes with $r$ which changes the $m_\mathrm{s} / m_\mathrm{\iota} $ term.  Therefore, $ v^2_\mathrm{rmax}$ changes with each star elemental type.  The H{\scriptsize{I}} RC is monitoring the same particle type in each galaxy from the RR to the SR.  Therefore, the H{\scriptsize{I}} RC was used to test Eq.~(\ref{eq:74}).

\begin{figure*}
\includegraphics[width=\textwidth, height=\textwidth]{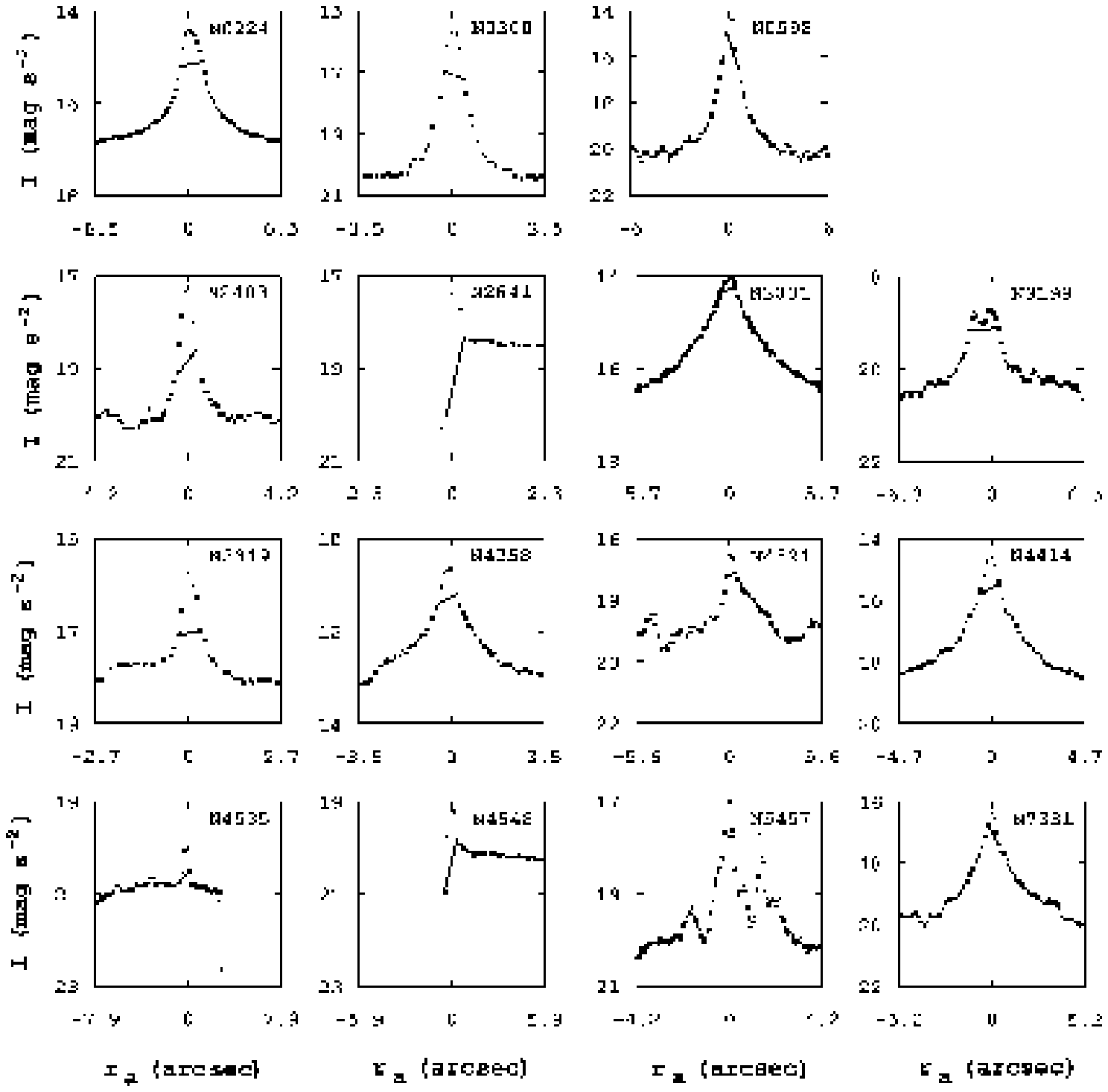}
\caption{\label{fig:10} Plots of the $I-r$ profile for the sample galaxies.  The straight line is drawn between the chosen pixels.  The slope on the line indicates the asymmetry.  The horizontal scale for each plot is approximately ten times the $r_\mathrm{a}$.}
\end{figure*}

The $r_\mathrm{\epsilon d}$ data listed in Table~\ref{tab:4} were calculated from $D_\mathrm{c}$ in Table~\ref{tab:3} and from the average angular (observed) radius $r_\mathrm{a}$ (arcsec).  The $r_\mathrm{a}$ for each sample galaxy was measured from the Hubble Space Telescope (HST) images listed in Table~\ref{tab:4}.  The Visual Basic program listed in Appendix~\ref{sec:program} was used to extract the $I-r$ curve and calculate $r_\mathrm{a}$.  The selection of each image was based upon being in the V band, upon having a sufficient area of the center of the galaxy in the image so $r_\mathrm{a}$ on both sides of the galaxy could be measured, and with sufficient exposure time to have the $I$ value at $r_\mathrm{a}$ sufficiently large.  The $I-r$ curve obtained from the HST images using the WFPC2 instrument are shown in Fig.~\ref{fig:10}.  NGC 2841 had only one side of the galaxy in the image.  

The straight line in each $I-r$ curve of Fig.~\ref{fig:10} is drawn between the pixels the Visual Basic program selected to calculate $r_\mathrm{a}$.  Note, the line is at differing $I$ levels among the galaxies and the slope of the line indicates asymmetry in $I$ value for the calculation of $r_\mathrm{a}$ in the same image.  Therefore, the selection of $r_\mathrm{a}$ is not based on selecting an isophotal value.  The $r_\mathrm{a}$ is not an isophotal based effective radius.  

\begin{table}
\caption{\label{tab:4} Data of the galaxies in Fig.~\ref{fig:10}. }
\begin{ruledtabular}
\begin{tabular}{llllc}
{Galaxy\footnotemark[1]}
&{      HST}& 
&{\phantom{0}$r_\mathrm{\epsilon d}$\footnotemark[2]}
&{line\footnotemark[3]} \\ 
& image &filter&\phantom{0}pc&n  m  p  s\\
\hline
N0224&U2LG0209T&F555W&\phantom{0}2.87$\pm$0.38&-  6  7  7\\
N0300&U6713701M&F547M&\phantom{0}3.44$\pm$0.97&-  2  5  3\\
N0598&U3MR0106M&F555W&\phantom{0}1.50$\pm$0.41&-  3  6  5\\
N2403&U6712504R&F547M&\phantom{0}6.61$\pm$1.56&-  3  4  4\\
N2841&U65M7008M&F555W&15.68$\pm$6.82&-  5  4  5\\
N3031&U32L0105T&F547M&\phantom{0}5.11$\pm$1.76\footnotemark[4]&3  5  5  6\\
N3198&U29R1602T&F606W&64.20$\pm$6.69&-  1  1  1\\
N3319&U29R1702T&F606W&19.05$\pm$6.45&-  2  4  2\\
N4258&U6712101R&F547M&13.41$\pm$3.87\footnotemark[4]&2  3  2  4\\
N4321&U2465107T&F555W&23.66$\pm$7.38&-  3  3  4\\
N4414&U67N8102R&F606W&39.95$\pm$8.58&-  3  2  3\\
N4535&U2782603T&F555W&31.25$\pm$7.65&-  2  2  3\\
N4548&U34L0303T&F555W&32.34$\pm$7.87&-  3  3  3\\
N5457&U6712704R&F547M&\phantom{0}5.32$\pm$3.25&-  5  5  6\\
N7331&U41V1802M&F450W&19.23$\pm$7.14&-  2  1  4\\
\end{tabular}
\end{ruledtabular}

\footnotetext[1]{``I'' means IC and ``N'' means NGC.}
\footnotetext[2]{The $r_\mathrm{\epsilon d} $ was measured as outlined in the text.  The error is $\pm$ 0.1 arcsec.}
\footnotetext[3]{The line is the ``n'' listed in Table~\ref{tab:2} and shown in Fig.~\ref{fig:4}, the ``m'' listed in Table~\ref{tab:5} and shown in Fig.~\ref{fig:11}, the ``p'' listed in Table~\ref{tab:6} and shown in Fig.~\ref{fig:12}, and the `s'' listed in Table~\ref{tab:7} and shown in Fig.~\ref{fig:13}.}
\footnotetext[4]{The $r_\mathrm{\epsilon d}$ values differ from Table~\ref{tab:1} because of differing distances used.}

\end{table}

\begin{figure}
\includegraphics[width=0.4\textwidth]{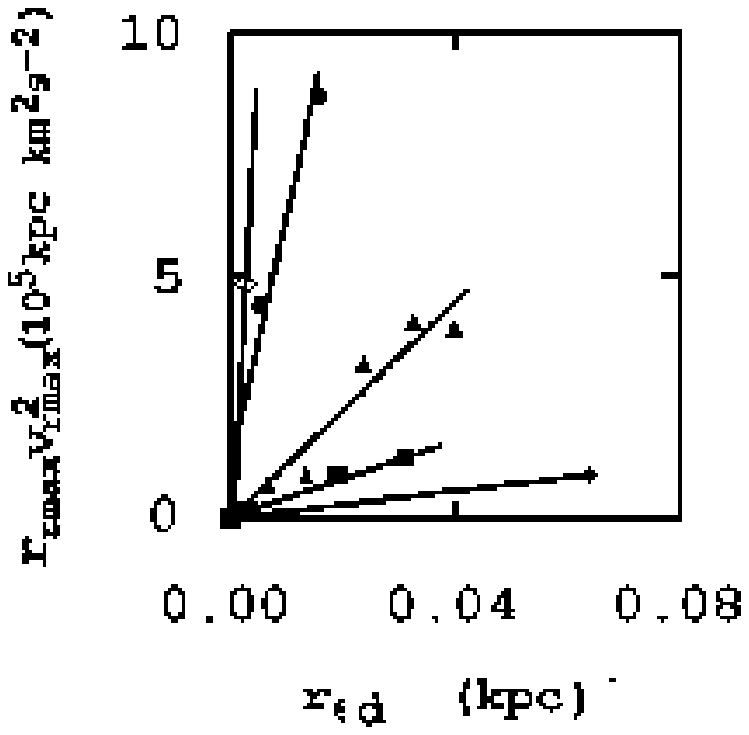}
\caption{\label{fig:11} Plot of the measured $ r_\mathrm{rmax} v^2_\mathrm{rmax}$ ($10^5 \,$kpc$\,$km$^2 \,$s$^{-2}$) versus the distance $r_\mathrm{\epsilon d}$ (kpc) along the major axis of a HST image to the first discontinuity or inflection point  using the $D_\mathrm{c}$ stated in Table~\ref{tab:4}.  The filled diamonds, filled squares, filled triangles, filled circles, and open diamond are data points along a linear $ r_\mathrm{rmax} v^2_\mathrm{rmax}$ - $r_\mathrm{\epsilon d}$ line using data from Tables~\ref{tab:3} and \ref{tab:4} and correspond to $m=1$, $m=2$, $m=3$, $m=5$, and $m=6$ as listed in Table~\ref{tab:5}, respectively.  The m=6 line is drawn between NGC 0224 and ($r_\mathrm{\epsilon d}$,$ r_\mathrm{rmax} v^2_\mathrm{rmax}$) $=$ (0,0).  The error bars were omitted for clarity.}
\end{figure}

Figure~\ref{fig:11} is a plot of $ r_\mathrm{rmax} v^2_\mathrm{rmax}$ versus $ r_\mathrm{\epsilon d}$.  Because the sample is small for six galaxy line classifications (four of the six line classifications have less than four galaxies), the ($r_\mathrm{\epsilon d}$,$ r_\mathrm{rmax} v^2_\mathrm{rmax}$) $=$ (0,0) point was included in all classifications.  Table~\ref{tab:5} lists the calculated values for the lines of Fig.~\ref{fig:11}.
\begin{table}
\caption{\label{tab:5} Data of the lines in Figures~\ref{fig:11}. }
\footnotesize
\begin{ruledtabular}
\begin{tabular}{llllll}
{$m$\footnotemark[1]}
&{line\footnotemark[2]}
&{Corr.\footnotemark[3]} 
&{F Test} 
&{$K_\mathrm{srr}$\footnotemark[4] \footnotemark[6]}
&{$K_\mathrm{irr}$\footnotemark[5] \footnotemark[6]} \\
& & & & 10$^7 \,$km$^2 \,$s$^{-2}$ & 10$^4 \,$kpc$\,$km$^2 \,$s$^{-2}$ \\
\hline
1&filled diamonds&ND\footnotemark[7]&ND&\phantom{0}0.14&\phantom{-}0\\
2&filled squares&0.99+&0.99+&\phantom{0}0.42$\pm$0.03&\phantom{-}0.3$\pm$0.5\\
3&filled triangles&0.97&0.94&\phantom{0}1.1\phantom{0}$\pm$0.1&-0.4$\pm$3\\
4&ND\\
5&filled circles&0.97&0.96&\phantom{0}5.2\phantom{0}$\pm$0.9&\phantom{-}9\phantom{.0}$\pm$8\\
6&open diamond&ND&ND&16.8&\phantom{-}0\\
\end{tabular}
\end{ruledtabular}
\footnotetext[1]{The integer denoting the place of the line in the order of increasing $K_\mathrm{seff}$.}
\footnotetext[2]{The identification of the line symbol in Figures~\ref{fig:11}.}
\footnotetext[3]{The correlation coefficient.}
\footnotetext[4]{The least squares fit slope of the $ r_\mathrm{rmax} v^2_\mathrm{rmax}$ - $r_\mathrm{\epsilon d}$ lines.}
\footnotetext[5]{The least squares fit intercept of the $ r_\mathrm{rmax} v^2_\mathrm{rmax}$ - $r_\mathrm{\epsilon d}$ lines.}
\footnotetext[6]{The lines are calculated using ($r_\mathrm{\epsilon d}$,$ r_\mathrm{rmax} v^2_\mathrm{rmax}$) $=$ (0,0) as a data point.}
\footnotetext[7]{``ND'' means no data because there are only two points (one data point).}
\end{table}

The lines are straight lines.  Therefore, the equation for each line is 
\begin{equation}
r_\mathrm{rmax} v^2_\mathrm{rmax} = K_\mathrm{srr} r_\mathrm{\epsilon d} + K_\mathrm{irr}
\label{eq:57g},
\end{equation}
where $K_\mathrm{srr}$ and $K_\mathrm{irr}$ are the slopes and intercepts of the lines in Fig.~\ref{fig:11}, respectively.

Since $M_\mathrm{rmax} >0$, the intercepts of the lines in Fig.~\ref{fig:11} nearly intersect at ($r_\mathrm{\epsilon d}$,$r_\mathrm{rmax} v^2_\mathrm{rmax} $) $=$ (0,0), and $K_\mathrm{irr}$ is determined by neighboring galaxies, 
\begin{equation}
M_\mathrm{rmax} = K_\mathrm{r \epsilon} \epsilon = K_\mathrm{r \epsilon} K_\mathrm{ \epsilon d} r_\mathrm{\epsilon d}
\label{eq:57d},
\end{equation}
where $K_\mathrm{r \epsilon}$ is the proportionality constant.

The slopes among the lines in Table~\ref{tab:5} obey the relation 
\begin{equation}
\log_{10} \left ( \frac{ K_\mathrm{srr}}{ \mathrm{km}^2 \, \mathrm{s}^{-2} } \right ) = K_\mathrm{snr} m + K_\mathrm{inr} 
\label{eq:57e},
\end{equation}
where $K_\mathrm{snr}= 0.40 \pm 0.02$ and $K_\mathrm{inr} = 5.79 \pm 0.06$  at 1 $\sigma$ are the least squares fit of the slope and intercept of the linear relation of Eq.~(\ref{eq:57e}), respectively.  The $m$ is an integer of one to six depending on the line from Table~\ref{tab:5}.  The correlation coefficient and F test for Eq.~(\ref{eq:57e}) are 0.99 and 0.99, respectively.  Note, the lowest slope corresponds to m=1 in Eq.~(\ref{eq:57e}) and the lines nearly intersect at a point.  The $K_\mathrm{srr}$ is nearly independent of the distance used to calculate the $ r_\mathrm{rmax} v^2_\mathrm{rmax} $ and $r_\mathrm{\epsilon d}$ if the same distance is used for both.  The values of log$_{10}$(e)$\approx 0.434$ is within 1.7 $\sigma$ of $K_\mathrm{snr}$.  

For the $ r_\mathrm{rmax} v^2_\mathrm{rmax} - r_\mathrm{\epsilon d}$ relationship, 
\begin{equation}
K_\mathrm{srr} = \frac{m_\mathrm{g}}{m_\mathrm{\iota}} G K_\mathrm{r \epsilon} K_\mathrm{\epsilon d} - \frac{m_\mathrm{s}}{m_\mathrm{\iota}} G_\mathrm{s} K_\mathrm{\epsilon d} 
\label{eq:57j}.
\end{equation}

Therefore, restating Eq.~(\ref{eq:57e}) as a strong Principle of Repetition yields
\begin{equation}
 K_\mathrm{srr} = K_\mathrm{ser} \, e^m 
\label{eq:57h},
\end{equation}
where $K_\mathrm{ser} = (4.8 \pm 0.9) \times 10^5 $ km$^2 \,$s$^{-2}$ at 1 $\sigma$.

Combining Eqs.~(\ref{eq:57g}) and (\ref{eq:57h}) yields 
\begin{equation}
r_\mathrm{rmax} v^2_\mathrm{rmax} = K_\mathrm{ser} \, e^m \, r_\mathrm{\epsilon d}
\label{eq:57ad},
\end{equation}
where $K_\mathrm{irr} =0$.

The standard deviation of the $\chi^2$ function are the error bars of $ r_\mathrm{\epsilon d} $ listed in Table~\ref{tab:4}.  The $\chi ^2 = 11$ and $P_\chi \approx 70\%$ between Eq.~(\ref{eq:57ad}) and the measured $ r_\mathrm{rmax} v^2_\mathrm{rmax}$ (not including the $ (r_\mathrm{\epsilon d},r_\mathrm{rmax} v^2_\mathrm{rmax}) =(0,0) $ point).  Since $P_\chi$ is greater than the $ 5\%$ significance level, this test does not invalidate Eq.~(\ref{eq:57ad}).  The $ r_\mathrm{rmax} v^2_\mathrm{rmax} $ is a function of $ r_\mathrm{\epsilon d}$ and $m$.  The $ r_\mathrm{\epsilon d}$ is a parameter of each galaxy, the $m$ is a galaxy classification parameter, and the $e$ implies the Principle of Repetition is applicable to the $ r_\mathrm{rmax} v^2_\mathrm{rmax}$ parameter, which is proportional to the effective mass at $r_\mathrm{rmax}$. 

Since $ r_\mathrm{rmax} v^2_\mathrm{rmax}$ is at the outer edge of a $m_\mathrm{s} / m_\iota$ strata, it may be used to compare parameters among galaxies.  Posit that $ r_\mathrm{rmax}$ and $ v^2_\mathrm{rmax}$ are related to $ r_\mathrm{\epsilon d}$, also.

\begin{figure}
\includegraphics[width=0.4\textwidth]{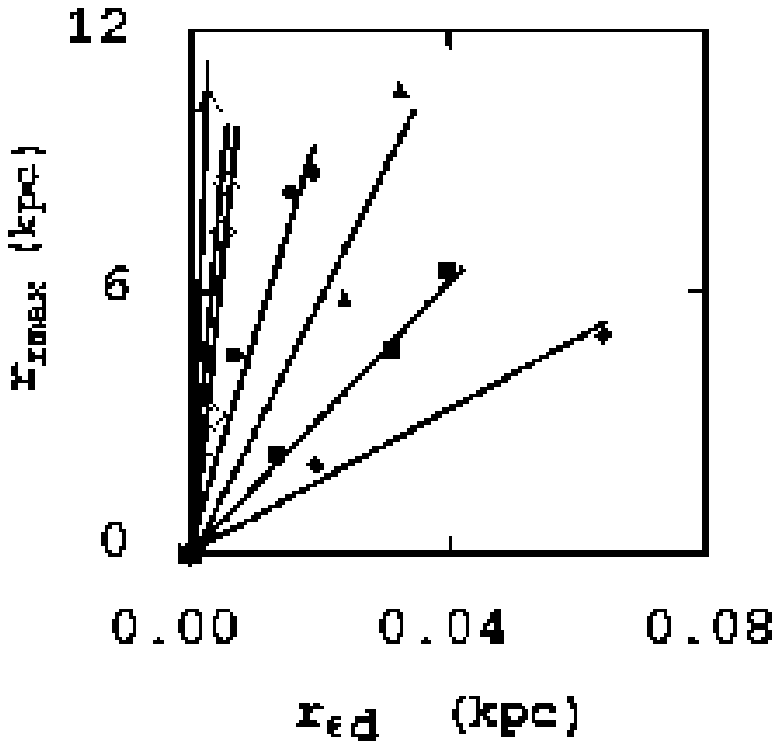}
\caption{\label{fig:12} Plot of the measured $ r_\mathrm{rmax}$ (kpc) versus the distance $r_\mathrm{\epsilon d}$ (kpc) along the major axis of a HST image to the first discontinuity or inflection point using the $D_\mathrm{c}$ stated in Table~\ref{tab:4}.  The filled diamonds, filled squares, filled triangles, filled circles, open diamonds, open square, and open triangle are data points along a linear $ r_\mathrm{rmax}$ - $r_\mathrm{\epsilon d}$ line using data from Tables~\ref{tab:3} and \ref{tab:4} and correspond to $p=1$, $p=2$, $p=3$, $p=4$, $p=5$, $p=6$, and $p=7$ as listed in Table~\ref{tab:6}, respectively.  The lines are drawn to include the point ($r_\mathrm{\epsilon d}$,$ r_\mathrm{rmax}$) $=$ (0,0).  The error bars were omitted for clarity.}
\end{figure}

Figure~\ref{fig:12} shows a plot of $r_\mathrm{rmax}$ versus $r_\mathrm{\epsilon d}$ for the sample galaxies in Tables~\ref{tab:3} and \ref{tab:4}.    Because the sample is small for seven galaxy integer $p$ classifications (all of the seven $p$ classifications have less than four galaxies), the ($r_\mathrm{\epsilon d}$,$ r_\mathrm{rmax} $) $=$ (0,0) point was included in all classifications.  The linear relationships are 
\begin{equation}
r_\mathrm{rmax} = K_\mathrm{srd} r_\mathrm{\epsilon d} + K_\mathrm{ird} 
\label{eq:57k},
\end{equation}
where $K_\mathrm{srd}$ and $ K_\mathrm{ird}$ are the slope and intercept of the $ r_\mathrm{rmax}$ - $r_\mathrm{\epsilon d}$ relationship, respectively, and the respective values are shown in Table~\ref{tab:6}.

\begin{table}
\caption{\label{tab:6} Data of the lines in Figures~\ref{fig:12}. }
\footnotesize
\begin{ruledtabular}
\begin{tabular}{llllll}
{$p$\footnotemark[1]}
&{line\footnotemark[2]}
&{Corr.\footnotemark[3]} 
&{F Test} 
&{$K_\mathrm{srd}$\footnotemark[4] \footnotemark[6]}
&{$K_\mathrm{ird}$\footnotemark[5] \footnotemark[6]} \\
& & & & & kpc\\
\hline
1&filled diamonds&0.99+&0.99+&\phantom{00}76 $\pm$ 9 &\phantom{-}0.2\phantom{0}$\pm$0.4\\
2&filled squares&0.99+&0.99+&\phantom{0}156 $\pm$ 8&\phantom{-}0.03$\pm$0.2\\
3&filled triangles&0.98&0.98&\phantom{0}310 $\pm$ 60&-0.3\phantom{0}$\pm$1\\
4&filled circles& 0.98&0.98&\phantom{0}459 $\pm$60&\phantom{-}0.6\phantom{0}$\pm$0.8\\
5&open diamonds&0.96&0.95&1500 $\pm$300&-0.6\phantom{0}$\pm$1\\
6&open square&ND\footnotemark[7]&ND&1620&\phantom{-}0\\
7&open triangle&ND&ND&3590&\phantom{-}0\\
\end{tabular}
\end{ruledtabular}
\footnotetext[1]{The integer denoting the place of the line in the order of increasing $K_\mathrm{srd}$.}
\footnotetext[2]{The identification of the line symbol in Figures~\ref{fig:12}.}
\footnotetext[3]{The correlation coefficient.}
\footnotetext[4]{The least squares fit slope of the $ r_\mathrm{rmax}$ - $r_\mathrm{\epsilon d}$ lines.}
\footnotetext[5]{The least squares fit intercept of the $ r_\mathrm{rmax}$ - $r_\mathrm{\epsilon d}$ lines.}
\footnotetext[6]{The lines are calculated using ($r_\mathrm{\epsilon d}$,$ r_\mathrm{rmax}$) $=$ (0,0) as one of the data points.}
\footnotetext[7]{``ND'' means no data because there are only two points (one data point).}
\end{table}

The relationship of the $ K_\mathrm{srd}$ values is 
\begin{equation}
\log_\mathrm{10} K_\mathrm{srd} = K_\mathrm{serd} p + K_\mathrm{ierd}
\label{eq:57f},
\end{equation}
where $K_\mathrm{serd} = 0.28 \pm .02$ and $ K_\mathrm{ierd} = 1.63 \pm 0.08$ at 1 $\sigma$.

Note, log$_{10}2 \approx 0.3010$.  The $p$ is a power of two rather than $e$.  This implies the strong Principle of Change is applicable and 
\begin{equation}
K_\mathrm{srd} = K_\mathrm{serdt} \, 2^p 
\label{eq:57l},
\end{equation}
where $K_\mathrm{serdt} = 36 \pm 8$ at 1 $\sigma$.

Combining Eqs.~(\ref{eq:57k}) and (\ref{eq:57l}) and setting $K_\mathrm{ird} =0$ yields 
\begin{equation}
r_\mathrm{rmax} = K_\mathrm{serdt} \, 2^p \, r_\mathrm{\epsilon d}
\label{eq:57ada}.
\end{equation}

The standard deviation of the $\chi^2$ function are the error bars of $ r_\mathrm{\epsilon d} $ listed in Table~\ref{tab:4}.  The $\chi ^2 = 10$ and $P_\chi \approx 76\%$ between Eq.~(\ref{eq:57ada}) and the measured $ r_\mathrm{rmax} $ (not including the $ (r_\mathrm{\epsilon d},r_\mathrm{rmax}) =(0,0) $ point).  Since $P_\chi$ is greater than the $ 5\%$ significance level, this test does not invalidate Eq.~(\ref{eq:57ada}).  The $ r_\mathrm{rmax}$ is a function of $ r_\mathrm{\epsilon d}$ and $p$.  The $ r_\mathrm{\epsilon d}$ is a parameter of each galaxy, the $p$ is a galaxy classification parameter, and the $2$ implies the Principle of Change is applicable to the $ r_\mathrm{rmax} $ parameter. 

Posit 
\begin{equation}
v^2_\mathrm{rmax} = K_\mathrm{srvd} r_\mathrm{\epsilon d} + K_\mathrm{irvd}
\label{eq:57m},
\end{equation}
where $K_\mathrm{srvd}$ and $ K_\mathrm{irvd}$ are the slope and intercept of the linear relation, respectively.

\begin{figure}
\includegraphics[width=0.4\textwidth]{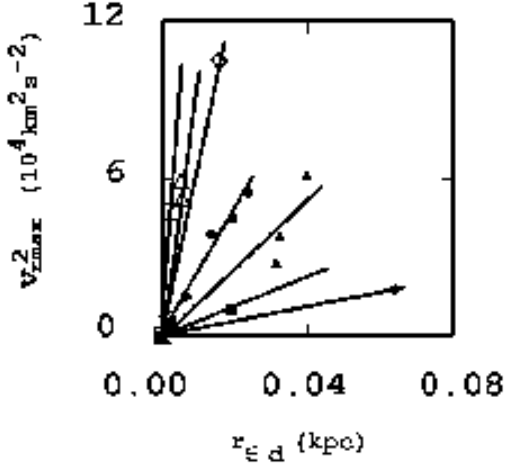}
\caption{\label{fig:13} Plot of the measured $ v^2_\mathrm{rmax}$ (km$^2 \,$s$^{-1}$) versus the distance $r_\mathrm{\epsilon d}$ (kpc) along the major axis of a HST image to the first discontinuity or inflection point using the $D_\mathrm{c}$ stated in Table~\ref{tab:3}.  The filled diamonds, filled squares, filled triangles, filled circles, open diamond, open triangles, and open square are data points along a linear $ v^2_\mathrm{rmax}$ - $r_\mathrm{\epsilon d}$ line using data from Tables~\ref{tab:3} and \ref{tab:4} and correspond to $s=1$, $s=2$, $s=3$, $s=4$, $s=5$, $s=6$, and $s=7$ as listed in Table~\ref{tab:7}, respectively.  The error bars were omitted for clarity.}
\end{figure}

Using the data from Tables~\ref{tab:3} and \ref{tab:4}, the plot of $ v^2_\mathrm{rmax}$ versus $r_\mathrm{\epsilon d}$ is shown in Fig.~\ref{fig:13} and the data for the lines is listed in Table~\ref{tab:7}.  Because the sample is small for seven galaxy integer $s$ classifications (five of the seven $s$ classifications have less than four galaxies), the ($r_\mathrm{\epsilon d}$,$ v^2_\mathrm{rmax}$) $=$ (0,0) point was included in all classifications.  The relationship of the slopes of the lines is 
\begin{equation}
\log_\mathrm{10} \left( \frac{K_\mathrm{srvd}}{\mathrm{km}^2 \, \mathrm{s}^{-2} \mathrm{kpc}^{-1}} \right) = K_\mathrm{servd} s + K_\mathrm{iervd}
\label{eq:57n},
\end{equation}
where $K_\mathrm{servd} = 0.31 \pm .02 $ and $ K_\mathrm{iervd} = 5.16 \pm 0.07 $ at 1 $\sigma$.

\begin{table}
\caption{\label{tab:7} Data of the lines in Figures~\ref{fig:13}. }
\footnotesize
\begin{ruledtabular}
\begin{tabular}{llllll}
{$s$\footnotemark[1]}
&{$s$\footnotemark[2]}
&{Corr.\footnotemark[3]} 
&{F Test} 
&{$K_\mathrm{srvd}$\footnotemark[4] \footnotemark[6]}
&{$K_\mathrm{irvd}$\footnotemark[5] \footnotemark[6]} \\
& & & & $10^6 $ kpc$\,$km$^2 \,$s$^{-2}$ & $10^3 $ km$^2 \, $s$^{-2}$ \\
\hline
1&filled diamonds&ND\footnotemark[7]&ND&\phantom{00}0.28&\phantom{-}0\\
2&filled squares&ND&ND&\phantom{00}0.54&\phantom{-}0\\
3&filled triangles&0.95&0.92&\phantom{00}1.3\phantom{0}$\pm$0.3&-1\phantom{.0}$\pm$ 7\\
4&filled circles&0.99&0.98&\phantom{00}2.3 $\pm$0.2&\phantom{-}1\phantom{.0}$\pm$ 3\\
5&open diamonds&0.99+&0.99+&\phantom{00}6.8 $\pm$0.1&-1\phantom{.0}$\pm$ 1\\
6&open triangles&0.99&0.97&\phantom{0}10.6 $\pm$0.04&\phantom{-}0.2$\pm$ 4\\
7&open square&ND&ND&\phantom{0}16&\phantom{-}0\\
\end{tabular}
\end{ruledtabular}
\footnotetext[1]{The integer denoting the place of the line in the order of increasing $K_\mathrm{srvd}$.}
\footnotetext[2]{The identification of the line symbol in Figures~\ref{fig:13}.}
\footnotetext[3]{The correlation coefficient.}
\footnotetext[4]{The least squares fit slope of the $ v^2_\mathrm{rmax}$ - $r_\mathrm{\epsilon d}$ lines.}
\footnotetext[5]{The least squares fit intercept of the $ v^2_\mathrm{rmax}$ - $r_\mathrm{\epsilon d}$ lines.}
\footnotetext[6]{The lines are calculated using ($r_\mathrm{\epsilon d}$,$ v^2_\mathrm{rmax}$) $=$ (0,0) as one of the data points.}
\footnotetext[7]{``ND'' means no data because there are only two points (one data point).}
\end{table}

The strong Principle of Change is applicable and 
\begin{equation}
K_\mathrm{srvd} = K_\mathrm{serds} \, 2^s 
\label{eq:57o},
\end{equation}
where $K_\mathrm{servds} = (1.6 \pm 0.3) \times 10^5$ km$^2 \,$s$^{-2} \,$kpc$^{-1}$ at 1 $\sigma$.

Combining Eqs.~(\ref{eq:57m}) and (\ref{eq:57o}) yields 
\begin{equation}
v^2_\mathrm{rmax} = K_\mathrm{servds} \, 2^s \, r_\mathrm{\epsilon d}
\label{eq:57ads},
\end{equation}
where $K_\mathrm{irvd} = 0$.

The standard deviation of the $\chi^2$ function are the error bars of $ r_\mathrm{\epsilon d} $ listed in Table~\ref{tab:4}.  The $\chi ^2 = 8.2$ and $P_\chi \approx 88\%$ between Eq.~(\ref{eq:57ads}) and the measured $ v^2_\mathrm{rmax}$ (not including the $ (r_\mathrm{\epsilon d},v^2_\mathrm{rmax}) =(0,0) $ point).  Since $P_\chi$ is greater than the $ 5\%$ significance level, this test does not invalidate Eq.~(\ref{eq:57ads}).  The $v^2_\mathrm{rmax} $ is a function of $ r_\mathrm{\epsilon d}$ and $s$.  The $ r_\mathrm{\epsilon d}$ is a parameter of each galaxy, the $s$ is a galaxy classification parameter, and the power of two implies the Principle of Change is applicable to the $ v^2_\mathrm{rmax}$ parameter. 

A feature of Figs.~\ref{fig:11}, \ref{fig:12}, and \ref{fig:13} is the lack of data points in the upper right of the plots (galaxies with relatively high $r_\mathrm{\epsilon d}$ and high $r_\mathrm{rmax}$, $v^2_\mathrm{rmax}$, and $r_\mathrm{rmax} v^2_\mathrm{rmax}$ values).

\subsection{SR}

In the bulge, strong gravitational and Space forces separate the regions into strata by particle $m_\mathrm{s} / m_\mathrm{\iota}$ type.  In the T$_\mathrm{r-s}$, the decreasing height $H_\mathrm{trs}$ of a galaxy's cylindrical shells results in a lower slope of the plot of the square of the rotation velocity $v_\mathrm{trs}^2$ in the T$_\mathrm{r-s}$ versus radius $r_\mathrm{trs}$ in the T$_\mathrm{r-s}$ ($v_\mathrm{trs}^2$ - $r_\mathrm{trs}$ curve) than in the RR.

In the SR, the $M$ and $\epsilon$ forces are less than in the KR.  Therefore, the tendency to separate into separate $m_\mathrm{s}/m_\mathrm{\iota}$ regions is less than in the KR.  Therefore, the regions are less distinct.  The SR could be modeled with these independent regions as rings as is often done to analyze rotation curves.  Since the change of $m_\mathrm{s} / m_\mathrm{\iota}$ from one ring to the next is small (unlike in the bulge), the form of the fundamental equation of motion for each ring will change little.  Also, since the elemental composition of the stars is changing, the border of the rings may be blurred.  For purposes herein, to calculate the mass $M_\mathrm{s}$ within a sphere with a radius $r_\mathrm{s}$ in the SR, the SR is considered as a whole with a smoothly varying rather than stepwise changing $m_\mathrm{s} / m_\mathrm{\iota}$.

The thin disk of the SR is modeled as a cylinder of density $ D_\mathrm{s}$ and height $H_\mathrm{s}$, then the $ M_\mathrm{s}$ is 
\begin{equation}
M_\mathrm{s}= \left( D_\mathrm{s} H_\mathrm{s} 2 \pi \right)r_\mathrm{s} + M_\mathrm{\Delta s} 
\label{eq:76},
\end{equation}
where 
\begin{equation}
M_\mathrm{\Delta s} \equiv M_\mathrm{trs} - \left( D_\mathrm{s} H_\mathrm{s} 2 \pi \right)r_\mathrm{rmax}
\label{eq:77}
\end{equation}
and $ M_\mathrm{trs}$ is the mass inward of the SR.

Because the rotation appears as a rigid body, the projection of the $M_\mathrm{s}$ into the RR may be real and not just a mathematical convenience.

Inserting Eq.~(\ref{eq:76}) into Eq.~(\ref{eq:49}) and having circular orbits~\cite[page 725]{binn} yields the rotation velocity $ v_\mathrm{s}$ in the SR 
\begin{eqnarray}
v_\mathrm{s}^2 &=& \left(\frac{m_\mathrm{g}}{m_\mathrm{\iota}} G D_\mathrm{s} H_\mathrm{s} 2 \pi \right)r_\mathrm{s} \nonumber \\*
& & + \frac{m_\mathrm{g} G M_\mathrm{\Delta s} - m_\mathrm{s} G_\mathrm{s} \epsilon}{m_\mathrm{\iota} r_\mathrm{s}} - \bm r_\mathrm{s} \bullet \bm a_\mathrm{o}  
\label{eq:78}.
\end{eqnarray}

As in the RR, the $r_\mathrm{s}^{-1}$ term is nearly constant.  By averaging from side-to-side of the galaxy as before, the $\bm r_\mathrm{s} \bullet \bm a_\mathrm{o} $ term is reduced.  However, unlike in the RR, $D_\mathrm{s} H_\mathrm{s} < D_\mathrm{srr} H_\mathrm{srr}$.  Therefore, $\bm r_\mathrm{s} \bullet \bm a_\mathrm{o}$ is more significant.  In cases where there are other close galaxies, close galaxies with moderate $\epsilon$ values, several close galaxies asymmetrically arranged, or a close galaxy with a high $\epsilon$, the residual of the side-to-side variation in the $\bm r_\mathrm{s} \bullet \bm a_\mathrm{o} $ term may be relatively significant (see \citet{hodg2}).  Therefore, 
\begin{equation}
v_\mathrm{s}^2 \approx K_\mathrm{ss} r_\mathrm{s} + K_\mathrm{si}  -  \bm r_\mathrm{s} \bullet \bm a_\mathrm{o} 
\label{eq:79},
\end{equation}
where $K_\mathrm{ss}$ and $K_\mathrm{si}$ are the slope and intercept of the linear relationship, respectively.

By reasoning similar to the SRR region, the maximum rotation velocity $v_\mathrm{smax}$ in the SR is also a linear function of $\epsilon$.  Therefore, for the intrinsic galaxy with $\bm r_\mathrm{s} \bullet \bm a_\mathrm{o} \approx 0$ 
\begin{equation}
v_\mathrm{smax}^2= K_\mathrm{s \epsilon slope} r_\mathrm{\epsilon d} + K_\mathrm{sas} 
\label{eq:80},
\end{equation}
where 
\begin{equation}
K_\mathrm{s \epsilon slope} = \frac{m_\mathrm{g}}{m_\mathrm{\iota}} G K_\mathrm{\epsilon s} - \frac{m_\mathrm{s}}{m_\mathrm{\iota}} G_\mathrm{s} K_\mathrm{\epsilon d} 
\label{eq:81}
\end{equation}
and the constant $K_\mathrm{\epsilon s } = M_\mathrm{smax}/ r_\mathrm{\epsilon d}$.

However, unlike the RR, the last rotation velocity data point measured may not be $ v_\mathrm{smax}$ and will be strongly influenced by $\bm r_\mathrm{s} \bullet \bm a_\mathrm{o}$.

If the impact of the $\bm r_\mathrm{s} \bullet \bm a_\mathrm{o}$ is small, the RC in the SR is linear and rising.  Another result of a small $\bm r_\mathrm{s} \bullet \bm a_\mathrm{o}$ is that the difference of rotation velocity at a given radius from one side to the other of a galaxy is small.  However, if $\bm r_\mathrm{s} \bullet \bm a_\mathrm{o}$ has a value comparable to the other terms of Eq.~(\ref{eq:79}) and $\nabla^2 ( \bm r_\mathrm{s} \bullet \bm a_\mathrm{o} ) \approx 0$, the $\bm r_\mathrm{s} \bullet \bm a_\mathrm{o}$ term also increases approximately linearly.  Therefore, the SR RC will be approximately flat with a slope depending on the relative value of the terms.  If $\nabla^2 ( \bm r_\mathrm{s} \bullet \bm a_\mathrm{o} )$ is significant, the average rotation curve will decline and there will be significant asymmetry in the RC at a given $r_\mathrm{s}$.  Since $ \bm {r}_\mathrm{s} \bullet \bm a_\mathrm{o}$ is in the plane of the galaxy, the asymmetry due to neighboring galaxies found by \citet{hodg2} is in accordance with Eq.~(\ref{eq:79}).

Consider two adjacent ring volumes in the disk with different elemental composition between the rings and with \emph{uniform} elemental composition in each ring.  Within each ring, Eq.~(\ref{eq:51}) applies and the RC will be Keplerian.  Because of the differing elemental composition, the $v_\mathrm{s}$ between the rings obey Eq.~(\ref{eq:79}) and the outer ring will have a greater $v_\mathrm{s}$ than the inner ring if $ \bm r_\mathrm{s} \bullet \bm a_\mathrm{o} =0$.  However, the differing $m_\mathrm{s}/m_\iota$ is small.  Therefore, the detection of the slight decline within each ring may be difficult.

\subsection{DR}

The DR is where particles are no longer rotating around the galaxy.  The galaxy forces are comparable to the forces from other galaxies.  Equation~(\ref{eq:49}) reduces to 
\begin{equation}
v_\mathrm{d}^2 = r_\mathrm{d} \ddot{r_\mathrm{d}}  -  \bm r_\mathrm{d} \bullet \bm a_\mathrm{o d} 
\label{eq:82},
\end{equation}
where 
\begin{eqnarray}
\bm a_{\mathrm{o d} } &=& \sum_{j=1}^\mathrm{ all \, galaxies} \frac{ m_\mathrm{g} G M_{j} - m_\mathrm{s} G_\mathrm{s} \epsilon_{j}}{m_\mathrm{\iota} r_{ij}^3} \bm r_{ij} \nonumber \\*
& & + \sum_{k=1}^\mathrm{ Sinks} \frac{ m_\mathrm{g} G M_{k} + m_\mathrm{s} G_\mathrm{s} \eta_{k}}{m_\mathrm{\iota} r_{ik}^3} \bm r_{ik} 
\label{eq:82a}
\end{eqnarray}
and the sum is now over all galaxies.

Equation~(\ref{eq:82a}) has the form of the Newtonian gravitational force with a anti-gravitational like force, both centered at the center of a galaxy.

\section{\label{sec:other}OTHER RELATIONSHIPS}

\subsection{\label{sec:sample}SAMPLE DATA}

The sample of galaxies was drawn from the sample of \citet{free}, who present a sample of 31 spiral galaxies with distances $D_\mathrm{c}$ (Mpc) calculated using Cepheid variables (``$D_\mathrm{z}$'' in \citet{free}).  Since $r_\mathrm{\epsilon d}$ was desired, the sample required each galaxy to have a HST, optical FITS image of the central few kpc of the galaxy.  NGC 0925 lacked an adequate HST image and, therefore, was excluded from the sample.  NGC 2841 \citep{macr} was added to the sample since the $D_\mathrm{c}$ determining method was consistent with \citet{free} and because NGC 2841 is a potential falsifier of the MOND model \citep{bott}. 

The data for the 31 selected galaxies are shown in Tables~\ref{tab:3}, \ref{tab:8}, and \ref{tab:9}.  The morphology type data were obtained from the NED database.  The luminosity class code $lc$ and de Vaucouleurs radius $R_{25}$ data were obtained from the LEDA database.  The H{\scriptsize{I}} (21 cm) line width $W_{20}$ (km$\,$s$^{-1}$) at 20\% of the peak value corrected for inclination data were calculated from ``incl'' and ``w20'' parameters from the LEDA database.  The $m_\mathrm{b}$ =  ``a magnitude in the visible range'' - ``extinction'' from the NED database.  The methods and equations used to calculate the $F_\mathrm{p}$, $s$, $u$, $w$, and $x$ values are described in Section~\ref{sec:result}.

\begin{table*}
\caption{\label{tab:8}Data for the chosen galaxies without rotation curves.}
\begin{ruledtabular}
\begin{tabular}{lllllrrrrrrc}
Galaxy &morphology\footnotemark[1] &\phantom{00imag}HST\footnotemark[2] & &lc\footnotemark[3]&$D_\mathrm{c}$\footnotemark[4]&$r_\mathrm{\epsilon d}$\\
& &\phantom{00}image&filter& &Mpc&$\times 10^{-3}$kpc\\
\hline
{I1613}&IB(s)m&U4040201R&F555W&9.8&0.65&1.2$\pm$\phantom{1}0.3\\

{I4182}&SA(s)m&U6G24404R&F606W&8.1&4.49&5.0$\pm$\phantom{1}2.2\\

{N1326A}&SB(s)m:&U34L3401O&F555W&7.8&16.14&14.4$\pm$\phantom{1}7.8\\

{N1365}&(R')SBb(s)b    Sy1.8&U2KV010XT&F547M&1&17.95&30.5$\pm$\phantom{1}8.7\\

{N1425}&SA(rs)b&U29R5P02T&F606W&3.1&21.88&31.9$\pm$10.6\\

{N2090}& SA:(rs)b &U29R0W02T&F606W&3.9&11.75&19.8$\pm$\phantom{1}5.7\\

{N2541}&SA(s)cd        LINER&U29R0Z02T&F606W&6&11.22&9.1$\pm$\phantom{1}5.4\\

{N3351}&SB(r)b;HII     Sbrst&U67N3002R&F555W&3&10.00&17.4$\pm$\phantom{1}4.9\\

{N3368}&SAB(rs)ab;Sy   LINER&U6EAN504M&F606W&3&10.52&22.2$\pm$\phantom{1}5.1\\

{N3621}&SA(s)d&U29R1I02T&F606W&5.8&6.64&10.0$\pm$\phantom{1}3.2\\

{N3627}&SAB(s)b;LINER  Sy2  &U29R1K02T&F606W&3&10.05&55.3$\pm$\phantom{1}4.9\\

{N4496A}&SB(rs)m&U2690102T&F555W&5.9&14.86&14.3$\pm$\phantom{1}7.2\\

{N4536}& SAB(rs)bc      HII  &U2DT0601&F555W&1.9&14.93&114.1$\pm$\phantom{1}7.2\\

{N4639}&SAB(rs)bc      Sy1.8&U2NU0202P&F555W&3.8&21.98&135.2$\pm$10.7\\

{N4725}&SAB(r)ab pec   Sy2  &U67N4602R&F606W&2&12.36&28.1$\pm$\phantom{1}6.0\\

{N5253}& Im pec;HII     Sbrst&U2E68101T&F606W&ND&3.15&3.9$\pm$\phantom{1}1.5\\
\end{tabular}
\end{ruledtabular}

\footnotetext[1] {Galaxy morphological from the NED database.}  
\footnotetext[2] {Galaxy Hubble Space Telescope image from MAST.}  \footnotetext[3] {Galaxy data from the LEDA database.}  
\footnotetext[4] {The distance $D_\mathrm{c}$ (Mpc) to the galaxy from~\citet{free} unless otherwise noted.} 
\end{table*}

\begingroup
\squeezetable
\begin{table*}
\caption{\label{tab:9}Additional data for the chosen galaxies.}
\begin{ruledtabular}
\begin{tabular}{lrrrrc}
Galaxy &$W_{20}$\footnotemark[1]&$R_{25}$\footnotemark[1]&$F_\mathrm{p}$\footnotemark[2]\phantom{10}&$m_\mathrm{b}$&$s$  $u$  $w$  $x$\\
&km$\,$s$^{-1}$&kpc&$\times 10^6$&mag. \\
\hline
Galaxies with rotation curves\\
N0224& 548$\pm$10&21.4$\pm$1.4&ND\footnotemark[3]\phantom{100}&4.26&7  8  -  9\\

N0300& 233$\pm$11&5.7$\pm$1.6&3.22$\pm$0.93&8.95&3  6  5  5\\

N0598& 243$\pm$\phantom{1}8&8.1$\pm$0.5&ND\phantom{100}&6.09&5  8  -  8\\

N2403&279$\pm$18&11.0$\pm$1.8&2.98$\pm$0.86&8.77&4  6  5  6\\

N2841& 655$\pm$\phantom{1}7&15.5$\pm$2.0&1.04$\pm$0.30&10.09&5  5  3  7\\

N3031& 517$\pm$\phantom{1}4&11.8$\pm$1.0&0.98$\pm$0.28&7.75&6  6  4  8\\

N3198& 343$\pm$\phantom{1}7&15.6$\pm$2.6&2.89$\pm$0.84&10.87&1  3  2  4\\

N3319& 263$\pm$13&11.4$\pm$1.9&2.01$\pm$0.61&11.48&2  4  3  5\\

N4258& 461$\pm$\phantom{1}5&19.7$\pm$3.3&3.68$\pm$1.07&9.10&4  6  4  7\\

N4321& 540$\pm$10&16.8$\pm$2.5&14.90$\pm$4.32&10.00&4  5  5  7\\

N4414& 509$\pm$\phantom{1}9&9.1$\pm$2.0&1.72$\pm$0.50&10.94&3  3  2  5\\

N4535& 423$\pm$12&15.9$\pm$1.4&6.41$\pm$1.86&10.59&3  4  4  6\\

N4548& 449$\pm$31&12.4$\pm$1.1&0.81$\pm$0.23&10.90&3  4  2  5\\

N5457& 502$\pm$12&30.1$\pm$6.7&3.14$\pm$0.91&8.31&6  7  5  9\\

N7331& 538$\pm$\phantom{1}7&22.9$\pm$3.4&1.00$\pm$0.29&10.02&4  5  3  7\\

Galaxies without rotation curves\\
{I1613}& 68$\pm$\phantom{1}6&1.5$\pm$0.3&ND\phantom{100}&9.86&1  5  -  2\\

{I4182}& 134$\pm$11&3.9$\pm$0.3&22.80$\pm$6.61&13.00&1  5  7  1\\

{N1326A}& 130$\pm$16&4.0$\pm$0.9&24.90$\pm$7.22&13.77&1  3  6  3\\

{N1365}&465$\pm$15&28.6$\pm$2.5&1.66$\pm$0.48&10.32&3  5  3  7\\

{N1425}& 392$\pm$10&19.2$\pm$2.5&3.48$\pm$1.01&11.87&3  4  3  5\\

{N2090}&319$\pm$10&9.8$\pm$2.9&1.60$\pm$0.46&11.99&2  4  3  4\\

{N2541}& 230$\pm$11&9.4$\pm$2.4&4.19$\pm$1.22&12.09&2  5  5  5\\

{N3351}& 424$\pm$\phantom{1}6&11.0$\pm$0.7&1.55$\pm$0.45&10.49&4  4  3  5\\

{N3368}& 439$\pm$14&11.9$\pm$1.0&32.70$\pm$9.48&10.06&4  4  6  6\\

{N3621}& 306$\pm$\phantom{1}6&11.3$\pm$2.5&0.05$\pm$0.01&9.79&3  5  1  6\\

{N3627}& 449$\pm$10&12.7$\pm$1.4&0.70$\pm$0.20&9.64&2  3  1  5\\

{N4496A}& 235$\pm$\phantom{1}7&8.2$\pm$0.5&33.20$\pm$9.63&11.93&2  4  6  5\\

{N4536}&394$\pm$14&15.7$\pm$1.7&19.50$\pm$5.66&11.16&1  2  4  3\\

{N4639}& 420$\pm$43&9.7$\pm$0.6&26.90$\pm$7.80&12.19&1  1  4  2\\

{N4725}& 510$\pm$14&19.7$\pm$1.3&19.70$\pm$5.71&10.08&4  5  5  6\\

{N5253}&101$\pm$14&2.2$\pm$0.2&0.58$\pm$0.173&10.68&1  4  4  4\\

\end{tabular}
\end{ruledtabular}

\footnotetext[1]{Galaxy data from the LEDA database.}  
\footnotetext[2]{The units of $F_\mathrm{p}$ are erg$\,$cm$^{-2} \,$s$^{-1} \,$Mpc$^{-2}$.}  
\footnotetext[3]{``ND'' means No Data because the Milky Way would be one of the close galaxies.}

\end{table*}
\endgroup

\begin{figure*}
\includegraphics[width=\textwidth, height=\textwidth]{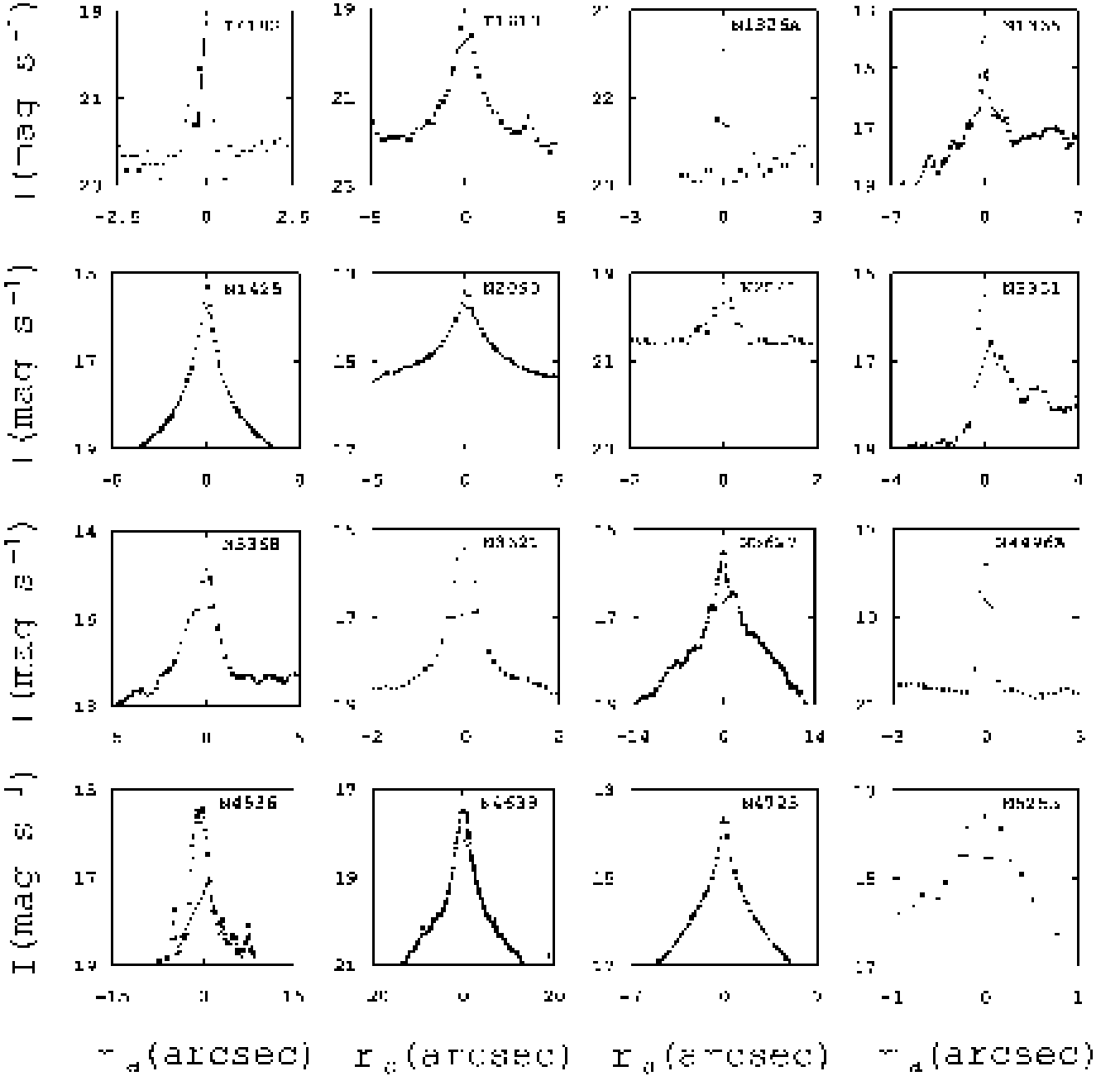}
\caption{\label{fig:14} Plots of the $I-r$ profile for the sample galaxies without a rotation curve.  The straight line is drawn between the chosen pixels.  The slope on the line indicates the asymmetry. }
\end{figure*}

The $r_\mathrm{\epsilon d}$ data listed in Table~\ref{tab:8} were calculated from $D_\mathrm{c}$ listed in Tables~\ref{tab:3}and \ref{tab:8} and from $r_\mathrm{a}$.  The $r_\mathrm{a}$ for each sample galaxy was measured from the HST images, which were obtained using the WFPC2 instrument, listed in Table~\ref{tab:8}.  The Visual Basic program listed in Appendix B was used to extract the $I-r$ curve and calculate $r_\mathrm{a}$.  The selection of each image was based upon being in the V band, upon having a sufficient area of the center of the galaxy in the image so $r_\mathrm{a}$ could be measured, and upon the $I$ value at $r_\mathrm{a}$ being sufficiently large.  Figure~\ref{fig:10} shows the $I-r$ curves for the galaxies with an RC.  The remaining $I-r$ curves are shown in Fig.~\ref{fig:14}.  The straight line in each $I-r$ curve of Fig.~\ref{fig:14} is drawn between the pixels the Visual Basic program selected to calculate $r_\mathrm{a}$.  Note, the line is at differing $I$ levels among the galaxies.  The slope of the line indicates asymmetry in $I$ value for the calculation of $r_\mathrm{a}$.  Therefore, the selection of $r_\mathrm{a}$ is not based on selecting an isophotal value.  The $r_\mathrm{a}$ is not an isophotal based effective radius.  

This sample has low surface brightness (LSB), medium surface brightness (MSB), and high surface brightness (HSB) galaxies; galaxies with a range of $W_{20}$ from 68 km s$^{-1}$ to 655 km$\,$s$^{-1}$; includes LINER, Sy, HII and less active galaxies; a $D_c$ range of from 0.65 Mpc to 21.98 Mpc; field and cluster galaxies; and galaxies with rising, flat, and declining RCs in the disk region.

\subsection{\label{sec:result}RESULTS}

\begin{figure}
\includegraphics[width=0.4\textwidth]{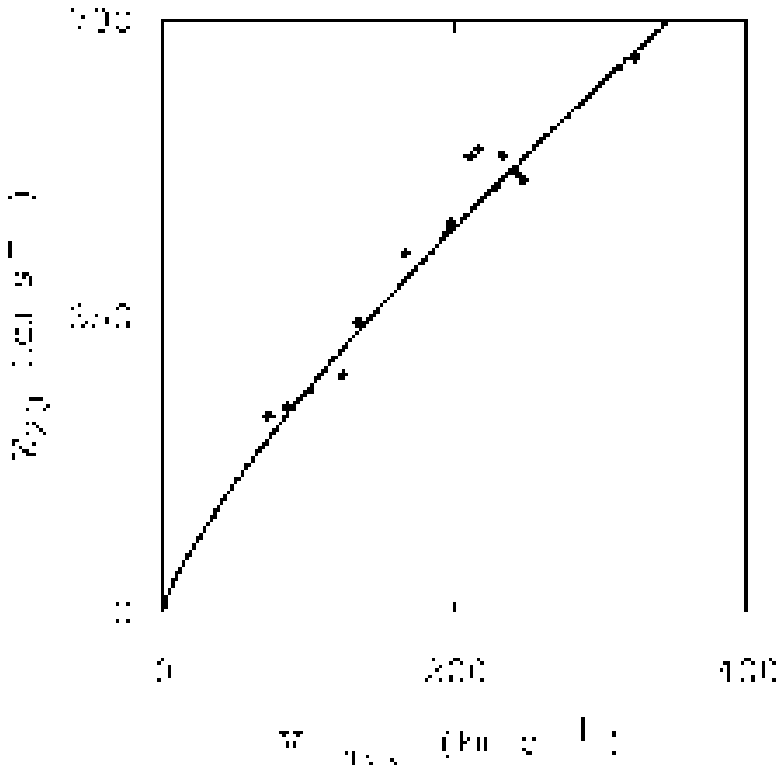}
\caption{\label{fig:15}Plot of $W_{20}$ versus $v_\mathrm{rmax}$.  The filled diamonds use the measured values of $W_{20}$ and $v_\mathrm{rmax}$.  The line is a plot of Eq.~(\ref{eq:101}).}
\end{figure}

Figure~\ref{fig:15} shows a plot of $W_{20}$ versus $v_\mathrm{rmax}$.  The filled diamonds use the measured values of $W_{20}$ and $v_\mathrm{rmax}$ of the galaxies with an RC.  The line is a plot of,
\begin{equation}
\frac{W_{20}}{\mathrm{km} \, \mathrm{s}^{-1}} = K_1 \, \left( \frac{v_\mathrm{rmax}}{\mathrm{km} \, \mathrm{s} ^{-1}} \right)^{K_2}\, \, (\pm 12\%)
\label{eq:101},
\end{equation}
where $K_1 = 7.3$ and $K_2 = 0.78$.

Equation~(\ref{eq:57ads}) relates $r_\mathrm{\epsilon d}$ and $v_\mathrm{rmax}$.  The $s$ for each galaxy in Table~\ref{tab:9} without a RC was chosen to provide the data point closest to the line of Eq.~(\ref{eq:101}).

The uncertainties of $ r_\mathrm{\epsilon d} $ listed in Table~\ref{tab:1} were used to calculate the standard deviation of the $\chi^2$ function.  The $\chi ^2 = 8$ and $P_\chi \approx 86\%$ between Eq.~(\ref{eq:101}) and the measured $ W_{20} $.  Since $P_\chi$ is greater than the $ 5\%$ significance level, this test does not invalidate Eq.~(\ref{eq:101}).  The $W_{20}$ is a function of $v_\mathrm{rmax}$ that is a function of $ r_\mathrm{\epsilon d}$ and $s$.  The $ r_\mathrm{\epsilon d}$ is a parameter of each galaxy, the $s$ is a galaxy classification parameter, and the ``$2$'' implies the Principle of Change is applicable to the $W_{20}$ parameter. 

The $W_{20}$ is used by the Tully-Fisher relationship \cite{tull77,tull00} to calculate total absolute magnitude $M_\beta$ (mag.) in band $\beta$
\begin{equation}
M_\beta = K_{4 \beta} + K_{5 \beta} \, \log(W_{20})
\label{eq:104},
\end{equation}
where $ K_{4 \beta}$ and $ K_{5 \beta}$ are constants that depend on the band under consideration.  Therefore, $M_\beta$ is a function of $s$ and $r_\mathrm{\epsilon d}$.

The correlation coefficients of $W_{20}$ versus $r_\mathrm{rmax}$ and $ r_\mathrm{rmax} v^2_\mathrm{rmax}$, which is proportional to the effective mass within a sphere of radius $r_\mathrm{rmax}$, are 0.38 and 0.80, respectively.

\begin{figure}
\includegraphics[width=0.4\textwidth]{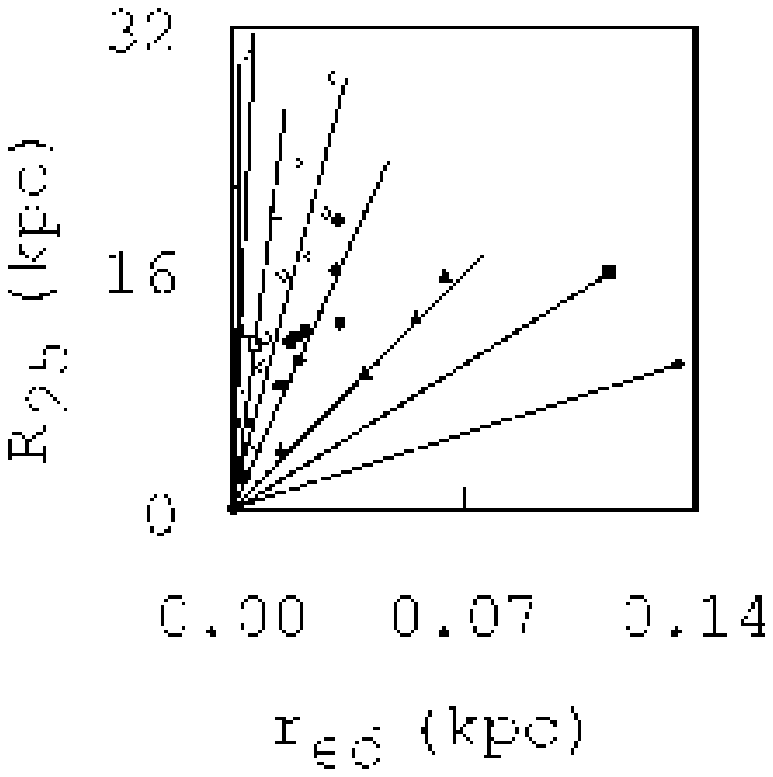}
\caption{\label{fig:16} Plot of the measured $ R_{25}$ (kpc) versus the distance $r_\mathrm{\epsilon d}$ (kpc) along the major axis of a HST image to the first discontinuity or inflection point using the $D_\mathrm{c}$ stated in Table~\ref{tab:3} and \ref{tab:8}.  The filled diamond, filled square, filled triangles, filled circles, open diamonds, open squares, open triangles, and open circle are data points along a linear $ R_{25}$ - $r_\mathrm{\epsilon d}$ line using data from Table~\ref{tab:9} and correspond to $u=1$, $u=2$, $u=3$, $u=4$, $u=5$, $u=6$, $u=7$, and $u=8$ as listed in Table~\ref{tab:10}, respectively.}
\end{figure}

Figure~\ref{fig:16} is a plot of $ R_{25}$ versus $ r_\mathrm{\epsilon d}$.  Each line in Fig.~\ref{fig:16} is a different ``classification''.  Because four of the eight classifications had less than four galaxies, the ($r_\mathrm{\epsilon d}$,$R_{25}$) $=$ (0,0) point was included in all classifications.  Table~\ref{tab:10} lists the calculated values for the lines of Fig.~\ref{fig:16}.
\begin{table}
\caption{\label{tab:10}Data of the lines in Figures~\ref{fig:16}. }
\footnotesize
\begin{ruledtabular}
\begin{tabular}{llllll}
{$u$\footnotemark[1]}
&{line\footnotemark[2]}
&{Corr.\footnotemark[3]} 
&{F Test} 
&{$K_\mathrm{sr25}$\footnotemark[4] \footnotemark[6]}
&{$K_\mathrm{ir25}$\footnotemark[5] \footnotemark[6]} \\
\hline
1&filled diamond&ND\footnotemark[7]&ND&\phantom{10}71&0\\
2&filled square&ND&ND&\phantom{0}138&0\\
3&filled triangles&0.99+&0.99+&\phantom{1}233$\pm$\phantom{10}9&1\phantom{.2}$\pm$1\\
4&filled circles&0.95&0.87&\phantom{1}490$\pm$\phantom{1}60&1\phantom{.2}$\pm$1\\
5&open diamonds&0.95&0.87&\phantom{1}800$\pm$100&1\phantom{.2}$\pm$2\\
6&open squares&0.97&0.95&1400$\pm$200&1\phantom{.2}$\pm$2\\
7&open triangles&0.99+&0.99+&5680$\pm$\phantom{1}80&0.2$\pm$0.3\\
8&open circle&ND&ND&7450&0\\
\hline
\end{tabular}
\end{ruledtabular}

\footnotetext[1]{The integer denoting the place of the line in the order of increasing $K_\mathrm{sr25}$.}  
\footnotetext[2]{The identification of the line symbol in Figures~\ref{fig:16}.}
\footnotetext[3]{The correlation coefficient.}  
\footnotetext[4]{The least squares fit slope of the $ R_{25} $ - $r_\mathrm{\epsilon d}$ lines.}  
\footnotetext[5]{The least squares fit intercept of the $ R_{25} $ - $r_\mathrm{\epsilon d}$ lines.}  
\footnotetext[6]{The lines are calculated using ($r_\mathrm{\epsilon d}$,$R_{25}$) $=$ (0,0) as a data point.}  
\footnotetext[7]{``ND'' means no data because there are only two points (one data point).}
\end{table}

The lines of Fig.~\ref{fig:16} are straight lines.  Therefore, the equation for each line is 
\begin{equation}
R_{25} = K_\mathrm{sr25} r_\mathrm{\epsilon d} + K_\mathrm{ir25}
\label{eq:105},
\end{equation}
where $K_\mathrm{sr25}$ and $K_\mathrm{ir25}$ are the slopes and intercepts of the lines in Fig.~\ref{fig:16}, respectively.

Since $K_\mathrm{ir25} \approx 0$, the slopes among the lines in Table~\ref{tab:10} obey the relation 
\begin{equation}
\log_{10} \left ( \frac{ R_{25}}{ r_\mathrm{\epsilon d} } \right ) = K_\mathrm{su} u + K_\mathrm{iu} 
\label{eq:106},
\end{equation}
where $K_\mathrm{su}= 0.30 \pm 0.02$ and $K_\mathrm{iu} = 1.52 \pm 0.08$  at 1$\sigma$ are the least squares fit of the slope and intercept of the linear relation of Eq.~(\ref{eq:106}), respectively.  The $u$ is an integer of one to eight depending on the line from Table~\ref{tab:10}.  The correlation coefficient and F test for Eq.~(\ref{eq:106}) are 0.99 and 0.98, respectively.  Note, the lowest slope corresponds to $u=1$ in Eq.~(\ref{eq:106}).  The $R_{25}/r_\mathrm{\epsilon d}$ is nearly independent of the distance used to calculate the $ R_{25} $ and $r_\mathrm{\epsilon d}$ if the same distance is used for both.  The value of log$_{10}$(2)$\approx 0.301$ is well within 1$\sigma$ of $K_\mathrm{su}$.  

Therefore, restating Eq.~(\ref{eq:106}) as a strong Principle of Change and combining Eqs.~(\ref{eq:105}) and (\ref{eq:106}) yields 
\begin{equation}
R_{25} = K_\mathrm{Rr} \, 2^u \, r_\mathrm{\epsilon d}
\label{eq:107},
\end{equation}
where $K_\mathrm{Rr} = (32 \pm 7) $ at 1$\sigma$ and $K_\mathrm{ir25} =0$.  Note the lack of data points in the upper right of Fig.~\ref{fig:16}.

The error bars of $ r_\mathrm{\epsilon d} $ listed in Tables~\ref{tab:4} and \ref{tab:8} were used to calculate the standard deviation of the $\chi^2$ function.  The $\chi ^2 = 17$ and $P_\chi \approx 96\%$ between Eq.~(\ref{eq:107}) and the measured $ R_{25} $ [not including the $ (r_\mathrm{\epsilon d},R_{25}) =(0,0) $ point].  Since $P_\chi$ is greater than the $ 5\%$ significance level, this test does not invalidate Eq.~(\ref{eq:107}).  The $R_{25}$ is a function of $ r_\mathrm{\epsilon d}$ and $u$.  The $ r_\mathrm{\epsilon d}$ is a parameter of each galaxy, the $u$ is a galaxy classification parameter, and the ``$2$'' implies the Principle of Change is applicable to the $ R_{25} $ parameter. 

If $\epsilon$ is the total energy entering a galaxy and luminosity ($10^{-0.4 \, M_j }$) is the total energy emitted from a galaxy, then $\epsilon \propto 10^{-0.4 \, M_j }$ in a stable galaxy, where $M_j$ is the total B band absolute magnitude of the $j^\mathrm{th}$ galaxy and B band luminosity is a fixed proportion of the total luminosity.  \citet{tull77} originally suggested luminosity is proportional to its effective mass of the galaxy.   Then the force $F_\mathrm{p}$ (erg$ \, $cm$^{-2} \, $s$^{-1} \,$Mpc$^{-2}$) exerted by the mass and $\rho$ of the neighboring galaxies on the target galaxy \citep{hodg1} along the polar axis is
\begin{equation}
\bm{F}_{\mathrm{p}} = K_\mathrm{fz} \sum_{j=1}^n 10^{-0.4 \, M_{j} } \frac{ 1}{ R_{ j}^3} \bm{Z}_{j}
\label{eq:108},
\end{equation}
where $K_\mathrm{fz}$ is the proportionality constant, the sum is over the closest galaxies, $n$ is the number of galaxies included in the calculation, $\bm{R}_{j}$ is the vector distance from the target galaxy to the $j^{th}$ close galaxy, and $\bm{Z}_j$ is component of $\bm{R}_{j}$ along the target galaxy's polar axis.  In the calculation herein, $K_\mathrm{fz} =1 $ and $n = 10$. 

The calculation followed \citet{hodg1}.  IC 1613, NGC 0224, and NGC 0300 were omitted from the calculation since the Milky Way is one of the closest galaxies and the $M$ of the Milky Way is uncertain.  The result of the calculation\footnote{The data and a sample calculation are available in a Microsoft Excel file by e-mail to scjh@citcom.net.} for 28 of the sample galaxies.  The $F_\mathrm{p}$ error was calculated using 0.2 mag error for $M_j$ and a 5\% error for distance.

\begin{figure}
\includegraphics[width=0.4\textwidth]{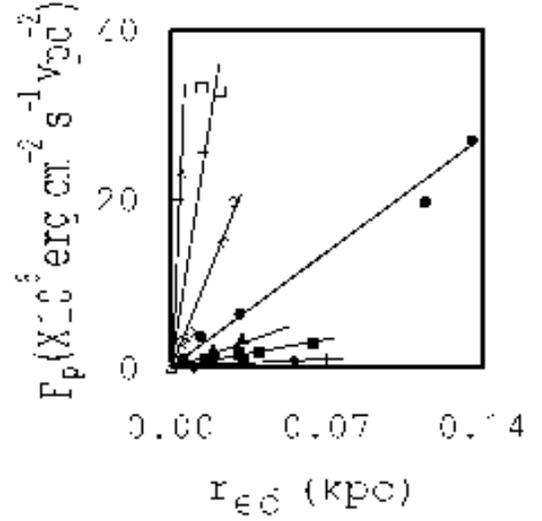}
\caption{\label{fig:17} Plot of the measured $ {F}_\mathrm{p} $ (erg$ \, $cm$^{-2} \, $s$^{-1} \,$ Mpc$^{-2}$) versus the distance $r_\mathrm{\epsilon d}$ (kpc) along the major axis of a HST image to the first discontinuity or inflection point using the $D_\mathrm{c}$ stated in Tables~\ref{tab:3} and \ref{tab:8}.  The filled diamonds, filled squares, filled triangles, filled circles, open diamonds, open squares, and open triangle are data points along a linear $ {F}_\mathrm{p}$ - $r_\mathrm{\epsilon d}$ line using data from Table~\ref{tab:9} and correspond to $w=1$, $w=2$,$ w=3$, $w=4$, $w=5$, $w=6$, and $w=7$ as listed in Table~\ref{tab:11}, respectively. }
\end{figure}

Figure~\ref{fig:17} is a plot of $ {F}_\mathrm{p}$ versus $ r_\mathrm{\epsilon d}$.  Each line in Fig.~\ref{fig:17} is a different ``classification''.  The ($r_\mathrm{\epsilon d}$,${F}_\mathrm{p} $) $=$ (0,0) point was included in all classifications.  Table~\ref{tab:11} lists the calculated values for the lines of Fig.~\ref{fig:17}.
\begin{table}
\caption{\label{tab:11}Data of the lines in Figures~\ref{fig:17}. }
\footnotesize
\begin{ruledtabular}
\begin{tabular}{llllll}
{$w$\footnotemark[1]}
&{line\footnotemark[2]}
&{Corr.\footnotemark[3]} 
&{F Test} 
&{$K_\mathrm{sf}$\footnotemark[4] \footnotemark[6]}
&{$K_\mathrm{iw}$\footnotemark[5] \footnotemark[6]} \\
\hline
1&filled diamonds&0.99&0.99&\phantom{40}13.2$\pm$\phantom{0}0.5&-0.04$\pm$0.03\\
2&filled squares&0.94&0.89&\phantom{40}42\phantom{.2}$\pm$\phantom{2}3&\phantom{-}0.7\phantom{0}$\pm$0.7\\
3&filled triangles&0.93&0.86&\phantom{40}97\phantom{.2}$\pm$\phantom{0}1&-0.07$\pm$0.04\\
4&filled circles&0.99&0.99&\phantom{0}185\phantom{.2}$\pm$\phantom{0}6&\phantom{-}0.4\phantom{0}$\pm$0.3\\
5&open diamonds&0.99&0.98&\phantom{0}680\phantom{.2}$\pm$30&-0.5\phantom{0}$\pm$0.6\\
6&open squares&0.93&0.91&1570\phantom{.2}$\pm$90&\phantom{-}3\phantom{.09}$\pm$2\\
7&open triangle&ND\footnotemark[7]&ND&4570&\phantom{-}0\\
\hline
\end{tabular}
\end{ruledtabular}

\footnotetext[1]{The integer denoting the value of $w$ in Eq.~\ref{eq:1011}.}  
\footnotetext[2]{The identification of the line symbol in Fig.~\ref{fig:17}.}  
\footnotetext[3]{The correlation coefficient.}  
\footnotetext[4]{The least squares fit slope of the lines in Fig.~\ref{fig:17}.  The units of $K_\mathrm{sf}$ are $\times 10^6 $ erg$ \, $cm$^{-2} \, $s$^{-1} \,$Mpc$^{-2} \,$kpc$^{-1}$.}  
\footnotetext[5]{The least squares fit intercept of the lines in Fig.~\ref{fig:17}.  The units of $K_\mathrm{iw}$ are $\times 10^6 $ erg$ \, $cm$^{-2} \, $s$^{-1} \,$Mpc$^{-2}$.}  
\footnotetext[6]{The lines are calculated using ($r_\mathrm{\epsilon d}$,${F}_\mathrm{p} $) $=$ (0,0) as a data point.}  
\footnotetext[7]{``ND'' means no data because there are only two points (one data point).}
\end{table}

The lines of Fig.~\ref{fig:17} are straight lines.  Therefore, the equation for each line is 
\begin{equation}
{F}_\mathrm{p} = K_\mathrm{sf} r_\mathrm{\epsilon d} + K_\mathrm{iw}
\label{eq:1010},
\end{equation}
where $K_\mathrm{sf}$ and $K_\mathrm{iw}$ are the slopes and intercepts of the lines in Fig.~\ref{fig:17}, respectively.

Unlike the previous $r_\mathrm{\epsilon d}$ relationships, $K_\mathrm{iw}$ deviates slightly from zero.  However, the slopes among the lines in Table~\ref{tab:11} obey the relation 
\begin{equation}
\log_{10} \left ( \frac{K_\mathrm{sf}}{\times 10^6 \, \mathrm{erg} \, \mathrm{cm}^{-2} \, \mathrm{s}^{-1} \, \mathrm{Mpc}^{-2} \,\mathrm{kpc}^{-1} } \right ) = K_\mathrm{sw} w + K_\mathrm{iw} 
\label{eq:1011},
\end{equation}
where $K_\mathrm{sw}= 0.41 \pm 0.01$ and $K_\mathrm{iw} = 0.73 \pm 0.05$ at 1$\sigma$ are the least squares fit of the slope and intercept of the linear relation of Eq.~(\ref{eq:1011}), respectively.  The $w$ is an integer of one to seven depending on the line from Table~\ref{tab:11}.  The correlation coefficient and F test for Eq.~(\ref{eq:1011}) are 0.998 and 0.996, respectively.  Note, the lowest slope corresponds to $w=1$ in Eq.~(\ref{eq:1011}).  The value of log$_{10}(\mathrm{e}) \approx 0.43$ is within 2$\sigma$ of $K_\mathrm{sw}$.  

Therefore, restating Eq.~(\ref{eq:1011}) as a strong Principle of Repetition and combining Eqs.~(\ref{eq:1010}) and (\ref{eq:1011}) yields 
\begin{equation}
{F}_\mathrm{p} = K_\mathrm{Fr} \, \mathrm{e}^w \, r_\mathrm{\epsilon d}
\label{eq:1012},
\end{equation}
where $K_\mathrm{Fr} = (4.5 \pm 0.8) \times 10^6 $ erg$ \, $cm$^{-2} \, $s$^{-1} \,$Mpc$^{-2} \,$kpc$^{-1}$ at 1$\sigma$ and $K_\mathrm{iw} =0$. 

The error bars of $ F_\mathrm{p} $ listed in Tables~\ref{tab:9} were used to calculate the standard deviation of the $\chi^2$ function.  The $\chi ^2 = 27$ and $P_\chi \approx 34\%$ between Eq.~(\ref{eq:1012}) and the measured ${F}_\mathrm{p} $ [not including the $ (r_\mathrm{\epsilon d}, {F}_\mathrm{p}) =(0,0) $ point].  Since $P_\chi$ is greater than the $ 5\%$ significance level, this test does not invalidate Eq.~(\ref{eq:1012}).  The ${F}_\mathrm{p}$ is a function of $ r_\mathrm{\epsilon d}$ and $w$.  The $ r_\mathrm{\epsilon d}$ is a parameter of each galaxy, the $w$ is a galaxy classification parameter, and the ``e'' implies the Principle of Repetition is applicable to the ${F}_\mathrm{p} $ parameter. 

Since $\epsilon = K_\mathrm{\epsilon d} \, r_\mathrm{\epsilon d}$, 
\begin{equation}
r_\mathrm{\epsilon d} = \frac{\epsilon}{ K_\mathrm{\epsilon d}} = \frac{ {F}_\mathrm{p}}{ K_\mathrm{Fr} \, \mathrm{e}^w }
\label{eq:1013},
\end{equation}
where $ K_\mathrm{\epsilon d}$ is the proportionality constant.

Since $\bm{Z}_{\mathrm{j}}$ may have any value, the continuous factors of the $r_\mathrm{\epsilon d}$ relationships may be due, in part, to $\bm{F}_\mathrm{p}$.

\begin{figure}
\includegraphics[width=0.4\textwidth]{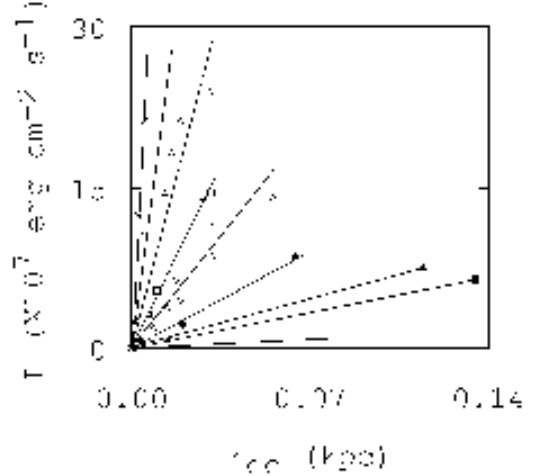}
\caption{\label{fig:18} Plot of the measured $ L$ (erg$\,$cm$^{-2} \,$s$^{-1}$) versus the distance $r_\mathrm{\epsilon d}$ (kpc) along the major axis of a HST image to the first discontinuity or inflection point using the $D_\mathrm{c}$ stated in Tables~\ref{tab:3} and \ref{tab:8}.  The filled diamond, filled squares, filled triangles, filled circles, open diamonds, open squares, open triangles, open circles, and ``X'' are data points along a linear $ L$ - $r_\mathrm{\epsilon d}$ line using data from Table~\ref{tab:9} and correspond to $x=1$, $x=2$, $x=3$, $x=4$, $x=5$, $x=6$, $x=7$, $x=8$, and $x=9$ as listed in Table~\ref{tab:12}, respectively.}
\end{figure}

Figure~\ref{fig:18} is a plot of galaxy luminosity $ L$ versus $ r_\mathrm{\epsilon d}$.  The $L$ value is calculated as
\begin{equation}
L = 10^{-0.4 \, M_\mathrm{V}}
\label{eq:1013a},
\end{equation}
where
\begin{equation}
M_\mathrm{V} = m_\mathrm{b} -25 +5 \, \log_{10}D_\mathrm{c}
\label{eq:1013b},
\end{equation}
where the ``-25'' and ``5'' in the equation are because the unit of $D_c$ is Mpc.

Each line in Fig.~\ref{fig:18} is a different ``classification''.  Because six of the nine classifications had less than four galaxies, the ($r_\mathrm{\epsilon d}$,$L$) $=$ (0,0) point was included in all classifications.  Table~\ref{tab:12} lists the calculated values for the lines of Fig.~\ref{fig:18}.
\begin{table}
\caption{\label{tab:12}Data of the lines in Figures~\ref{fig:18}. }
\footnotesize
\begin{ruledtabular}
\begin{tabular}{llllrr}
{$x$\footnotemark[1]}
&{line\footnotemark[2]}
&{Corr.\footnotemark[3]} 
&{F Test} 
&{$K_\mathrm{sLr}$\footnotemark[4] \footnotemark[6]}
&{$K_\mathrm{iLr}$\footnotemark[5] \footnotemark[6]} \\
\hline
1&filled diamond&ND\footnotemark[7]&ND&26\phantom{$\pm$000}&0.0\phantom{$\pm$0.0}\\
2&filled squares&0.99+&0.99+&48$\pm$\phantom{00}0&0.0$\pm$0.0\\
3&filled triangles&0.99+&0.99+&68$\pm$\phantom{00}1&-0.1$\pm$0.2\\
4&filled circles&0.99+&0.99+&133$\pm$\phantom{00}5&-0.1$\pm$0.2\\
5&open diamonds&0.97&0.92&285$\pm$\phantom{0}20&0.1$\pm$0.7\\
6&open squares&0.99+&0.99+&474$\pm$\phantom{0}18&0.2$\pm$0.4\\
7&open triangles&0.97&0.94&851$\pm$100&2\phantom{.0}$\pm$2\phantom{.0}\\
8&open circles&0.99+&0.99+&2100$\pm$100&-0.2$\pm$0.3\\
9&X&0.99+&0.99+&4000$\pm$200&0.3$\pm$0.6\\
\hline
\end{tabular}
\end{ruledtabular}

\footnotetext[1]{The integer denoting the place of the line in the order of increasing $K_\mathrm{sLr}$.}  
\footnotetext[2]{The identification of the line symbol in Figures~\ref{fig:18}.}  
\footnotetext[3]{The correlation coefficient.}  
\footnotetext[4]{The least squares fit slope (erg$\,$cm$^{-2} \,$s$^{-1} \,$kpc) of the $ L $ - $r_\mathrm{\epsilon d}$ lines. }  \footnotetext[4]{The least squares fit intercept (erg$\,$cm$^{-2} \,$s$^{-1}$) of the $ L $ - $r_\mathrm{\epsilon d}$ lines.}  
\footnotetext[6]{The lines are calculated using ($r_\mathrm{\epsilon d}$,$L$) $=$ (0,0) as a data point.}  
\footnotetext[7]{``ND'' means no data because there are only two points (one data point).}
\end{table}

The lines of Fig.~\ref{fig:18} are straight lines.  Therefore, the equation for each line is 
\begin{equation}
L = K_\mathrm{sLr} r_\mathrm{\epsilon d} + K_\mathrm{iLr}
\label{eq:105a},
\end{equation}
where $K_\mathrm{sLr}$ and $K_\mathrm{iLr}$ are the slopes and intercepts of the lines in Fig.~\ref{fig:18}, respectively.

Since $K_\mathrm{iLr} \approx 0$, the slopes among the lines in Table~\ref{tab:12} obey the relation 
\begin{equation}
\log_{10} \left ( \frac{ K_\mathrm{sLr}}{ \mathrm{erg}\,\mathrm{cm}^{-2} \,\mathrm{s}^{-1} \,\mathrm{kpc} } \right ) = K_\mathrm{sx} x + K_\mathrm{ix} 
\label{eq:106a},
\end{equation}
where $K_\mathrm{sx}= 0.274 \pm 0.007$ and $K_\mathrm{ix} = 1.08 \pm 0.04$  at 1$\sigma$ are the least squares fit of the slope and intercept of the linear relation of Eq.~(\ref{eq:106a}), respectively.  The $x$ is an integer of one to nine depending on the line from Table~\ref{tab:12}.  The correlation coefficient and F test for Eq.~(\ref{eq:106a}) are 0.99 and 0.99, respectively.  Note, the lowest slope corresponds to $x=1$ in Eq.~(\ref{eq:106a}).  The $L/r_\mathrm{\epsilon d}$ is not independent of the distance used to calculate the $ L $ and $r_\mathrm{\epsilon d}$.  The value of log$_{10}$(2)$\approx 0.301$ is within 4$\sigma$ of $K_\mathrm{sx}$.  There are only three galaxies on the $x=1$ and $x=2$ lines.  If the $x=1$ and $x=2$ lines are ignored, the $K_\mathrm{sx}= 0.29 \pm 0.03$ that is within 1$\sigma$ of log$_{10}$(2).

Therefore, restating Eq.~(\ref{eq:106a}) as a strong Principle of Change and combining Eqs.~(\ref{eq:105a}) and (\ref{eq:106a}) yields 
\begin{equation}
L = K_\mathrm{Lr} \, 2^x \, r_\mathrm{\epsilon d}
\label{eq:107a},
\end{equation}
where $K_\mathrm{Lr} = (8.9 \pm 2.0) $ erg$\,$cm$^{-2} \,$s$^{-1} \,$kpc at 1$\sigma$ and $K_\mathrm{ir25} =0$.  Note the lack of data points in the upper right of Fig.~\ref{fig:18}.

The error bars of $ r_\mathrm{\epsilon d} $ listed in Tables~\ref{tab:4} and \ref{tab:8} were used to calculate the standard deviation of the $\chi^2$ function.  The $\chi ^2 = 30$ and $P_\chi \approx 44\%$ between Eq.~(\ref{eq:107a}) and the measured $ L $ [not including the $ (r_\mathrm{\epsilon d},L) =(0,0) $ point].  NGC 4639 had a $\chi ^2 = 18$.  Since $P_\chi$ is greater than the $ 5\%$ significance level, this test does not invalidate Eq.~(\ref{eq:107a}).  The $L$ is a function of $ r_\mathrm{\epsilon d}$ and $x$.  The $ r_\mathrm{\epsilon d}$ is a parameter of each galaxy, the $x$ is a galaxy classification parameter, and the ``$2$'' implies the Principle of Change is applicable to the $ L $ parameter. 

The ${F}_\mathrm{p} $ is calculated from the magnitude, which is derived from $\epsilon$, and the vector position of other Sources.  Since ${F}_\mathrm{p} $ and M are independent variables, $ r_\mathrm{\epsilon d}$ of a galaxy depends on the $M_j$, hence $\epsilon$, of all galaxies.  Also, other galaxy parameters are related to $ r_\mathrm{\epsilon d}$.  This suggests the $\epsilon$ and the position in Euclidian space of all the Sources (center of galaxies) completely determines the parameters of a target galaxy.

\section{\label{sec:lines}The many lines in the parameter relations with $r_\mathrm{\epsilon d}$ are not random.}

The many lines in each plot in Figs.~\ref{fig:4}, \ref{fig:11}, \ref{fig:12}, \ref{fig:13}, \ref{fig:16}, \ref{fig:17}, and \ref{fig:18} may suggest the relationships portrayed may be just random data points.  The null hypothesis to be tested  is ``the data points are indeed random points''.  This null hypothesis may be tested by following the procedure used in discovering the Eqs.~(\ref{eq:57ac}), (\ref{eq:57ad}), (\ref{eq:57ada}), (\ref{eq:57ads}), (\ref{eq:107}), (\ref{eq:1012}), and (\ref{eq:107a})as follows:
(1) Generate 15 sets of two random numbers between zero and one.  Call one of the $i^\mathrm{th}$ set the independent variable ($X_i$) and the other the dependant variable ($Y_i$), where $i$ varies from one to 15.
(2) The equation to be tested is
\begin{equation}
Y_{\mathrm{calc} i} = K_1 \, 2^{t_i} \, X_i
\label{eq:57adt},
\end{equation}
where $K_1$ is the proportionality constant and $t_i$ is the integer classification for the $i^\mathrm{th}$ set.
(3) A trial consists of generating the random numbers for the 15 sets and subjecting the sets to the slope, correlation coefficient, and $\chi^2$ calculations and tests the galaxy samples passed.
(4) Calculate the $K_1$ value for each trial as the minimum value of $K_1 = Y_i/(2\,X_i)$ with an $X_i > 0.4$.
(5) Calculate the $t_i$ value for each set of ($X_i , Y_i$) values as
\begin{equation}
t_i = \mathrm{ROUND} \left[ \log_2 \, \left( \frac{Y_i}{k_1 \, X_i} \right) \right]
\label{eq:57adu},
\end{equation}
where ``ROUND'' means to round the value in the braces to the nearest integer.
(6) If any one of the $t_i > 7$, the trial failed.
(7) For calculating the slopes and correlation coefficients, if any subset lacks a sufficient number of values, that subset will be ignored.
(8) If the correlation coefficient of any $t_i$ subset of ($X_i ,Y_i$) values, including the ($X_0,Y_0$) = (0,0) point, is less than 0.95, the trial failed.
(9) If the linear slope of the relation between the $t_i$ and slope, including the ($X_0,Y_0$) = (0,0) point, of the $t_i$ subsets $\neq \, (0.30 \pm 0.03)$, the trial failed.
(10) If the correlation coefficient of the line in (9) above $< 0.95$, the trial failed.
(11) Calculate the $Y_{\mathrm{calc} i}$ according to Eq.~(\ref{eq:57adt}). If the $P_\chi < 0.05$ between the $Y_i$ and $Y_{\mathrm{calc} i}$ for the 15 sets, the trail failed.
(12) maintain a count of the number $N_\mathrm{\pi}$ of sets of ($X_i , Y_i$) with $X_i^2 + Y_i^2 \leq 1$ and the number $N_\mathrm{XltY}$ of sets of ($X_i , Y_i$) with $X_i < Y_i$.
(13) Redo the trial with another 15 sets of random numbers for 30,000 trials.

The results are: (1) The $N_\mathrm{\pi} /(15 \times 30,000) = 0.78528 \approx \pi/ 4$ and $N_\mathrm{XltY} /(15 \times 30,000) = 0.49999$.  Therefore, the random number generator performed satisfactorily. 
(2) 22 of the 30,000 trials passed (0.07\%), the remainder of the trials failed.  Therefore, the null hypothesis is rejected with a 99+\% confidence.

\section{\label{sec:disc}DISCUSSION}

Space has a physical presence, a density, and an ability to exert a force on the $\alpha$ of hods (matter) without being matter.  The Space in this Paper looks like the ether with added properties.  The Michelson-Morley experiments~\citep{mich} applied to the velocity of photons.  To achieve the results of Michelson-Morley's experiments, photons must have zero surface area presented to their direction of travel.  Hence, photons must be hods arranged in a column.  However, recent re-examination of the Michelson-Morley data~\cite{cons} and more accurate measurements~\cite{mull} suggests a preferred reference frame may exist.  The CUM suggests these findings indicate ripples on the hods of photons provide a dragging force on the photon.  Therefore, in addition to testing the ether model, Lorentzian relativity, and Einstein's special relativity, the Michelson-Morley type experiment may be testing the $\rho$, ${\bm\nabla}\rho $, and d($\bm \nabla \rho$)/d$t$ in our solar system.  The ratio of the diameter of the earth's orbit to the distance from the center of the Galaxy suggests the Michelson-Morley experiment may require greater precision and appropriate timing to test the $\rho$, ${\bm\nabla}\rho $ and d($\bm \nabla \rho$)/d$t$.  

\citet{fer4} discovered a linear relationship between circular velocity $v_c$ beyond $R_{25}$ and bulge velocity dispersion $\sigma_\mathrm{c}$ (see her Fig. 1).  \citet{baes} expanded on the data.  NGC 0598 was a clear outlier.  Galaxies with $\sigma_\mathrm{c}$ less than 70 km$\,$s$^{-1}$ ($v_c$ less than 150 km$\,$s$^{-1}$) also deviate from the linear relation.  Also, galaxies with significant non-circular motion in the HI gas such as NGC 3031 were omitted in \citet{fer4}.  The $v_c$ for NGC 0598 used in \citet{fer4} was 135 km$\,$s$^{-1}$, which is the highest data point of a rising RC~\cite{corb}.  Most of the other curves \citet{fer4} used were flat.  For flat RCs $v_c \approx v_{rmax}$.  If the $v_c = v_{rmax} = 85$ km$\,$s$^{-1}$ is used by \citet{fer4} for NGC 0598, NGC 0598 will fit on her plotted line.  Thus, the range of $v_c - {\sigma}_\mathrm{c}$ relation may extend to 27 km$\,$s$^{-1} \, < \,\sigma_\mathrm{c} \, < \, 350$ km$\,$s$^{-1}$.  NGC 3031 has a declining RC, $R_{25} \approx 11$ arcmin (from the LEDA database), $v_\mathrm{c}(R_{25}) \approx 190$ km$\,$s$^{-1}$ \cite{rots}, and $\sigma_\mathrm{c} = 174 \pm 17$ km$\,$s$^{-1}$ \cite{merr2}.  NGC 3031 is not on the plotted line of \citet{fer4}.  However, if $ v_\mathrm{c} = v_{rmax} = 243 \pm 14$ km$\,$s$^{-1}$ is used, NGC 3031 does fit on the plotted line.

Multiple slopes of the $C_\mathrm{a}$ versus $E_\mathrm{a}$ (erg cm$^{-2}$ s$^{-1}$) relation from \citet{hodg1} (see \citet{hodg1} for the definition and calculation of $C_\mathrm{a}$ and $E_\mathrm{a}$) also obey Eq.~(\ref{eq:57}).  Separate the CET1 zone into two zones.  The first zone (CET1a) has NGC 3621 and NGC 5253 which have low $E_\mathrm{a}$ and a slope of $-2.28 \times 10^{-2}$ erg$^{-2}$ cm$^{2}$ s$^{1}$.  The second zone (CET 1b) has NGC 3031 and NCG 2090 which have higher $E_\mathrm{a}$ and a slope of $8.87 \times 10^{-4}$ erg$^{-1}$ cm$^{2}$ s$^{1}$.  The slope values were arranged in ascending order with CET1b, CET2, CET3, and CET4 corresponding to n=1, n=2, n=3, and n=4, respectively.  For the $C_\mathrm{a}$ versus $E_\mathrm{a}$ relationship, $U_\mathrm{nit}$ is erg$^{-1}$ cm$^{2}$ s$^{1}$, $K_\mathrm{sn} = 0.585 \pm 0.005$  and $K_\mathrm{in} = -0.54 \pm 0.02$ at 1 $\sigma$.  The correlation coefficient and F test for Eq.~(\ref{eq:57}) are 0.99 and 0.99, respectively. 

The linear potential model (LPM) explored by \citet[and references therein]{mann} is derived from relativistic considerations.  The LPM proposes a gravitational potential $V(r)$ of the form $V(r) = - \beta / r +\gamma r / 2$ where $\beta$ and $\gamma$ are constants.  The potentials in the CUM are proportional to $r^{-1}$.  RCs with $v^2_\mathrm{srr} \propto r_\mathrm{srr}$ or $v^2_\mathrm{frr} \propto r^2_\mathrm{frr}$ are derived by geometric and particle type variations after the force of the $M_\mathrm{kmax}$ is balanced by the Space force.  However, both models conclude the intrinsic $v_\mathrm{srr}$ and $v_\mathrm{s}$ are a linear function of $r$.  Hence, both predict rising rotation curves.  The LPM derives flat curves by mass distribution but ultimately has rising curves.  If LPM is to derive declining curves, it is by the influence of other galaxies which have a linear $r$ potential.  The LPM does not predict the relations with $r_\mathrm{\epsilon d}$ or $v^2_\mathrm{frr} \propto r^2_\mathrm{frr}$.

The DM model (DMM) places large amounts of unknown, non-luminous matter in a halo around a galaxy to explain rising rotation curves.  Whereas, the DMM requires the DM term to become larger for rising rotation curves, CUM requires the $\bm r \bullet \bm a_\mathrm{o}$ term to become smaller.  The disagreement between current CDM theory and observation is most evident in galaxies with low surface brightness (LSB) such as NGC 0300 and rising rotation curves such as NGC 0300, NGC 0598, and NGC 3319.  These galaxies fit in the CUM.  In addition, the DMM would consider $v_\mathrm{rmax}$ as merely another point on the rotation curve.  The DMM suggests mass is infalling from outside the galaxy.  The CUM suggests matter is outflowing as gas and inflows as denser particles.  The DMM is not really a predictive model but rather a parameterization of the difference between observation and Newtonian expectation.  Because of the number of parameters in the DMM, it is very difficult to falsify.  However, the large number of parameters makes the DMM cumbersome and, therefore, replaceable.  

\citet[and references therein]{padm} explored the possibility of a scalar field acting as dark matter.  The CUM suggests the massless (mass density = 0) scalar $\rho$ field at small scales causes an intrinsic rising RC.  The DMM would require a large amount of DM.  The $\rho$ field acts like large amounts of DM at galaxy diameter scales.  Therefore, the CUM suggests the amount of DM in the universe is zero.  Also, this Paper suggest the $\rho$ field causes Newtonian kinematic measurements to considerably underestimate mass in a galaxy.  For example, Eq.~(\ref{eq:52}) with $M_\mathrm{eff} = M_\bullet$ suggests the larger $M_\mathrm{tckmax}$ is the more accurate measure of the amount of mass in the central region of galaxies.

The CUM's $( m_\mathrm{g} M_\mathrm{g} - m_\mathrm{s} \epsilon)/r$ term has the form of a subtractive change in acceleration.  The CUM suggests this causes the species of matter to change with radius which causes the appearance of a gravitational acceleration change with radius.  In the MOND model~\cite{bott}, the Newtonian gravitational acceleration is measured in our solar system, which is in the low $r$ end of the SR.  The MOND data comparison is usually in the far RR and SR regions.  Hence, the existence of a differing $\epsilon / r$ or an apparent modified gravitational acceleration is expected by the CUM model.  NGC2841 is a candidate to falsify MOND.  A distance greater than 19 Mpc, compared to the measured $D_\mathrm{c}$ of 14.1 Mpc~\cite{macr}, is needed for MOND~\cite{bott}.  However, NGC2841 data fits the CUM.  Further, given a relatively short radius range of the MOND application, the $\epsilon / r$ change may be approximated as a constant.  Also, since we are near the $v_\mathrm{rmax}$ in the Galaxy and since the CUM holds the relationship of $v^2_\mathrm{rmax} - \epsilon$ as a universal relationship, the MOND proposition that the modified gravitational acceleration is a universal constant follows.  If in the SR there is another acceleration from $\bm r \bullet \bm a_\mathrm{o}$ to balance the $\epsilon$ decline, then the MOND's flat rotation curve can result.  MOND also predicts that if the rotation curve is flat, it is axis symmetric.  Flat and declining rotation curves are asymmetric \cite[and references therein]{hodg2}.  Rising rotation curves are more symmetric.  Therefore, the CUM reduces to MOND where $\Delta \bm r \bullet \bm a_\mathrm{o} \neq 0$ and $\nabla ^ 2 (\bm r \bullet F_\mathrm{o}) =0$ in the SR. 

The model that places a SBH at the center of galaxies says nothing about the RC outside the KR.  Also, this model of the center of galaxies seems to be invalidated by observations of gas near the center of the Galaxy~\cite[page 595]{binn}\cite{koni}, by the apparent inactivity of the SBH~\cite{baga,baga2,naya,zhao}, by the observed relations with other galaxy parameters~\cite{fer6, gebh, fer4, mclu, fer5, merr, wand, grah,grah2,grah3}, and by the apparent need for a feedback control mechanism~\cite{merr2}.

The DMM adds several parameters per galaxy.  Both the LPM and MOND add one parameter to Newtonian mechanics and another mildly varying parameter to achieve their fits and have been applied to only close galaxies.  The CUM adds one parameter ($\epsilon$) per galaxy and one term per $m_\mathrm{s}$ species to Newtonian dynamics, but needs other galaxies and their distances to further explain a galaxy's dynamics.  The $\epsilon$ of other galaxies modify the equations to allow flat and declining RCs and to allow asymmetries in RCs.  Neither the LPM nor MOND address the relationships outside the RR and SR.  The DMM, LPM, MOND, and SBH galaxy models either are invalidated by or fail to explain the data examined herein.

I consider it very attractive that the CUM, at this early stage of model development, has brought together heretofore apparently unrelated observations and has discovered new galaxy parameter relations. 

\section{\label{sec:conc}CONCLUSION}

A changing universe model (CUM) was developed in accordance with a new set of Fundamental Principles.  The CUM posits the stuff of our universe is continually erupting into our universe from sources at the center of galaxies.  The stuff of our universe is matter and a matterless scalar whose gradient exerts a force on the cross section of matter.  An energy continuity equation (Section~\ref{sec:ener}) was derived.   Equations describing the forces of gravity (Section~\ref{sec:grav}) and Space (Section~\ref{sec:spac}) were derived.  A description of the structure of particles (Section~\ref{sec:part}) was presented to provide an explanation of the phenomena observed at the Galaxy center.  An explanation of Galaxy structure was presented.    The CUM qualitatively suggests: (1) That our universe is flat.  (2) The characteristic of the right angle is orthogonality.  (3) That total kinetic energy equals total potential energy in our universe.  (4) That $\bm{ \nabla}_\mathrm{m} \rho$, $\alpha$, $\tau$, and $\iota$ are constants of the universe.  (5) That Space is causally connected and coherent.  (6) Why there are three, and only three, Space dimensions.  (7) Why there are three, and only three, families of particles.

To test the proposition, a single equation was derived using the CUM that calculates the rotation velocity $v$ of particles along the major axis of spiral galaxies.  The equation adds one parameter for each mature galaxy and one parameter for each particle (atom nucleus and smaller) species to Newtonian dynamics.  The equation is used to correlate the measured mass $M_\bullet$ of the theorized, central supermassive black hole (SBH) with other galaxy parameters and to describe the rotation velocity curve (RC) of spiral galaxies.  The application of the CUM suggests galaxies are divided into regions.  The CUM suggests the intrinsic rotation curve in the disk region is rising and suggests flat and declining disk rotation curves are cause by neighboring galaxies.  Each region of a galaxy has its own physics.  The single equation, through the relative value of terms, reduces to a description of the physics of the RC of each region.  Four new parameters were defined: (1) the radius $r_\mathrm{\epsilon d}$ to a discontinuity in the surface brightness versus radius curve, (2) the rotation velocity $v_\mathrm{rmax}$ at the maximum extent of the rapidly rising region of the RC, (3) the radius $r_\mathrm{rmax}$ to the $v_\mathrm{rmax}$, and (4) the Source strength $\epsilon$ of each galaxy.

In a sample of 20 galaxies with the radius of influence of the central black hole resolved, this Paper predicted a linear $M_\bullet \, - \, r_\mathrm{\epsilon d}$ relation and found $M_\bullet = K_\mathrm{se} \, r_\mathrm{\epsilon d} \, e^\mathrm{n}$, where $K_\mathrm{se} = (3.7 ^{+0.0} _{- 0.9}) \times 10^8 \, M _\odot \,$kpc$^{-1}$, $e \approx 2.718$ (the natural log base), and $n$ is an integer from one to six depending on the galaxy.  Since the sample included E, SO, and S galaxies, the derived equation blends smoothly across the galaxy morphologies of the sample.

In a sample of 15 spiral galaxies with H{\scriptsize{I}} RCs and Cepheid based distances, this Paper predicted a linear $r_\mathrm{rmax} \, v^2_\mathrm{rmax} \, - \, r_\mathrm{\epsilon d}$ relation and found $r_\mathrm{rmax} \, v^2_\mathrm{rmax} = K_\mathrm{ser} \, r_\mathrm{\epsilon d} \, e^\mathrm{m}$, where $K_\mathrm{ser} = (4.8 \pm 0.9) \times 10^5 $ km$^2 \,$s$^{-2}$ and $m$ is an integer from one to six depending on the galaxy.  Also, this Paper predicted a linear $r_\mathrm{rmax} \, - \, r_\mathrm{\epsilon d}$ relation and found $r_\mathrm{rmax} = K_\mathrm{serdt} \, r_\mathrm{\epsilon d} \, 2^\mathrm{p}$, where $K_\mathrm{serdt} = 36 \pm 8$ and $p$ is an integer from one to seven depending on the galaxy.  Also, this Paper predicted a linear $v^2_\mathrm{rmax} \, - \, r_\mathrm{\epsilon d}$ relation and found $v^2_\mathrm{rmax} = K_\mathrm{servds} \, r_\mathrm{\epsilon d} \, 2^\mathrm{s}$, where $K_\mathrm{servds} = (1.6 \pm 0.3) \times 10^5$ km$^2 \,$s$^{-2} \,$kpc$^{-1}$ and $s$ is an integer from one to seven depending on the galaxy.  Further, for each of the four linear relationships, linear relationships between the respective parameter and $r_\mathrm{\epsilon d}$ for each of the integer classifications were also found.  The CUM predicts the square of the rotation velocity is linearly related to $r$ in a portion of the RR and is linearly related to $r^2$ in another portion of the RR.  

For a sample of 31 spiral galaxies with a wide range of characteristics, the H{\scriptsize{I}} (21 cm) line width $W_{20}$ (km$\,$s$^{-1}$) at 20\% of the peak value corrected for inclination has been found to be tightly related to $v_\mathrm{rmax}$ by $W_{20}/\mathrm{km} \, \mathrm{s}^{-1} = K_1 \, (v_\mathrm{rmax}/\mathrm{km} \, \mathrm{s}^{-1})^{K_2}$, where $K_1 = 7.3$, $K_2 = 0.78$, and the uncertainty in $W_{20}$ is $\pm 12\%$ with $P_\chi \approx 86\%$.  Since $W_{20}$ is related to the total absolute magnitude $M_\beta$ (mag.) in band $\beta$ by the Tully-Fisher relationship~\citep{tull77,tull00}, $M_\beta$ is also related to $ r_\mathrm{\epsilon d}$.  The de Vaucouleurs radius $R_{25}$ (kpc) is proportional to $ r_\mathrm{\epsilon d}$ by $R_{25} = K_\mathrm{Rr} \, 2^u \, r_\mathrm{\epsilon d}$ with $P_\chi \approx 96\%$, where $K_\mathrm{Rr} = (32 \pm 7) $ at 1$\sigma$ and $u$ is an integer from one to eight depending on the galaxy.  For 28 of the 31 sample galaxies, the force $F_\mathrm{p}$ of the $\epsilon$, which is proportional to the luminosity, of neighboring galaxies directed along the polar axis of the target galaxy is related to $r_\mathrm{\epsilon d}$ by $F_\mathrm{p} = K_\mathrm{Fr} \, \mathrm{e}^w \, r_\mathrm{\epsilon d}$ with $P_\chi \approx 34\%$, where $K_\mathrm{Fr} = (4.5 \pm 0.8) \times 10^6 $ erg$ \, $cm$^{-2} \, $s$^{-1} \,$Mpc$^{-2} \,$kpc$^{-1}$ at 1$\sigma$ and $w$ is an integer from one to seven depending on the galaxy.  This suggests the continuous factors of the $r_\mathrm{\epsilon d}$ relationships are due, in part, to $F_\mathrm{p}$.  For the 31 sample galaxies, the luminosity $L$ in the visible band (from the NED database) is proportional to $r_\mathrm{\epsilon d}$ by $L = K_\mathrm{Lr} \, 2^x \, r_\mathrm{\epsilon d}$ with $P_\chi \approx 44\%$, where $K_\mathrm{Lr} = (8.9 \pm 2.0) $ erg$\,$cm$^{-2} \,$s$^{-1} \,$kpc at 1$\sigma$ and $x$ is an integer from one to nine depending on the galaxy.  

The $r_\mathrm{\epsilon d}$ is related to galaxy dynamics parameters, luminosity parameters, and an isophotal radius parameter by the Principle of Change.  Also, $r_\mathrm{\epsilon d}$ is related to $F_\mathrm{p}$ and effective mass parameters by the Principle of Repetition.

The CUM is not invalidated by the data presented herein.

\begin{acknowledgments}
This research has made use of the NASA/IPAC Extragalactic Database (NED) which is operated by the Jet Propulsion Laboratory, California Institute of Technology, under contract with the National Aeronautics and Space Administration.

This research has made use of the LEDA database (http://leda.univ-lyon1.fr).

This research has made use of data obtained from the Multimission Archive at the Space Telescope Science Institute (MAST).  STScI is operated by the Association of Universities for Research in Astronomy, Inc., under NASA contract NAS5-26555.  Support for MAST for non-HST data is provided by the NASA Office of Space Science via grant NAG5-7584 and by other grants and contracts.

This research has made use of the NASA's Skyview facility located at NASA Goddard Space Flight Center.

I acknowledge and appreciate the financial support of Cameron Hodge, Stanley, New York, and Maynard Clark, Apollo Beach, Florida, while I was working on this project.
\end{acknowledgments}

\appendix

\section{\label{sec:princ}PRINCIPLES}

The Principal of Fundamental Principles is that a Fundamental Principle, and the models developed from it, is a meaningful and useful principle that applies to all levels of physical systems.  To be meaningful is to be able to be used to describe and predict the outcome of experiments.  To be useful is to be able to be used to cause desired outcomes.  An outcome of an experiment includes placing bounds on what (parameter or event), where (position coordinates), and when (time coordinate).  The level of a physical system refers to the size of the domain of applicability over which a set of physical theories applies such as galaxies versus atoms and Newtonian versus special relativity.

Corollary I is if a candidate to be a Fundamental Principle is found to not apply in some system then it is not a Fundamental Principle.  Corollary II is if a principle is found in all of physical systems, then it is likely to apply to larger and smaller levels and to new concepts.

The Principle of Superposition, the Correspondence Principle, and Principle of Minimum Potential Energy are such Fundamental Principles.  The Correspondence Principle is an interpolation of the Fundamental Principles.  Corollary II is an extrapolation of the Fundamental Principle.

A ``scientific model'' (theory) is derived from the transcendent idea of the Fundamental Principle and is applicable to a defined domain.  Since life and social systems are physical systems, by Corollary II and the Cosmological Principle, the transcendent idea of the Fundamental Principles must also apply to life and to social systems.  The more fundamental scientific models have larger domains.  A proposal becomes a scientific model when a deduction from it can be found to be factual (is observed).  This concept does not require a candidate model to include ``falsifiable predictions'' and does not invalidate a scientific model because of the existence of falsifying observations.  For instance, Newtonian dynamics is a valid scientific model.  Observations in the domain including relative velocities approaching the speed of light falsify Newtonian dynamics.  However, this only limits the Newtonian dynamics domain.  Religious ideology, models based on belief, and philosophy models may be scientific models provided they are useful and meaningful with restricted what, where, and when bounds.  To survive, a scientific model must compete for attention.  The concept of a scientific model survives because the human mind is limited in the ability to maintain a catalog of the nearly infinite number of possible observation.  Scientific models with empty or more limited domains have little usefulness and meaningfulness.

The Copernican Principle, which states we are not privileged observers of the universe, is accepted.  In quantum mechanics our presence may change the outcome of experiments.  However, this is true for any observer.  If, in some set of observations, we appear privileged, the privilege must be incorporated in the model.  For example, we are in a galaxy disk and are close to a sun.  We are in a highly unique and privileged area.  Just because we are carbon based does not imply all intelligent life is carbon based.

The Anthropic Principle is accepted to the extent that what is observed must have been created and have evolved to the present.  What is observed must be able to be observed.  Note this statement of the Anthropic Principle omits the requirement that it depend on an assumption of ``life'' and ``intelligence'' since life and intelligence are inadequately defined.  The existence of life, social systems, and intelligence are observations of our universe and, therefore, must be able to exist.  An unobserved parameter may or may not be able to be observed.  Therefore, the negative model candidates are not useful.  

Some argue \cite[and references therein]{smol} the Anthropic Principle cannot be part of science since it cannot yield falsifiable predictions.  In this statement of Fundamental Principles, what is observed can be used to define the domain of scientific models.

The Cosmological Principle is accepted to the extent that the physics must be the same at all positions in the universe.  Therefore, the theories must explain the conditions at all positions in the universe.  For example, the physics that states all objects fall to earth was found to be limited when Galileo noted moons rotating around Jupiter.  Newton's restatement of physics was more cosmological.  However, the Cosmological Principle is not extended to imply the universe is isotropic and homogeneous.  Near a star is a unique position.  The physics must explain other positions.  The physical theories must explain any isotropies and anisotropies.

The Principle of Change states that each system is built of previous systems by a minimum step change.  When the size of a structure of an object becomes limited, a new structure comprising a combination of existing elements can occur.  Change is a building on existing structures.  Destruction of objects into its parts does not foster the Principle of Change.  Since the minimum step Change is a split from one to two, a corollary (strong Principle of Change) is that whatever exists will split into two parts.  The two parts may be a continuation of what was plus a new system or may be what was incorporated into the two new systems.  Particles became a hydrogen atom followed by evolution of other atoms.  Atoms became molecules.  A model that requires a large step where there are possible intervening steps is not observed and is forbidden.

A corollary of the Principle of Change is that all things have a beginning.

The Principle of Repetition states that there are two ways to repeat a Change: (1) If conditions allow an observable change, then the change will occur again under similar conditions.  (2) The repeated Changes have a common cause (reproduction).  A corollary is that if two systems have the same observable results, then similar conditions exist or the systems were reproduced.  A strong statement of the Principle of Repetition is that the amount of increase of a parameter by the Principle of Repetition depends on the size of the parameter.  Destruction of objects to have ``room'' for ``the new'' is a Repetition because the only objects that can be built from the pieces are a Repetition of objects.

The Principle of Negative Feedback states that any system with a relatively narrow parameter relationships must evolve from a broader system and must have a negative feedback loop to maintain the narrow parameters and achieve balance between the Change and Repetition processes.  Otherwise, the system is unstable and transitory.  The effects of the unstable system will cease to exist without consequential fallout or permanent change.  We observe objects that have limited size.  So, there is a limiting force or negative feedback condition controlling the size of each object.  So too must black holes have a size limitation and a negative feedback condition.

The Principle of Minimum Potential Energy can be stated as a Principle of Minimum Action, which states the path of least energy expenditure will be followed during the change from one state to another.  Combining this with the Principle of Negative Feedback yields the conclusion that the feedback energy equations are linear, of the first time derivative, and of the first time integral.  In an equation none, one, or two of these elements may be zero.  Since a higher power feedback equation would result in the same controlled output and consume more energy, the minimum equations are sufficient.  Therefore, the assumption of ``linear'' in formulating descriptive, negative feedback energy equations is sufficient.

The Principle of Geometric Rules states that the observed geometric relationships apply in all levels of systems.  Hence, the conservation of energy/mass must be related to geometric rules we observe in our universe.  Hence, $\pi$ = circumference / radius in two dimensions must be the same number in three dimensions.  $\pi$ is not only a universal constant in our universe, it is a constant in all the universes.  The division by two is another universal concept.  The division by two for each dimension into equal angles yields the right angle.

\section{\label{sec:program}VISUAL BASIC 5.0 PROGRAM FOR FINDING $r_\mathrm{a}$.}

This program starts with the appropriate information extracted from the HST FITS file and the operator has input the required additional data.

\begin{verbatim}
Option Explicit
Const Pi As Single = 3.141592654
'% integer
'& long integer
'! single precision
'# double precision
'$ string
Dim i%, j%, x%, y%, k&, l&, m&, p%  
Dim d!(802, 802), dmax!, Xmax!, Ymax! 
Dim XL!, YL!, RAp#, DECp#, Dist#, PA! 
Dim Ap#, b0#, cp#, Xc#, DEC0#   
Dim XLv&, YLv&, XLvLast%, YLvLast%, RA0# 
Dim crVal1#, crVal2#, crPix1#, crPix2#
Dim NAXIS1%, NAXIS2%, ORIENTAT!, CD1_1#
Dim dmaxI!, XmaxI!, YmaxI!, CD1_2#, CD2_1#
Dim VegaZP#, ExpTime#, Area#, CD2_2#  
Dim Instrument$, Filter$, IData!(2270)
Dim AData#(2270), Sb!, TSlope!, R1p!
Dim SbO!, CountL%, CountH%, R1n!, Rea1!
Dim Rea2!, Sxy#, Sx#, Sx2#, Sy#
'crVal1, crVal2, crPix1, crPix2, ExpTime
'Instrument, Filter,
'NAXIS1, NAXIS2, ORIENTAT, CD1_1, CD1_2
'CD2_1, CD2_2
'are as defined in the HST FITSFILE header.
'VegaZP is the zero point of the surface
'brightness calculation, input by operator
'd(X,Y) is the data number (DN) 
'of the pixel
'at the X and Y image coordinate.
'Since the galaxy may be asymmetric,
'd(X,Y) is not averaged or 
're-calculated statistically.

Private Sub cmdDoOne_Click()
    PA = CInt(Text1.Text)	'position
' angle of galaxy, operator input
    Area = Abs(CD1_1 * CD1_2) * 3600 _
* 3600
'find center of gal. by highest pixel
' value close to input coordinates
    Xmax = Val(Text2.Text)
    Ymax = Val(Text3.Text)
    dmax = d(Xmax, Ymax)
DoAgain2:
    For x = Xmax - 20 To Xmax + 20
    For y = Ymax - 20 To Ymax + 20
        If x > 0 And y > 0 And x < _
NAXIS1 And y < NAXIS2 Then
        If d(x, y) > dmax Then
            dmax = d(x, y)
            Xmax = x
            Ymax = y
            GoTo DoAgain2
        End If
        End If
    Next
    Next
    Text2.Text = CStr(Xmax)
    Text3.Text = CStr(Ymax)
    PA = PA - ORIENTAT
' PA is now angle from X axis 
'counterclockwise on image and Y axis
'is up from bottom
DoAgain:
    If PA >= 180 Then    'not matter
' for tan function
        PA = PA - 180
        GoTo DoAgain
    End If
    If PA < 0 Then
        PA = PA + 180
        GoTo DoAgain
    End If
    If PA > 89.3 And PA < 90 Then  
'TAN(90) is problem
        PA = 89.3
    End If
    If PA < 90.7 And PA > 90 Then
        PA = 90.7
    End If
'record general information in Data1
' which is an Excel file
    Data1.Refresh
    Data1.Recordset.MoveFirst

    Data1.Recordset.Edit
    Data1.Recordset.Fields("NAME").Value_
 = Left(File1.filename, _
Len(File1.filename) - 4)
    Data1.Recordset.Update
    Data1.Recordset.MoveNext
    Data1.Recordset.Edit
    Data1.Recordset.Fields("NAME").Value _
= Label2.Caption + " deg."
    Data1.Recordset.Update
    Data1.Recordset.MoveNext
    Data1.Recordset.Edit
    Data1.Recordset.Fields("NAME").Value_
= Label3.Caption + " deg"
    Data1.Recordset.Update
    Data1.Recordset.MoveNext
    Data1.Recordset.Edit
    Data1.Recordset.Fields("NAME").Value_
= "IMAGE PA:  " + Format(PA, "##0.#")
    Data1.Recordset.Update
    Data1.Recordset.MoveNext
    Data1.Recordset.Edit
    Data1.Recordset.Fields("NAME").Value _
= "CNT: " + Format(d(CInt(Text2.Text), _
CInt(Text3.Text)), "######")
    Data1.Recordset.Update
    Data1.Recordset.MoveNext
    Data1.Recordset.Edit
    Data1.Recordset.Fields("NAME")_
.Value _
= "CNT X: " + Format(Val(Text2.Text)_
, "###")
    Data1.Recordset.Update
    Data1.Recordset.MoveNext
    Data1.Recordset.Edit
    Data1.Recordset.Fields("NAME")_
.Value _
= "CNT Y: " + Format(Val(Text3.Text)_
, "###")
    Data1.Recordset.Update
    Data1.Recordset.MoveNext
    Data1.Recordset.Edit
    Data1.Recordset.Fields("NAME").Value _
= "INSTRUMENT: " + Instrument
    Data1.Recordset.Update
    Data1.Recordset.MoveNext
    Data1.Recordset.Edit
    Data1.Recordset.Fields("NAME").Value_
 = "FILTER: " + Filter
    Data1.Recordset.Update
    Data1.Recordset.MoveNext
    Data1.Recordset.Edit
    Data1.Recordset.Fields("NAME").Value_
= "VegaZP:" + CStr(VegaZP)
    Data1.Recordset.Update
    Data1.Recordset.MoveNext
    Data1.Recordset.Edit
    Data1.Recordset.Fields("NAME").Value _
= "EXPOSURE TIME:" + CStr(ExpTime)_
    Data1.Recordset.Update
'initialize
    Data1.Recordset.MoveFirst
    For i = 1 To 2270
        IData(i) = 0
        AData(i) = 0
        Data1.Recordset.Edit
        Data1.Recordset.Fields("INDEX")_
.Value = 0
        Data1.Recordset.Fields("X"_
).Value = 0
        Data1.Recordset.Fields("Y")_
.Value = 0
        Data1.Recordset.Fields("DEGREES")_
.Value = 0
        Data1.Recordset.Fields("VALUE")_
.Value = 0
        Data1.Recordset.Update
        Data1.Recordset.MoveNext
    Next
    Data1.Recordset.MoveFirst
'coordinates of gal. center
    DEC0 = crVal1 + ((Val(Text2.Text) _
- crPix1) * CD1_1) + ((Val(Text3.Text)_
 - crPix2) * CD1_2)
    RA0 = crVal2 + ((Val(Text2.Text)_
 - crPix1) * CD2_1) + ((Val(Text3.Text)_
 - crPix2) * CD2_2)
'a change of location of center on 
'the image
    If Val(Text2.Text) > 0 Then
        crPix1 = Val(Text2.Text)
    End If
    If Val(Text3.Text) > 0 Then
        crPix2 = Val(Text3.Text)
    End If
    p = 0
    If PA < 45 Or PA > 135 Then
        For XL = 1 To NAXIS1 - 1 _
Step 0.25
            XLv = Int(XL)
            YLv = crPix2 + ((XL - crPix1)_
 * (Tan(PA * Pi / 180)))
            If YLv > 1 And YLv < NAXIS2 _
- 1 Then
                Record
            End If
        Next
    Else
        For YL = 1 To NAXIS2 - 1 _
Step 0.25
            YLv = Int(YL)
            XLv = crPix1 + ((YL - crPix2)_
 * (Tan((90 - PA) * Pi / 180)))
            If XLv > 1 And XLv < NAXIS1_
 - 1 Then
                Record
            End If
        Next
    
    End If
'record last record
    Data1.Recordset.Edit
    Data1.Recordset.Fields("X").Value_
 = CSng(NAXIS1)
    Data1.Recordset.Fields("Y").Value_
 = CSng(NAXIS2)
    Data1.Recordset.Fields("VALUE")_
.Value = 0
    Data1.Recordset.Fields("DEGREES")_
.Value = 0
    Data1.Recordset.Update
    Data1.Recordset.Close
    
    Rea
    Label4.Caption = "DONE: " _
+ File1.filename
End Sub

Sub Record()
    If XLv = XLvLast And YLv _
= YLvLast Then 'prevent 
'duplication of pixels
        GoTo NotDo
    End If
    XLvLast = XLv
    YLvLast = YLv
    p = p + 1
    IData(p) = d(XLv, YLv)
'calculate the Right Assension and
' DEClination of the pixel
    RAp = RA0 + ((crPix1 - XLv) * CD1_1_
) + ((crPix2 - YLv) * CD1_2)
    DECp = DEC0 + ((crPix1 - XLv) _
* CD2_1) + ((crPix2 - YLv) * CD2_2)
    Ap = RAp - RA0  'left in degrees
' for troubleshoot ease
    b0 = 90 - DEC0
    cp = 90 - DECp
'calculate the cos(angle from the 
'center to the pixel)
    Dist = (Sin(b0 * Pi / 180) * _
Sin(cp * Pi / 180) * Cos(Ap * Pi / 180))_
 + (Cos(b0 * Pi / 180) _
* Cos(cp * Pi / 180))

'calculate arccos of Dist
    If Abs(Dist) > 0.99999999999995_
 Then
        Dist = 0
    Else
        AData(p) = (Atn(-Dist / _
Sqr(-Dist * Dist + 1)) + 2 * Atn(1))_
 * 180 / Pi
    End If

'dist >0 unless changed
    If XLv < crPix1 And YLv <=_
 crPix2 Then_
        AData(p) = -AData(p)
    End If
    If XLv >= crPix1 And YLv <_
 crPix2 Then_
        AData(p) = -AData(p)
    End If
    If PA > 90 Then
        AData(p) = -AData(p)
    End If
'record DN pixel values
    Data1.Recordset.Edit
    Data1.Recordset.Fields("INDEX")_
.Value = CSng(p)
    Data1.Recordset.Fields("X")_
.Value = CSng(XLv)
    Data1.Recordset.Fields("Y")_
.Value = CSng(YLv)
    Data1.Recordset.Fields("VALUE")_
.Value = IData(p)
    Data1.Recordset.Fields("DEGREES")_
.Value = AData(p)
    Data1.Recordset.Update
    Data1.Recordset.MoveNext
NotDo:
End Sub

Function Slope2(A As Long) As Double
    If A = 0 Then
        Sxy = 0
        Sx = 0
        Sy = 0
        Sx2 = 0
        l = 0
    Else
        Sxy = Sxy + (IData(A) 
* AData(A))
        Sx = Sx + AData(A)
        Sy = Sy + IData(A)
        Sx2 = Sx2 + (AData(m + k)
 * AData(m + k))
        l = l + 1
        If Abs((l * Sx2) - (Sx * Sx))_
 < 0.000000001 Then
            Slope2 = 0
        Else
            Slope2 = ((l * Sxy) _
- (Sx * Sy)) / ((l * Sx2) - (Sx * Sx))
        End If
    End If
End Function

Sub Rea()
'calculates and records Rea data

'find index number (m) of the gal. 
'center, the center already chosen
redom:
    IData(0) = 0
    For k = 1 To p
        If IData(0) < IData(k) And _
IData(k) <= dmax Then
            IData(0) = IData(k)
            AData(0) = AData(k)
            m = k
        End If
    Next
    For k = 1 To p
        If k <= m And AData(k) > 0 Then
            AData(k) = -AData(k)
        End If
        If k > m And AData(k) < 0 Then
            AData(k) = -AData(k)
        End If
    Next
're-record data as surface brightness 
'and angular data in arc seconds
    Data1.Refresh
    Data1.Recordset.MoveFirst
    For k = 1 To p
        Data1.Recordset.Edit
        Data1.Recordset.Fields("INDEX")_
.Value = CSng(k)
        If IData(k) < 1 Then
            Data1
.Recordset.Fields("VALUE")_
.Value = CSng(25)
        Else
            Data1
.Recordset.Fields("VALUE")_
.Value = CSng(VegaZP - 2.5 * _
(Log(CDbl(IData(k)) / ExpTime / Area)_
 / Log(10#)))
        End If
        Data1
.Recordset.Fields("DEGREES")_
.Value = CSng(3600 * AData(k))
        Data1.Recordset.Update
        Data1.Recordset.MoveNext
     Next
    Data1.Recordset.Close

'pos going
'find starting k
    k = 0
    Do While IData(m + k) > 0.95 _
* IData(m)
        k = k + 1
    Loop
    If IData(m + k) < 0.8 * IData(m_
) Then k = -1
    CountH = 0
    TSlope = -10000000
    Sb = -20000000
redoSlopeP3:
    SbO = Slope2(0) 'reset slope
' numbers
    For j = 1 To 2  'initialize
' slope
        k = k + 1
        If IData(m + k) < _
IData(m + k - 1) Or k = 0 Then
            SbO = Slope2(m + k)
        Else
            j = j - 1
        End If
    Next
    CountL = 0
    Do While CountL < 10 And_
 IData(m + k) > 50 And (m + k + 2) > 1
        k = k + 1
        If IData(m + k) > 5 Then 
'for false data pixel
            Sb = Slope2(m + k)
            If SbO < Sb Then 
                CountL = CountL + 1
                CountH = 0
                SbO = Sb
            Else
                CountH = CountH + 1
                CountL = 0
                SbO = Sb
            End If
        Else
            IData(m + k) = 60
        End If
    Loop
    R1p = m + k - CountL

'neg going
'find starting k
    k = 0
    Do While IData(m + k) > 0.95 _
* IData(m)
        k = k - 1
    Loop
    If IData(m + k) < 0.8 * IData(m)_
 Then k = 1
    CountH = 0
    TSlope = 10000000
    Sb = 20000000
redoSlopen3:
    SbO = Slope2(0) 'reset slope
' numbers
    For j = 1 To 2  'initialize
' slope
        k = k - 1
        If IData(m + k) < _
IData(m + k + 1) Or k = 0 Then
            SbO = Slope2(m + k)
        Else
            j = j - 1
        End If
    Next
    CountH = 0
    Do While CountH < 10 And _
IData(m + k) > 50 And (m + k - 2) > 1
        k = k - 1
        If IData(m + k) > 5 Then 
'for false data pixel
            Sb = Slope2(m + k)
            If SbO < Sb Then 
                CountL = CountL + 1
                CountH = 0
                SbO = Sb
            Else
                CountH = CountH + 1
                CountL = 0
                SbO = Sb
            End If
        Else
            IData(m + k) = 60
        End If
    Loop
    R1n = m + k + CountH

    Rea1 = Abs(AData(R1p) - AData(R1n))
    If Abs(IData(R1p + 2)) < 10 Then
        Rea1 = Abs(AData(m) -_
 AData(R1n)) * 2
        IData(R1p) = 0
    End If
    If Abs(IData(R1n)) < 10 Then
        Rea1 = Abs(AData(m) -_
 AData(R1p)) * 2
        IData(R1n) = 0
    End If
'record found r_a in Data2 which 
'is an Excel file
    Data2.Refresh
    Data2.Recordset.MoveFirst
    Do While Data2
.Recordset.Fields("USED").Value = "Y"
        Data2.Recordset.MoveNext
    Loop
    If Data2.Recordset.Fields("USED")_
.Value <> "Y" Then
        Data2.Recordset.Edit
        If IData(m) < 1 Then
            Data2
.Recordset.Fields("PEAK").Value = 0_
        Else
            Data2.Recordset
.Fields("PEAK").Value = CSng(VegaZP - _
2.5 * (Log(CDbl(IData(m)) /_
 ExpTime / Area) / Log(10#)))
        End If
        If IData(R1n) < 1 Then
            Data2.Recordset.Fields_
("R1SBDPN").Value = 0
        Else
            Data2.Recordset.Fields_
("R1SBDPN").Value = CSng(VegaZP - 2.5_
 * (Log(CDbl(IData(R1n)) / _
ExpTime / Area) / Log(10#)))
        End If
        If IData(R1p) < 1 Then
            Data2.Recordset.Fields_
("R1SBDPP").Value = 0
        Else
            Data2.Recordset.Fields_
("R1SBDPP").Value = CSng(VegaZP -_
 2.5 * (Log(CDbl(IData(R1p)) /_
ExpTime / Area) / Log(10#)))
        End If
        Data2.Recordset_
.Fields("USED").Value = "Y"
        Data2.Recordset.Fields_
("SOURCE").Value = Left(File1_
.filename, Len(File1.filename) - 4)
        Data2.Recordset.Fields_
("2REA1").Value = Rea1
        Data2.Recordset.Fields_
("2REA2").Value = Rea2
        Data2.Recordset.Fields_
("PEAKDEG").Value = AData(m)
        Data2.Recordset.Fields_
("M").Value = m
        Data2.Recordset.Fields_
("R1N1VALUE").Value = AData(R1n)
        Data2.Recordset.Fields_
("R1P1VALUE").Value = AData(R1p)
        Data2.Recordset.Fields_
("NOPIX").Value = p
        Data2.Recordset.Fields_
("PA").Value = PA
        Data2.Recordset.Fields_
("INSTRUMENT").Value = Instrument
        Data2.Recordset.Fields_
("FILTER").Value = Filter
        Data2.Recordset.Fields_
("VEGAZP").Value = VegaZP
        Data2.Recordset.Fields_
("EXPTIME").Value = ExpTime
        Data2.Recordset.Update
    Else
        Stop	'got a problem 
'with the data file
    End If
    Data2.Recordset.Close
End Sub
\end{verbatim}


\end{document}